\documentclass[a4paper,fleqn,usenatbib]{mnras}

\usepackage[T1]{fontenc}
\usepackage{ae,aecompl}

\usepackage{graphicx}	
\usepackage{pdflscape}
\usepackage[flushleft]{threeparttable}
\usepackage{hyperref}

\title[Baryons in DM Haloes I]{Baryonic Distributions in Galaxy Dark Matter Haloes I: 
New Observations of Neutral and Ionized Gas Kinematics}

\author[E. E. Richards et al.]{
Emily E. Richards,$^{1}$\thanks{E-mail: er7@indiana.edu}
L. van Zee,$^{1}$\thanks{E-mail: lvanzee@indiana.edu}
K. L. Barnes,$^{1}$
S. Staudaher,$^{2}$
D. A. Dale,$^{2}$
\newauthor
T. T. Braun,$^{1}$
D. C. Wavle,$^{1}$
J. J. Dalcanton,$^{3}$
J. S. Bullock$^{4}$
and R. Chandar$^{5}$
\\
$^{1}$Indiana University, 727 East 3rd Street, Swain West 318, Bloomington, IN 47405, USA\\
$^{2}$University of Wyoming, 1000 E. University, Dept 3905, Laramie, WY 82071, USA\\
$^{3}$University of Washington, Box 351580, U.W., Seattle, WA 98195, USA\\
$^{4}$University of California, Irvine, Department of Physics \&~Astronomy, 
4129 Frederick Reines Hall, Irvine, CA 92697, USA\\
$^{5}$University of Toledo, 2801 West Bancroft Street, Toledo, OH 43606, USA
}

\date{Accepted XXX. Received YYY; in original form ZZZ}

\pubyear{2016}

\begin{document}
\label{firstpage}
\pagerange{\pageref{firstpage}--\pageref{lastpage}}
\maketitle

\begin{abstract}
We present a combination of new and archival neutral hydrogen (HI) observations and 
new ionized gas spectroscopic observations for sixteen galaxies in the statistically 
representative EDGES kinematic sample. HI rotation curves are derived from new 
and archival radio synthesis observations from the Very Large Array (VLA) as well as 
processed data products from the Westerbork Radio Synthesis Telescope (WSRT). 
The HI rotation curves are supplemented with optical spectroscopic integral field 
unit (IFU) observations using SparsePak on the WIYN 3.5 m telescope to constrain the 
central ionized gas kinematics in twelve galaxies. The full rotation curves of each galaxy 
are decomposed into baryonic and dark matter halo components using 3.6$\mu$m images from the 
{\it Spitzer Space Telescope} for the stellar content, the neutral hydrogen data for the 
atomic gas component, and, when available, CO data from the literature for the molecular 
gas component. Differences in the inferred distribution of mass are
illustrated under fixed stellar mass-to-light ratio (M/L) and maximum disc/bulge assumptions
in the rotation curve decomposition.
\end{abstract}




\section{Introduction}
\label{sec:intro}

The past three decades have led to many advances in our understanding of galaxy rotation curves 
and a staggering accumulation of observational data. Radio synthesis observations of the 21 cm
line of neutral hydrogen have historically been and continue to be the primary method for tracing the
outer gravitational potential of galaxies 
\citep[e.g.][]{velfields,Begeman91,Broeils,Sanders96,ursamajhi,things,whisp,btfrgas}. 
The advent of integral field spectroscopic (IFS) data has further improved
our ability to measure galaxy kinematics, especially in the central regions 
\citep[e.g.][]{atlas3dI,manga,califagaskin}. Using galaxy kinematics to decompose the rotation curves 
has long been a primary method for investigating the mass components of galaxies 
\citep[e.g.][]{bosmaphd,RubinFordSB,univrc}. The dynamically inferred missing matter from these early observations 
is now part of the $\Lambda$CDM cosmology, which provides a strong framework within which one can trace 
the evolution of small perturbations in the early universe to the diverse range of morphological types 
found in nearby galaxies \citep[e.g.][]{SFWlargescale,diversemorph,diskformcdm}. 

Although progress is being made in producing realistic galaxies at z = 0 in $\Lambda$CDM simulations
\citep[e.g.][]{eaglesubgrid}, well known tensions between the models and observations still exist.
For example, there is a persistent discrepancy between the diversity of rotation curves in observed 
galaxies for a given maximum rotation velocity, $V_{\rm max}$, and the little variation seen in simulated 
galaxies with the same $V_{\rm max}$. Low mass dwarf galaxies, in particular, typically have much
lower circular velocities in the inner regions than expected from $\Lambda$CDM, leading to a significant
mass deficit \citep{eagledwarfrc}. Rotation curve decomposition analysis continues to be a powerful 
observational tool for constraining theoretical predictions of the distribution of mass on galaxy scales.

Despite the mountain of kinematic observations, there are still challenges in constraining and 
interpreting the distributions of mass components in galaxies. 
The stellar mass, in particular, is difficult to estimate due to uncertainties in mass-to-light 
ratio (M/L) leading to the use of maximum disc fits to rotation curves \citep{maxdisk} despite evidence 
which suggests discs are submaximal \citep[e.g.][]{tfsubmax,diskmasssubmax}. 
Much attention has been given to determining the stellar M/L so that the
degeneracy between the scaling of the stellar contribution and dark matter halo model in
rotation curve decomposition analysis may be broken. There is relatively good consensus among different
methods estimating the stellar M/L at 3.6$\mu$m, which suggest a M/L of about 0.5 
\citep[e.g.][]{LMCML,btfrml,ml5btfr}. Estimates of the stellar M/L in the K-band result in a larger range
of values between $\sim$0.3-0.6, depending on the method used \citep[e.g.][]{etgtf,diskmassdm,Just15}.
Even with constraints on the stellar M/L, uncertainties in the distance estimates still hinder our
ability to accurately resolve the true baryonic contribution in the context of rotation curve
decomposition analysis (see Section~\ref{sec:distances}).

These challenges can cause considerable uncertainties in the results of individual galaxies.
Therefore, it is ideal to interpret such results in the context of a larger statistical sample.
We utilize a statistical sample of about 40 nearby galaxies defined from the Extended Disk Galaxy Explore 
Science (EDGES) Survey (\citealt{edges}) in an effort to investigate the distribution of mass in galaxies 
in the context of galaxy formation and evolution. 
Data for the EDGES Survey includes deep 3.6$\mu$m 
observations from the {\it Spitzer Space Telescope} of 92 galaxies spanning a wide range of morphology
(S0 to Im), luminosity (-14 $>$ M$_{\rm B}$ $>$ -21) and environment (cluster, group and isolated).
Galaxies in EDGES were selected to have distances between 2 -- 20 Mpc and include the Ursa Major
cluster, but exclude Virgo. We have defined a kinematic sub-sample from EDGES which includes all galaxies
that have intermediate inclination angles (between 30\degr~and 68\degr) estimated from optical axial ratios 
so that both accurate rotation curves and surface density profiles may be determined. The complete
kinematic sample preserves the unbiased representative nature of the full EDGES sample allowing
us to investigate correlations between the distribution of baryonic and non-baryonic matter in 
a statistical manner.

In this study, we are presenting galaxies in the kinematic sample for which we have new HI radio synthesis
observations from the VLA and new ionized gas kinematics from the SparsePak IFU
on the WIYN 3.5 m telescope. The observational data products are discussed in 
Section~\ref{sec:obsdata}. Rotation curve decomposition results are shown in 
Section~\ref{sec:results}. Section~\ref{sec:discussion} provides a discussion of the main
results which are summarized in Section~\ref{sec:conclusion}. Finally, notes on the individual galaxies
are provided in Appendix~\ref{sec:notes}.

\section{Observational Data}
\label{sec:obsdata}

In this paper, we present new and archival HI synthesis observations and new spectroscopic IFU
observations from SparsePak on the WIYN 3.5 m telescope to constrain the neutral and ionized
gas kinematics in sixteen galaxies. We use additional multiwavelength observations
to probe the stellar and gas content, including deep near-infrared (NIR) images taken at 3.6$\mu$m 
from the {\it Spitzer Space Telescope} to trace the extended stellar populations. Moderate depth 
optical broadband $B$ and $R$ and narrowband H$\alpha$ provide information about the dominant stellar 
populations and star formation activity. Finally, archival molecular gas observations 
complement the HI to better estimate the total gas content in the galaxies.
Tables~\ref{tab:obsprops}, \ref{tab:corrprops} and \ref{tab:radprops} provide a summary of observed, 
corrected and radial properties derived from this multifrequency dataset. All reported magnitudes are 
calculated using the Vega system. A discussion of distance estimates used
in the present study is given below in addition to brief summaries of the data acquisition and processing.

\subsection{Distance Estimates}
\label{sec:distances}
Many of the forthcoming results, including the primary results from the mass 
decomposition analysis, depend on the adopted distance to each galaxy.
For the mass decomposition, an accurate determination of the total baryon 
content in galaxies requires an accurate distance so that the mass surface densities are 
scaled correctly when they are converted into circular rotational velocities.
Uncertainty in the distance is often absorbed into the uncertainty of the stellar M/L, 
particularly for more massive galaxies where the total baryon mass is dominated by the stellar
component. In order to be able to compare the distribution of mass in galaxies at fixed stellar
M/L, we must rely on relatively accurate distance estimates to remove at least some of this
uncertainty.

For nearby galaxies, we cannot rely on Hubble flow distance estimates,
as their peculiar velocities may be large relative to their systemic velocities.
Therefore, we are limited to using independent distance estimates derived from methods
such as Type Ia supernovae (SNIa) and surface brightness fluctuations (SBF).
Six galaxies in the present study have distance estimates from these more
robust independent methods (see Table~\ref{tab:corrprops}). For the remaining 10
galaxies we adopt the Luminous Tully-Fisher Relation (LTFR) from \cite{ltfr}, so that
the galaxies are on a consistent distance scale. The sample from which the LTFR was
derived is a better match to the kinematic sample than other Tully-Fisher samples
\citep[e.g.][]{tfd}. It also does not introduce as much circularity into the analysis
as the baryonic Tully-Fisher relation (BTFR; \citealt{btfr}) since it uses luminosity 
at 3.6$\mu$m (Section~\ref{sec:3.6data}) rather than baryonic mass and is, 
therefore, independent of stellar M/L.

\begin{figure}
\includegraphics[width=\columnwidth]{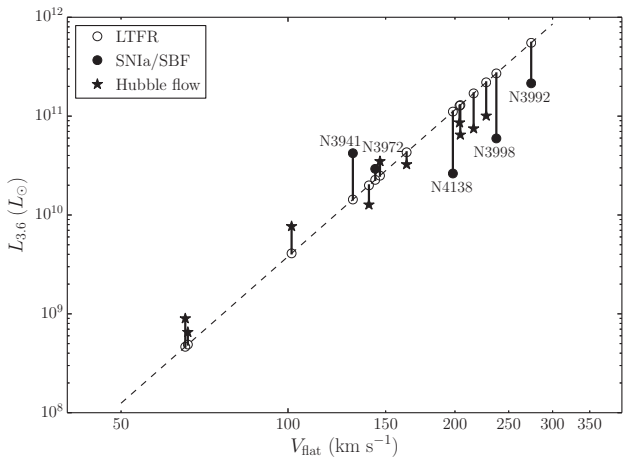}
\caption{Luminous Tully-Fisher Relation at 3.6$\mu$m from McGaugh \&~Schombert (2015b) (dashed line)
used to define distances to galaxies in the present sample with no independent distance estimate
from the literature (see text for details). Open circles indicate where the galaxies in this 
sample lie on the LTFR (with the exception of UGC~07639) given their estimated $V_{\rm flat}$. 
The filled circles show where the galaxies with independent distance estimates from
the literature would lie on this plot using their SNIa (NGC~3972 and NGC~3992) or SBF
(NGC~3941, NGC~3998 and NGC~4138) distances. Filled stars show where the Hubble flow distances
would lie for comparison.}
\label{fig:ltfr}
\end{figure}

Fig.~\ref{fig:ltfr} shows the LTFR for this sample along with the placement of 
the independent and Hubble flow distance estimates for comparison. The mean
offset is $\sim$0.3 dex with the largest differences between distance
estimates being on the order of $\sim$0.6-0.7 dex for NGC~3998 and NGC~3992. This
is likely due to the greatest limitation of the LTFR distance estimates, which is
the ability to define $V_{\rm flat}$. In general, $V_{\rm flat}$ was chosen from
an intermediate to outer radial range free from warps (Section~\ref{sec:kinematics}). 
$V_{\rm flat}$ was additionally
calculated following the method of \cite{ml5btfr} which uses a simple automated
algorithm to calculate the mean of the outermost rotation curve points until
it is no longer flat to within $\sim$5\%. The $V_{\rm flat}$ values calculated in
this manner are in excellent agreement with the estimated values reported in
Table~\ref{tab:corrprops}, except for the cases of NGC~3941, NGC~3992 and NGC~5055
where we have chosen to adopt a more interior value for $V_{\rm flat}$ to avoid
the influence of possible warps. An accurate rotation curve was not able to be
derived from UGC~07639 (see Section~\ref{sec:u7639}), so it is not included in
Fig.~\ref{fig:ltfr}. It is important to note that the galaxies which show the largest
discrepancy between the two distance estimates are the galaxies which have the most
trustworthy independent distances and the least trustworthy determinations
of $V_{\rm flat}$.

\subsection{New and Archival HI Observations}
\label{sec:hiobs}

Radio synthesis observations using the 21 cm line of HI were obtained with the 
VLA\footnotemark\footnotetext{The Very Large Array is operated by the National Radio 
Astronomy Observatory, which is a facility of the National Science Foundation 
operated under cooperative agreement by Associated Universities, Inc.} in C 
configuration in 2013 July - September for eight galaxies in the kinematic sample of the
EDGES survey.
All eight galaxies targeted for new observations either did not have archival HI synthesis 
observations or the existing data
were not adequate for rotation curve decomposition analysis. A complete analysis of one of these
galaxies, NGC~5005, is published in \cite{paperI}. In the current study, we present the new HI data 
for the seven remaining galaxies.

The VLA data had an initial frequency 
resolution of 7.8125 kHz channel$^{-1}$ (1.65 km s$^{-1}$ channel$^{-1}$). 
The standard flux calibrator 3C286 was observed at least once during every 
observing block, and a phase calibrator was observed approximately every 40 minutes, 
so that the data may be flux and phase calibrated. Galaxies were scheduled in one,
two, or four hour observing blocks during nighttime or early evening.

The data were loaded into AIPS\footnotemark\footnotetext{The Astronomical Image 
Processing System (AIPS) has been developed by the NRAO.} and processed following
the methods outlined in \cite{paperI}. Solar RFI was flagged on all channels simultaneously,
but was mostly not a problem. Three data cubes were created for each galaxy
with different 
Robust weighting parameters for varying spatial resolution and a channel averaging 
of 3 for a resulting velocity resolution of $\sim$5 km s$^{-1}$ channel$^{-1}$. 
When necessary, a lower velocity resolution cube (15 km s$^{-1}$ channel$^{-1}$) was 
created to probe the lowest column density of HI gas.
Synthesis imaging parameters for the relevant datacubes for each galaxy are presented
in Table~\ref{tab:cubeprops}. NGC~4389, UGC~07639, and UGC~08839 each had one hour blocks
that were discarded due to RFI issues. The time on source given in Table~\ref{tab:cubeprops} does 
not include these blocks. Many of the galaxies' observing blocks were scheduled at similar local
sidereal times on different nights resulting in a beam that is not round. This is
especially true for NGC~5608. 

In addition to the new VLA observations, eight of the galaxies for which we acquired 
ionized gas kinematics (see Section~\ref{sec:spak}) have archival HI observations adequate 
for rotation curve decomposition analysis. We used the processed data products from the 
Westerbork Survey of HI in 
Spiral Galaxies (WHISP; \citealt{whisp}) for the galaxies NGC~3726 and NGC~4051. We also
used processed data for NGC~4138, which is published in \cite{n4138jbh}. 
We pulled NGC~3675, NGC~3953, NGC~3992, NGC~5033, and NGC~5055 from the VLA
archive and processed following standard practice. All except NGC~3992 were processed in AIPS.
Since NGC~3992 was observed during the EVLA transition, the observations were processed in CASA
(\citealt{casa}). As is evident in Fig.~\ref{fig:n3953sum}, the bandpass of the archival VLA data
was not large enough to encompass the full frequency range of HI emission in NGC~3953. 
Processed data for NGC~3953 is available from WHISP, but the observations are low
signal-to-noise (S/N) due to the fact that NGC~3953 is detected outside WSRT's primary
beam half-power width. The kinematics were easier to constrain in the complete half 
of the archival VLA data than in the low S/N WHISP data (see Section~\ref{sec:n3953} for more detail).

\begin{landscape}
\begin{table}
\centering
\caption{Observed Properties}
\label{tab:obsprops}
\begin{threeparttable}
\begin{tabular}{lcrrrccrrrc}
\hline
Galaxy & Morphological\tnote{a} & $m_{B}$ & $m_{R}$ & $m_{\rm 3.6}$ &
$\log_{10}$(H$\alpha$ flux) & Equivalent & HI flux & $W_{\rm 20}$ & 
CO flux\tnote{b} & CO flux\tnote{c} \\
  & Type &  &  &  & (erg s$^{-1}$ cm$^{-2}$) & Width & (Jy km s$^{-1}$) & 
(km s$^{-1}$) & (Jy km s$^{-1}$) & Reference \\
  &  &  &  &  &  & (\AA) &  &  &  & \\
\hline
NGC~3486 & SABc & 11.14$\pm$0.03 & 10.20$\pm$0.03 & 7.67$\pm$0.01 &
-11.16$\pm$0.31 & 38.2$\pm$3.3 & 147$\pm$29 & 268 & 57$\pm$1 & 1 \\
NGC~3675 & SAb & 11.04$\pm$0.02 & 9.57$\pm$0.02 & 6.50$\pm$0.01 & 
-11.49$\pm$0.14 & 8.62$\pm$0.44 & 57.9$\pm$11.6 & 442 & 1440$\pm$280 & 2 \\
NGC~3726 & SABc & 11.11$\pm$0.02 & 10.02$\pm$0.02 & 7.44$\pm$0.01 &
-11.25$\pm$0.13 & 23.8$\pm$0.9 & 90.6$\pm$18.1 & 288 & 720$\pm$180 & 2,3 \\
NGC~3941 & SB0 & 11.45$\pm$0.02 & 9.97$\pm$0.02 & 7.18$\pm$0.01 &
-12.26$\pm$0.06 & 1.93$\pm$0.05 & 14.3$\pm$2.9 & 285 & -- & 4 \\
NGC~3953 & SBbc & 11.08$\pm$0.02 & 9.71$\pm$0.02 & 6.73$\pm$0.01 &
-11.20$\pm$0.03 & 8.76$\pm$0.20 & 39.3$\pm$0.8\tnote{d} & 442\tnote{d} & 1790$\pm$450 & 2,3 \\
NGC~3972 & SAbc & 13.14$\pm$0.02 & 11.94$\pm$0.02 & 9.17$\pm$0.02 & 
-12.16$\pm$0.13 & 17.2$\pm$0.8 & 16.3$\pm$3.3 & 283 & ... & ... \\
NGC~3992 & SBbc & 10.94$\pm$0.02 & 9.57$\pm$0.02 & 6.77$\pm$0.01 &
-11.45$\pm$0.12 & 9.87$\pm$0.22 & 75.5$\pm$15.1 & 484 & -- & 3 \\
NGC~3998 & SA0 & 11.64$\pm$0.02 & 10.09$\pm$0.02 & 7.16$\pm$0.01 &
-11.88$\pm$0.07 & 6.29$\pm$0.25 & 8.85$\pm$1.77 & 592 & -- & 4 \\
NGC~4051 & SABbc & 11.08$\pm$0.02 & 9.94$\pm$0.02 & 7.07$\pm$0.01 &
-11.18$\pm$0.09 & 27.1$\pm$1.4 & 39.4$\pm$7.9 & 265 & 740$\pm$140 & 2,3 \\
NGC~4138 & SA0 & 12.29$\pm$0.02 & 10.82$\pm$0.02 & 7.96$\pm$0.01 &
-11.93$\pm$0.09 & 9.05$\pm$0.36 & 20.6$\pm$0.3\tnote{e} & 340\tnote{e} & 100$\pm$30 & 2 \\
NGC~4389 & SBbc & 12.64$\pm$0.02 & 11.46$\pm$0.02 & 8.77$\pm$0.02 &
-11.82$\pm$0.06 & 24.7$\pm$0.5 & 6.96$\pm$1.39 & 192 & ... & ... \\
UGC~07639 & Im & 14.32$\pm$0.02 & 13.44$\pm$0.02 & 11.42$\pm$0.05 &
-13.36$\pm$0.14 & 4.10$\pm$0.25 & 2.54$\pm$0.51 & 66.4 & ... & ... \\
NGC~5033 & SAc & 11.01$\pm$0.02 & 9.73$\pm$0.02 & 6.60$\pm$0.01 & 
-11.17$\pm$0.11 & 21.3$\pm$0.5 & 213$\pm$43 & 489 & 2469$\pm$258 & 3 \\
NGC~5055 & SAbc & 9.44$\pm$0.02 & 8.14$\pm$0.02 & 5.36$\pm$0.01 &
-10.67$\pm$0.12 & 15.6$\pm$0.8 & 454$\pm$91 & 407 & 3812$\pm$276 & 3 \\
UGC~08839 & Im & 16.04$\pm$0.02 & 15.21$\pm$0.02 & 13.19$\pm$0.10 &
-13.61$\pm$0.18 & 11.4$\pm$1.1 & 26.9$\pm$5.4 & 117 & ... & ... \\
NGC~5608 & Im & 13.96$\pm$0.02 & 13.23$\pm$0.02 & 11.23$\pm$0.04 &
-12.56$\pm$0.16 & 22.2$\pm$1.5 & 12.6$\pm$2.5 & 131 & ... & ... \\
\hline
\end{tabular}
\begin{tablenotes}
\item {\it Note.} -- The apparent magnitudes are measured values and are not corrected for 
extinction. The reported apparent magnitudes and $\log_{10}$(H$\alpha$ flux) are measured 
within $R_{\rm 25}$.
\item[a] RC3; \cite{rc3}
\item[b] -- not detected; ... not observed
\item[c] CO flux references: (1) CARMA STING (Rui Xue, private communication); 
(2) FCRAO (\citealt{fcrao}); (3) BIMA SONG (\citealt{bimasongII}); 
(4) ATLAS$^{\rm 3D}$ (\citealt{atlas3dco})
\item[d] Integrated HI flux and $W_{\rm 20}$ values from \cite{ursamajhi}.
\item[e] Integrated HI flux and $W_{\rm 20}$ values from \cite{n4138jbh}.
\end{tablenotes}
\end{threeparttable}
\end{table}
\end{landscape}

\begin{landscape}
\begin{table}
\centering
\caption{Distance and Extinction Corrected Properties}
\label{tab:corrprops}
\begin{threeparttable}
\begin{tabular}{lccccccccrrrr}
\hline
Galaxy & $V_{\rm flat}$ & Distance & Distance\tnote{a} & $A_{B}$\tnote{b} & $M_{B}$ & 
($B-R$)$_{\rm 0}$ & ($B$-3.6)$_{\rm 0}$ & ($R$-3.6)$_{\rm 0}$ & SFR\tnote{c} & Total Stellar\tnote{d} &
Total HI & Total H$_2$ \\
  & (km s$^{-1}$) & (Mpc) & Reference &  &  &  &  &  & (M$_{\sun}$ yr$^{-1}$) & Mass & Mass & Mass \\
  &  &  &  &  &  &  &  &  &  & (10$^9$ M$_{\sun}$) & (10$^9$ M$_{\sun}$) & (10$^9$ M$_{\sun}$)\\
\hline
NGC~3486 & 139.9 & 10.6 & 1 & 0.078 & -19.16$\pm$0.03 & 0.91$\pm$0.04 & 3.39$\pm$0.03 & 2.48$\pm$0.03 &
0.50$\pm$0.36 & 9.94$\pm$1.99 & 3.90$\pm$0.78 & 0.05$\pm$0.001 \\
NGC~3675 & 204.5 & 15.3 & 1 & 0.071 & -20.10$\pm$0.02 & 1.44$\pm$0.03 & 4.47$\pm$0.02 & 3.03$\pm$0.02 &
0.49$\pm$0.16 & 64.9$\pm$13.0 & 3.19$\pm$0.64 & 2.64$\pm$0.51 \\
NGC~3726 & 146.5 & 10.5 & 1 & 0.060 & -19.13$\pm$0.02 & 1.07$\pm$0.03 & 3.61$\pm$0.02 & 2.54$\pm$0.02 &
0.40$\pm$0.12 & 12.5$\pm$2.5 & 2.35$\pm$0.47 & 0.62$\pm$0.16 \\
NGC~3941 & 130.9 & 12.2 & 2 & 0.076 & -19.14$\pm$0.02 & 1.45$\pm$0.03 & 4.19$\pm$0.02 & 2.74$\pm$0.02 &
0.05$\pm$0.007 & 21.1$\pm$4.2 & 0.50$\pm$0.10 & -- \\
NGC~3953 & 227.7 & 22.7 & 1 & 0.109 & -20.88$\pm$0.02 & 1.33$\pm$0.03 & 4.24$\pm$0.02 & 2.92$\pm$0.02 &
2.1$\pm$0.1 & 109$\pm$22 & 4.77$\pm$0.95 & 7.24$\pm$1.82 \\
NGC~3972 & 143.7 & 25.2 & 3 & 0.051 & -19.02$\pm$0.02 & 1.18$\pm$0.03 & 3.92$\pm$0.03 & 2.74$\pm$0.03 &
0.28$\pm$0.08 & 14.6$\pm$2.9 & 2.44$\pm$0.49 & -- \\
NGC~3992 & 274.5 & 22.7 & 4 & 0.106 & -21.04$\pm$0.02 & 1.33$\pm$0.03 & 4.06$\pm$0.02 & 2.74$\pm$0.02 &
1.2$\pm$0.3 & 107$\pm$21 & 9.17$\pm$1.83 & -- \\
NGC~3998 & 237.5 & 14.1 & 2 & 0.059 & -19.40$\pm$0.02 & 1.53$\pm$0.03 & 4.42$\pm$0.02 & 2.89$\pm$0.02 &
0.17$\pm$0.03 & 29.7$\pm$5.9 & 0.41$\pm$0.08 & -- \\ 
NGC~4051 & 163.6 & 11.7 & 1 & 0.047 & -19.39$\pm$0.02 & 1.12$\pm$0.03 & 3.96$\pm$0.02 & 2.84$\pm$0.02 &
0.58$\pm$0.12 & 21.6$\pm$4.3 & 1.27$\pm$0.25 & 0.79$\pm$0.15 \\
NGC~4138 & 198.3 & 13.8 & 2 & 0.051 & -18.54$\pm$0.02 & 1.45$\pm$0.03 & 4.28$\pm$0.02 & 2.83$\pm$0.02 &
0.14$\pm$0.03 & 13.1$\pm$2.6 & 0.92$\pm$0.18 & 0.15$\pm$0.04 \\
NGC~4389 & 101.5 & 7.7 & 1 & 0.053 & -17.00$\pm$0.02 & 1.16$\pm$0.03 & 3.82$\pm$0.03 & 2.66$\pm$0.03 &
0.06$\pm$0.01 & 2.03$\pm$0.41 & 0.10$\pm$0.02 & -- \\
UGC~07639 & 19.2 & 7.1 & 5 & 0.042 & -15.34$\pm$0.02 & 0.86$\pm$0.03 & 2.86$\pm$0.05 & 1.99$\pm$0.05 &
0.001$\pm$0.0005 & 0.22$\pm$0.04 & 0.03$\pm$0.01 & -- \\
NGC~5033 & 216.1 & 18.8 & 1& 0.042 & -20.56$\pm$0.02 & 1.26$\pm$0.03 & 4.37$\pm$0.02 & 3.11$\pm$0.02 & 
1.5$\pm$0.4 & 84.6$\pm$16.9 & 17.7$\pm$3.5 & 6.85$\pm$0.72 \\
NGC~5055 & 203.9 & 9.1 & 1 & 0.064 & -20.60$\pm$0.02 & 1.27$\pm$0.03 & 4.02$\pm$0.02 & 2.74$\pm$0.02 &
1.1$\pm$0.3 & 63.2$\pm$12.6 & 8.86$\pm$1.77 & 2.48$\pm$0.18 \\
UGC~08839 & 66.0 & 11.1 & 1 & 0.090 & -15.68$\pm$0.02 & 0.79$\pm$0.03 & 2.76$\pm$0.10 & 1.97$\pm$0.10 &
0.002$\pm$0.001 & 0.25$\pm$0.05 & 0.78$\pm$0.16 & -- \\
NGC~5608 & 65.3 & 7.5 & 1 & 0.034 & -15.75$\pm$0.02 & 0.72$\pm$0.03 & 2.70$\pm$0.04 & 1.98$\pm$0.04 &
0.01$\pm$0.004 & 0.23$\pm$0.05 & 0.17$\pm$0.03 & -- \\
\hline
\end{tabular}
\begin{tablenotes}
\item {\it Note.} -- $M_{B}$ and colours are extinction corrected. The extinction 
correction for the NIR is assumed to be negligible. The $M_{B}$ value also includes 
a small correction based on an extrapolation of the observed stellar disc. Colours are
measured within $R_{\rm 25}$.
\item[a] Distance references: (1) Distances determined using the LTFR at 3.6$\mu$m from
\cite{ltfr}: $\log_{10}$~$L_{3.6}$~$=$~(-0.28$\pm$0.36)~$+$~(4.93$\pm$0.17)~$\log_{10}$~$V_{\rm flat}$.
(2) SBF (\citealt{sbfdist}); (3) SNIa (\citealt{n3972distref}); (4) SNIa (\citealt{n3992distref});
(5) SBF (\citealt{u07639distref})
\item[b] As calculated by \cite{galextinct}.
\item[c] Measured within $R_{\rm 25}$ and calculated using the calibration given in \cite{sfrcal}.
\item[d] Calculated using M/L$_{3.6}$ = 0.5$\pm$0.1.
\end{tablenotes}
\end{threeparttable}
\end{table}
\end{landscape}

\begin{table*}
\centering
\caption{Radial Properties} 
\label{tab:radprops}
\begin{threeparttable}
\begin{tabular}{lccccccccc}
\hline
Galaxy & \multicolumn{2}{c}{$h_{\rm R}$\tnote{a}} & \multicolumn{2}{c}{$D_{\rm 25}$\tnote{b}} & 
\multicolumn{2}{c}{$D_{\rm HI}$\tnote{c}} & $C_{\rm 28}$\tnote{d} & $\log_{10}$(EW) gradient & ($B-R$) gradient\\
  & (arcsec) & (kpc) & (arcsec) & (kpc) & (arcsec) & (kpc) &  & (arcmin$^{-1}$) & (arcmin$^{-1}$)\\
\hline
NGC~3486 & 39.0 & 2.00 & 371 & 19.1 & 640 & 32.9 & 3.6 & -0.15 & -0.19 \\
NGC~3675 & 54.1 & 4.01 & 346 & 25.7 & 600 & 44.5 & 3.9 & -0.37 & -0.01 \\
NGC~3726 & 38.8 & 1.98 & 351 & 17.8 & 511 & 26.0 & 2.3 & 0.22 & -0.12 \\
NGC~3941 & 25.4 & 1.50 & 184 & 10.9 & 408 & 24.1 & 4.8 & -0.15 & -0.05 \\
NGC~3953 & 39.1 & 4.30 & 371 & 40.9 & 400\tnote{e} & 44.0 & 3.4 & -0.06 & -0.09 \\
NGC~3972 & 25.0 & 3.05 & 202 & 24.7 & 216 & 26.4 & 4.4 & -0.15 & -0.24 \\
NGC~3992 & 47.6 & 5.24 & 411 & 45.3 & 705 & 77.6 & 2.9 & 0.02 & -0.05 \\
NGC~3998 & 35.9 & 2.45 & 180 & 12.3 & 282 & 19.3 & 5.9 & -0.24 & -0.16 \\
NGC~4051 & 39.7 & 2.25 & 314 & 17.8 & 271 & 15.3 & 5.8 & -0.18 & 0.02 \\
NGC~4138 & 21.6 & 1.45 & 157 & 10.5 & 258 & 17.3 & 4.5 & -1.41 & 0.08 \\
NGC~4389 & 21.7 & 0.81 & 149 & 5.58 & 153 & 5.71 & 2.2 & -0.90 & 0.04 \\
UGC~07639 & 29.8 & 1.03 & 118 & 4.06 & 105 & 3.61 & 3.5 & -2.35 & 0.38 \\
NGC~5033 & 61.4 & 5.60 & 541 & 49.3 & 960 & 87.5 & 5.6 & 0.10 & -0.22 \\
NGC~5055 & 110 & 4.85 & 805 & 35.5 & 1950 & 86.0 & 4.3 & -0.07 & -0.02 \\
UGC~08839 & 34.1 & 1.84 & 68.4 & 3.68 & 340 & 18.3 & 3.2 & -0.26 & -0.14 \\
NGC~5608 & 22.1 & 0.80 & 142 & 5.18 & 201 & 7.31 & 3.0 & 0.62 & -0.35 \\
\hline
\end{tabular}
\begin{tablenotes}
\item[a] Total average disc scale length measured at 3.6$\mu$m.
\item[b] Measured at 25 mag arcsec$^{-2}$ in $B$.
\item[c] Measured at 10$^{20}$ atoms cm$^{-2}$.
\item[d] Calculated as 5$\log_{10}$($R_{\rm 80}$/$R_{\rm 20}$), where $R_{\rm 20}$ is the radius which 
contains 20 per cent and $R_{\rm 80}$ is the radius which contains 80 per cent of the total luminosity 
as measured from ellipse photometry (e.g. \citealt{KentC28}).
\item[e] From \cite{ursamajhi}.
\end{tablenotes}
\end{threeparttable}
\end{table*}

\begin{table*}
\centering
\caption{HI Synthesis Image Parameters}
\label{tab:cubeprops}
\begin{tabular}{llclrrcrc}
\hline
Galaxy & Telescope/ & Time & Image & Velocity & Robust & Beam & Beam & Noise \\
  & Project Code & On Source & Name & Resolution & Weighting & Size & Position Angle & (mJy beam$^{-1}$) \\
  &  & (hours) &  & (km s$^{-1}$) &  & (arcsec) & (deg) &  \\
\hline
NGC~3486 & VLA & 6.6 & low & 5.0 & 5 & 25.5$\times$17.7 & 55.8 & 0.47 \\
         & 13A-107 & & high & 5.0 & -0.5 & 16.4$\times$13.4 & 71.5 & 0.65 \\
NGC~3675 & VLA & 5.1 & medium & 10.3 & 0.5 & 21.2$\times$18.1 & 83.3 & 0.53 \\
         & AP225 & & high & 10.3 & -0.5 & 16.1$\times$14.4 & 84.4 & 0.64 \\
NGC~3726 & WSRT & 12.0 & 30\arcsec~smooth & 4.1 & -- & 30.0$\times$30.0 & 0.0 & 3.83 \\
         & WHISP & & full res. & 4.1 & -- & 15.8$\times$11.8 & 0.0 & 3.07 \\
NGC~3941 & VLA & 6.5 & low & 5.0 & 5 & 21.9$\times$18.6 & 58.2 & 0.46 \\
         & 13A-107 & & medium & 5.0 & 0.5 & 17.6$\times$15.2 & 71.8 & 0.51 \\
NGC~3953 & VLA & 1.2 & medium & 5.2 & 0.5 & 49.9$\times$47.1 & -68.5 & 1.31 \\
         & AV237 &   & high & 5.2 & -0.5 & 44.3$\times$42.6 & -72.7 & 1.44 \\
NGC~3992 & VLA & 5.1 & low & 3.3 & 5 & 32.3$\times$17.5 & 107 & 0.94 \\
         & 10B-207 & & high & 3.3 & -0.5 & 20.2$\times$12.7 & 89.2 & 1.33 \\
NGC~3998/ & VLA & 6.7 & spectral binned & 14.9 & 5 & 31.0$\times$21.0 & 68.9 & 0.32 \\
NGC~3972  & 13A-107 & & low & 5.0 & 5 & 29.9$\times$18.3 & 73.6 & 0.50 \\
NGC~4051 & WSRT & 12.0 & 30\arcsec~smooth & 4.1 & -- & 29.5$\times$29.0 & 0.0 & 2.80 \\
         & WHISP & & full res. & 4.1 & -- & 14.3$\times$10.2 & 0.0 & 2.49 \\
NGC~4138 & VLA/AB678 & 5.6 & low & 5.2 & 5 & 18.5$\times$20.8 & 75.0 & 0.55 \\
NGC~4389 & VLA & 5.1 & low & 5.0 & 5 & 30.3$\times$17.8 & 68.6 & 0.60 \\
         & 13A-107 & & high & 5.0 & -0.5 & 19.8$\times$12.3 & 75.9 & 0.70 \\
UGC~07639 & VLA & 5.8 & low & 5.0 & 5 & 17.9$\times$17.6 & -27.4 & 0.46 \\
          & 13A-107 & & high & 5.0 & -0.5 & 14.2$\times$12.4 & -45.6 & 0.60 \\
NGC~5033 & VLA & 14.0 & medium & 20.7 & 0.5 & 21.1$\times$19.9 & 80.5 & 0.26 \\
         & AP270 & & high & 20.7 & -0.5 & 15.9$\times$14.6 & -89.9 & 0.32 \\
NGC~5055 & VLA & 9.2 & taper & 10.3 & 5 & 39.5$\times$36.5 & 22.2 & 0.30 \\
         & AT172,AT185 & & high & 10.3 & -0.5 & 9.4$\times$8.1 & 46.9 & 0.44 \\
UGC~08839 & VLA & 5.9 & low & 5.0 & 5 & 36.4$\times$17.2 & 52.0 & 0.47 \\
          & 13A-107 & & high & 5.0 & -0.5 & 20.7$\times$13.0 & 55.5 & 0.62 \\
NGC5608 & VLA & 6.9 & low & 5.0 & 5 & 33.6$\times$18.1 & 64.0 & 0.59 \\
        & 13A-107 & & high & 5.0 & -0.5 & 18.3$\times$12.7 & 68.1 & 0.73 \\
\hline
\end{tabular}
\end{table*}

\subsection{WIYN SparsePak}
\label{sec:spak}

Many of the large barred spiral galaxies in the kinematic sample lack detectable neutral hydrogen
in their centres (e.g. NGC~3992; Fig.~\ref{fig:n3992sum}). This is likely caused 
by a phase change from atomic to molecular gas or bar-driven transport of gas to the centre 
where it is concentrated, and in some cases, consumed by nuclear star formation (\citealt{barCO}).
The desire to recover kinematic information in
the centres of such galaxies motivated us to acquire optical IFS observations of the ionized gas. 
Further, even the non-barred galaxies which 
have detectable HI in their centres benefit from the improved spatial resolution that the IFS data provide.
Thus, the only galaxies not targeted for SparsePak observations are irregular type galaxies 
which have little or no ionized gas emission visible in the narrowband H$\alpha$ images.

We present ionized gas velocity fields for the central regions of twelve galaxies 
obtained using the SparsePak IFU (\citealt{spak}) on the WIYN 3.5 m telescope 
in 2014 April, 2015 March, and 2015 April. The SparsePak 
IFU is composed of eighty-two 5\arcsec~diameter fibers arranged in a fixed 
70\arcsec~$\times$ 70\arcsec~square. Observations were acquired using
the same setup as described in \cite{paperI} with a wavelength range from
6480~\AA~to 6890~\AA~and a resulting velocity resolution of 13.9 km s$^{-1}$ pixel$^{-1}$.
Typical integration times ranged from 3$\times$480s for galaxies dominated by bright star forming
regions to 3$\times$1200s for galaxies dominated by diffuse ionized gas.

\begin{figure*}
\includegraphics[width=\textwidth]{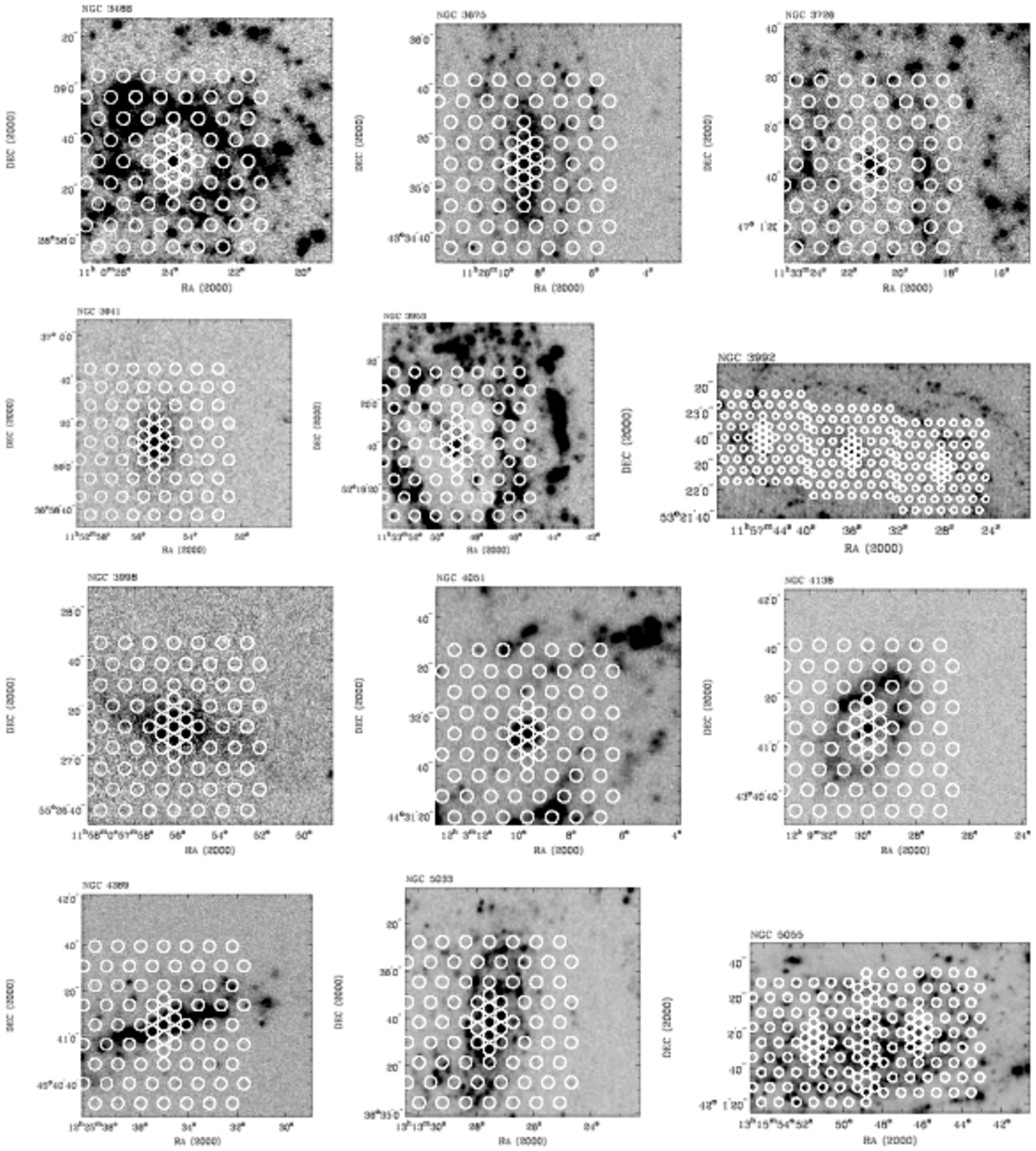}
\caption{SparsePak fiber footprints for one pointing (dithers not shown) overlaid on the 
narrowband H$\alpha$ images. Fibers designated as 'sky' are just outside the field-of-view.}
\label{fig:spakfibers}
\end{figure*}

As discussed in \cite{paperI}, the SparsePak array was aligned on the sky with a position 
angle of 0\degr~for simplicity and consistency between galaxies. We used a three pointing 
dither pattern to spatially fill in gaps between fibers. We took three exposures for each 
dither pointing to be able to detect faint diffuse ionized gas, not just bright star forming regions. 
We used additional pointings when necessary to cover the full extent of the central regions on the sky.
NGC~3992 required three separate pointings with dithers and NGC~5055 required two.
Observations of blank sky were also taken to remove sky line contamination more 
accurately, as all of the galaxies are much more extended than the SparsePak field-of-view
(See Fig.~\ref{fig:spakfibers}). 

The SparsePak data were processed using standard tasks in the HYDRA package within 
IRAF\footnotemark \footnotetext{IRAF is distributed by NOAO, which is operated by the 
Association of Universities for Research in Astronomy, Inc., under cooperative 
agreement with the National Science Foundation.} following the same procedure as described in
\cite{paperI}. The data were bias-subtracted and 
flattened, and the IRAF task DOHYDRA was used to fit and extract apertures
from the IFU data. The spectra were wavelength calibrated using a wavelength solution 
created from ThAr lamp observations. The three exposures for each dither pointing 
were cleaned of cosmic rays. The individual images were sky subtracted using a 
separate sky pointing scaled to the strength of the 6577 \AA~sky line, which is 
close to the redshifted H$\alpha$ and [NII]~$\lambda$6584 lines in some fibers. 
The cleaned, sky subtracted images were averaged together to increase S/N and 
then flux calibrated using observations of 
spectrophotometric standards from \cite{spakstandards}. Although the skies were
merely transparent, not photometric, we apply a flux calibration to remove instrumental
signatures and to permit measurement of relative line strengths.

Relative line strengths of [NII]~$\lambda$6584 to H$\alpha$ were measured from Gaussian fits 
to the emission lines for the central fiber's spectrum and integrated spectrum 
for each galaxy to reveal the nature of the dominant ionizing source.
The integrated spectrum for each galaxy was corrected for rotation using the 
Doppler shift of the centroid of the H$\alpha$ emission line in each galaxy's spectra.
With the exception of the S0
galaxies NGC~3941 and NGC~3998, the integrated spectra show [NII]~$\lambda$6584/H$\alpha$ values 
with a median of 0.5, indicative of thermal emission processes most likely due to star formation 
activity (\citealt{bpt}). We additionally find relatively narrow full width half maximum (FWHM) 
values on the order of 80 km s$^{-1}$ for these integrated spectra. 
All of the galaxies presented in this study with the exception of NGC~3726, NGC~4051, NGC~4389
and NGC~5055 show large [NII]~$\lambda$6584/H$\alpha$ values, ranging between 2 and 4.5, as well 
as broad FWHM values on the order of 100-350 km s$^{-1}$ in the spectra of their central fibers.
This is not unexpected given that NGC~3486, NGC~3941, NGC~4051, NGC~4138 and NGC~5033 are classified
as Seyferts, NGC~3675, NGC~3953, NGC~3992 and NGC~5055 have been classified as transition objects,
and NGC~3998 is classified as low-ionization nuclear emission-line region (LINER) by \cite{HoAGN}.

\begin{figure*}
\includegraphics[width=\textwidth]{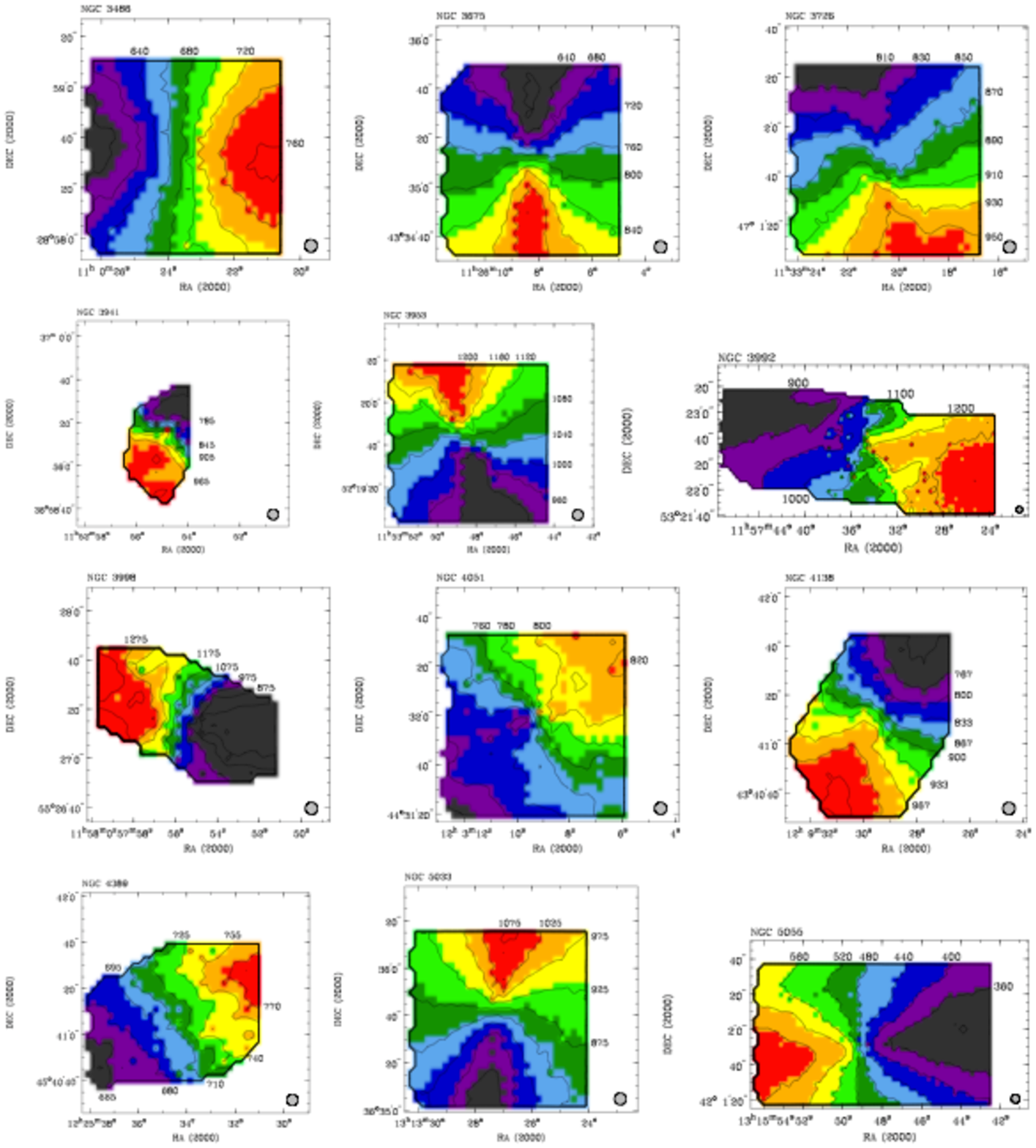}
\caption{Ionized gas velocity fields from the SparsePak data.}
\label{fig:spakvels}
\end{figure*}

The flux calibrated, sky subtracted spectra were cross-correlated with a template 
emission line spectrum to extract the luminosity weighted mean recessional velocity 
at every position using the IRAF task FXCOR. To minimize systematic effects, the 
cross-correlation analysis uses a high S/N template spectrum from within
each galaxy. This results in robust measures of the relative velocity offset for each fiber.
Spectra with broad or double-peaked emission line profiles resulted in broad 
cross-correlation peaks that were nonetheless of high significance and had a 
well-defined centroid. The velocity values from FXCOR were 
then placed into a grid that mapped the SparsePak fiber locations. To fill in the 
missing spacings, the velocity field was then interpolated using the average value 
from the nearest eight pixels. Fig.~\ref{fig:spakvels} displays
the resulting ionized gas velocity fields for the twelve galaxies.

\subsection{Archival CO}
\label{sec:codata}

When available, archival molecular gas observations of CO are used to supplement
the HI surface density measurements. Archival CO data from the BIMA Survey of 
Nearby Galaxies (BIMA SONG; \citealt{bimasongII}) were acquired to supplement the 
HI observations of NGC~5033 and NGC~5055.
The BIMA SONG CO intensity maps have a typical spatial resolution of about 6.0 arcsec and
are mapped in red contours in Figs.~\ref{fig:n5033sum} and~\ref{fig:n5055sum}. 
More detailed maps can be found in \cite{bimasongII}.
NGC~3992 is in BIMA SONG, but was not detected in CO line emission. It was formally detected
in one pointing with the KPNO 12 m, but the molecular gas component for NGC~3992 is not 
included in the present study due to lack of information about the spatial distribution.
Surface density measurements from the FCRAO Extragalactic CO Survey (\citealt{fcrao})
were used to estimate the molecular gas distribution in NGC~3675, NGC~3726, NGC~3953, NGC~4051 and NGC~4138. 
A CO intensity map of NGC~3486 was acquired from the CARMA STING survey (Rui Xue, private communication). 
NGC~3941 and NGC~3998 were observed as part of ATLAS$^{\rm 3D}$ (\citealt{atlas3dco}), but
were not formally detected above 3$\sigma$. 
The three dwarf irregular galaxies in this sample, UGC~07639, UGC~08839, and NGC~5608, as well as
NGC~4389 have no CO observations available in the literature.

\subsection{Spitzer 3.6$\mu$m Data}
\label{sec:3.6data}

In this analysis, we take advantage of 3.6$\mu$m imaging observations from the Infrared Array Camera 
(IRAC) on the {\it Spitzer Space Telescope} to provide an unbiased census of the
underlying stellar populations in galaxies. Specifically, the 3.6$\mu$m
images are relatively insensitive to radial changes in M/L that might
be introduced by either internal dust obscuration or changes in the dominant stellar population.
The M/L ratio at 3.6$\mu$m is nearly independent of star formation history 
\citep[e.g.][]{nirml,3.6ML,3.6model}, and emission at 3.6$\mu$m is predominantly stellar and not 
greatly affected by 
PAH emission at 3.3$\mu$m (\citealt{3.6noPAH}). This allows many of the uncertainties associated
with the stellar mass distribution to be minimized.

The 3.6$\mu$m imaging observations were acquired in 2011-2013 as part of {\it Spitzer's} Cycle
8. Observations of each galaxy were conducted in a grid-like mapping pattern of dithered pointings 
to cover a total area 5 times the optical radius $R_{\rm 25}$. Each pixel in the resulting map has 
a total exposure time of 1800 s, which provides a sensitivity of a few$\times$0.01 M$_{\sun}$ pc$^{-2}$. 
After standard processing, the individual 
pointings were combined into maps using the MOsaicker and Point source EXtractor 
(MOPEX) software. Images were drizzled to improve the resolution from the native pixel
scale of 1.2 arcsec pixel$^{-1}$ to 0.75 arcsec pixel$^{-1}$ (\citealt{drizzle}). 
The maps were fit for a first order sky subtraction to remove any gradient in the 
sky level. Foreground stars were replaced by interpolated values based
on nearest neighbor intensities.

Aperture photometry on the 3.6$\mu$m images was carried out using the GALPHOT\footnotemark 
\footnotetext{GALPHOT is a collection of scripts in the IRAF STSDAS environment first developed by 
W. Freudling and J. J. Salzer. The current version has been further enhanced by members of the 
Cornell Extragalactic Group and is maintained by M. P. Haynes.} package for IRAF following the
analysis steps in \cite{paperI}. Concentric, con-eccentric ellipses were used to measure
surface magnitudes and the integrated magnitude at $R_{\rm 25}$ (Section~\ref{sec:optdata}),
which is given in Table~\ref{tab:obsprops}. No foreground extinction correction was 
used for the NIR since the expected value is negligible in these high galactic latitude fields. 
The total stellar mass presented in Table~\ref{tab:corrprops} was determined from the total 
stellar luminosity at 3.6$\mu$m using a M/L of 0.5$\pm$0.1. The luminosity 
was calculated from the absolute magnitude at 3.6$\mu$m, which includes a correction based 
on an extrapolation of the 3.6$\mu$m surface brightness on the order of 0.01 - 0.03 mag. 
The ellipse photometry was also used to 
estimate the concentration of light using radii containing 20 per cent and 80 per cent of the total luminosity.
The values were estimated using the curve of growth, following the method and definition in
\cite{KentC28} (see Table~\ref{tab:radprops}). In this definition, values of $C_{\rm 28}$ range from 2 to 5, where
disc-dominated galaxies have values between 3 and 4, elliptical or spheroidal systems have values 
$>$ 4, and galaxies with low central surface brightness or low internal velocity dispersions 
have values $<$ 3 (\citealt{c28}). As expected from their classifications (Table~\ref{tab:obsprops}),
most of the galaxies in this sample roughly follow the concentration trends. 
However, NGC~3998,
NGC~4051 and NGC~5033 have abnormally high $C_{\rm 28}$ values. This could be related to
nuclear activity as NGC~3998 is classified as a LINER and NGC~4051 and NGC~5033 are classified as
Seyferts (\citealt{HoAGN}). In these cases, the nuclear activity causes the central few 
pixels in the images to be unusually bright, skewing the luminosity-weighted concentration 
measurement to higher values.

\subsection{Optical Data}
\label{sec:optdata}

While the {\it Spitzer} 3.6$\mu$m images give robust estimates of the total stellar distribution
in galaxies, we turn to optical imaging to provide insight into the dominant stellar populations. 
In particular, comparison of broadband $B$ and $R$ images reveal changes in age or metallicity while
narrowband H$\alpha$ images provide information about current star formation activity.
Optical imaging observations were taken with the WIYN 0.9 m telescope at Kitt Peak National 
Observatory in 2013--2015 with the Half Degree Imager (HDI) and S2KB imager. 
Narrowband imaging 
was done using a filter with FWHM of 60 \AA~centered at a wavelength 
of 6580 \AA. A filter with similar FWHM and slightly offset central wavelength was 
used for continuum subtraction. The H$\alpha$ images were taken with total exposure 
times of 2 $\times$ 20 min in the 6580 \AA~filter, and 20 min in the 
narrowband continuum filter. Broadband imaging was done with three exposures of 900 s 
and 300 s with the $B$ and $R$ filters, respectively. Additional short exposures 
were acquired when necessary to replace saturated pixels in the cores of bright galaxies. 
The optical images were reduced and analysed with IRAF following the
steps listed in \cite{paperI}.

Similar to the 3.6$\mu$m image analysis, aperture photometry using concentric, con-eccentric ellipses 
was carried out to derive surface brightness magnitudes for the $B$- and $R$-band images for each galaxy.
Table~\ref{tab:obsprops} gives measured integrated magnitudes derived from the aperture photometry. The
integrated apparent magnitudes tabulated here are measured at the radius at 
which the $B$-band surface brightness equals 25 mag arcsec$^{-2}$ ($R_{\rm 25}$) and are not corrected for
foreground extinction. The absolute $B$-band magnitude listed in Table~\ref{tab:corrprops} has been 
corrected for extinction (\citealt{galextinct}) and includes a correction based on an extrapolation 
of the $B$-band surface brightness profile. Integrated colours tabulated in Table~\ref{tab:corrprops} 
are also measured at $R_{\rm 25}$ and have been extinction corrected.

Aperture photometry was performed on the H$\alpha$ images to measure H$\alpha$ fluxes and derive 
equivalent widths (EW). The current total star formation rate (SFR) is derived from the H$\alpha$ flux 
measured at $R_{\rm 25}$ using the calibration given in \cite{sfrcal}. The reported SFRs have not
been corrected for [NII]~$\lambda$6584 flux which falls within the narrowband filter. In some 
galaxies, the [NII]~$\lambda$6584 line may be a large contributor to the total measured flux 
(e.g. Section~\ref{sec:spak}). This is most likely the case for NGC~3941 and NGC~3998, whose 
integrated spectra show dominant [NII]~$\lambda$6584 lines.
In other galaxies (i.e. NGC~5005; \citealt{paperI}), internal extinction may play a larger role.
Therefore, the SFRs in Table~\ref{tab:corrprops} should taken as indicative only. The EW of the 
H$\alpha$ emission line is used as a tracer of the specific star formation
rate. It is calculated by dividing the H$\alpha$ flux by the continuum flux density measured from the 
H$\alpha$ narrowband continuum filter. It, therefore, serves as an indicator of the strength of the 
current SFR relative to the past average SFR. A larger EW value would indicate a larger current SFR 
relative to the continuum, or past average star formation.

Radial profiles were additionally created to examine the surface brightness profiles as well as
gradients in colour and EW. Table~\ref{tab:radprops} gives estimates of the EW gradient and $B-R$
colour gradient from a linear fit to the radial profiles. Often, the radial trends in these
profiles are not linear, but the sign of the value reported in Table~\ref{tab:radprops} provides
a rough picture of the star formation history and changes in the dominant stellar population.
A negative gradient in $B-R$ colour is indicative of a transition from redder in the bulge-dominated
nuclear region to bluer in the disc where most of the recent star formation is occurring, as is 
expected for spiral galaxies (e.g. \citealt{galcolors}). A corresponding EW gradient would have
a positive sign with higher EW values at larger radii in the disc where there is more current
star formation relative to past star formation.
Interesting features of each galaxy's surface brightness profile, colour and EW gradients
are described in more detail in Section~\ref{sec:notes}.

\section{Results}
\label{sec:results}

In this section, we focus on the galaxy kinematics and rotation curve decomposition analysis.
Neutral and ionized gas rotation curves are shown in Fig.~\ref{fig:rc}, and kinematic properties are summarized
in Table~\ref{tab:kinprops}. Rotation curve decompositions are shown in Fig.~\ref{fig:rcdecomp}
and results of the decomposition are given in Table~\ref{tab:dmprops}.

\subsection{Neutral and Ionized Gas Kinematics}
\label{sec:kinematics}

Data products from both new and archival HI observations were analysed using standard tools
distributed as part of the {\sc gipsy} software package (\citealt{gipsy}). Integrated intensity maps
and intensity-weighted mean velocity fields were extracted from the data cubes following the
process outlined in \cite{paperI}. 

Rotation curves for each galaxy were derived by fitting a series of concentric tilted rings to the 
HI velocity fields and SparsePak ionized gas velocity fields separately ({\sc gipsy} task {\sc rotcur}).
The ring centre coordinates ($x_{\rm 0}$, $y_{\rm 0}$), systemic velocity ($V_{\rm sys})$,
inclination angle ({\it i}), position angle (P.A.), and rotation velocity 
($V_{\rm rot}$) were iteratively fit following the method detailed in \cite{paperI}. The expansion 
velocity was always fixed to zero. Initially, the rings were allowed to vary 
in position and inclination angle as a function of radius to account for warps. Then, a smoothly 
varying or constant distribution in {\it i} and P.A. was adopted to derive the underlying bulk 
rotation and prevent spurious second-order effects in the rotation curve. The final azimuthally 
averaged rotation curve was derived with all other ring parameters fixed. Fig.~\ref{fig:rc} 
shows the rotation curve results for each galaxy in this sample.

As in \cite{paperI}, uncertainties in the neutral gas rotational
velocities are estimated to be a combination of non-circular thermal gas motions and inclination 
errors. Uncertainties due to thermal gas motions were estimated to be 10 per cent of the average 
velocity dispersion values as measured from a velocity dispersion map created from the highest
usable resolution HI data. Uncertainty due to kinematic asymmetries was also taken 
into account by assuming an error equal to 0.25 times the absolute difference between the circular 
velocities of the approaching and receding sides (\citealt{dmthings}). Uncertainty of the inclination
angle is typically on the order of $\pm$5\degr, which translates to 5-20 per cent error on the circular
velocity. This is the dominant source of uncertainty in the rotation velocities.
For this reason, only uncertainties due to inclination angle errors were calculated for the ionized
gas rotational velocities.

There is generally good agreement between the neutral and ionized gas kinematics
when they overlap. However, there are a few cases where the ionized gas peaks at larger rotational
velocities than is seen in the HI (e.g. NGC~3675 and NGC~3998 in Fig.~\ref{fig:rc}). These
discrepancies can usually be explained through a combination of beam smearing in the HI and
effects of non-circular motions on the ionized gas from nuclear activity. In the cases of NGC~3675, 
NGC~5033 and NGC~5055, the largest discrepancy between the ionized and neutral gas rotation curves
occurs within about two resolution elements of the HI observations. The lower spatial resolution of 
the HI data has the effect of making the inner slope of the rotation curve appear shallower with
lower rotational velocities. A discussion of how the uncertainty in the central rotational
velocities affects the derived distribution of mass is given below.

\begin{figure*}
\includegraphics[width=\textwidth]{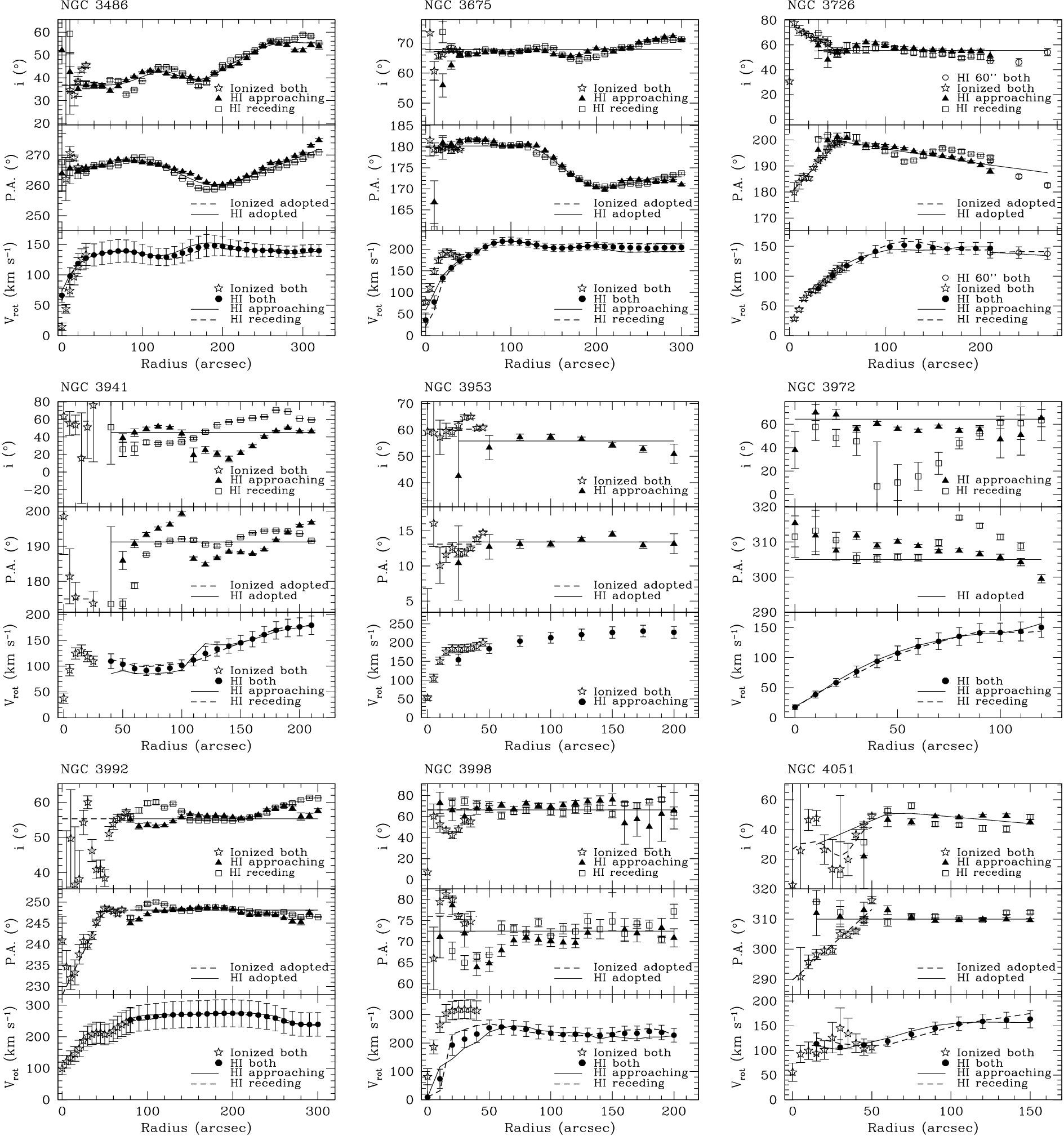}
\caption{Rotation curves for VLA HI and SparsePak ionized gas derived from tilted rings 
using the {\sc gipsy} task {\sc rotcur}. The panels for each galaxy include the inclination angle 
(top), position angle (middle) and circular rotation velocity (bottom). Inclination and
position angle fits to the approaching ({\it filled triangles}) and receding
({\it open squares}) sides and the adopted fits to both sides ({\it solid line})
are derived from the HI velocity field. Both sides of the ionized gas velocity field are 
fit at the same time ({\it open stars}). The {\it dashed line} shows the adopted ionized gas fits. 
Circular rotational velocities for the HI ({\it filled circles}) and SparsePak ionized gas ({\it open stars}) 
are derived using the adopted inclination and position angles at each radius.
The {\it dotted} and {\it dash-dotted}
lines show the separate circular rotational velocities derived for the approaching and receding sides 
of the HI velocity field, respectively, using the same adopted inclination and position angles.}
\label{fig:rc}
\end{figure*}

\setcounter{figure}{3}
\begin{figure*}
\includegraphics[width=\textwidth]{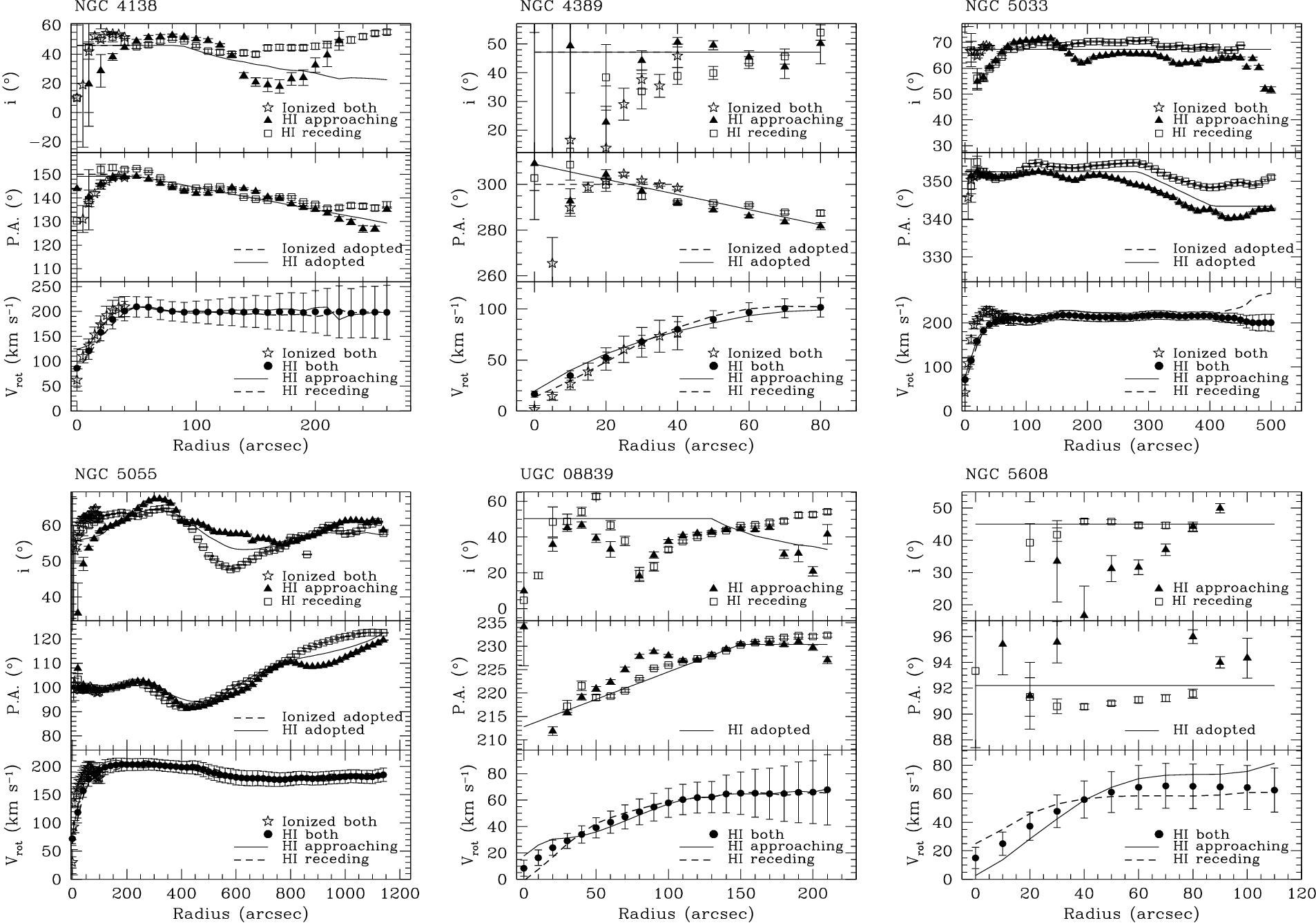}
\caption{Continued.}
\end{figure*}

\begin{table*}
\centering
\caption{Kinematic Properties}
\label{tab:kinprops}
\begin{threeparttable}
\begin{tabular}{lccccc}
\hline
Galaxy & \multicolumn{2}{c}{Kinematic centre} & Systemic & Adopted\tnote{$\dagger$} & Adopted\tnote{$\dagger$} \\
  & RA & Dec & velocity & position angle & inclination angle \\
  & (J2000) & (J2000) & (km s$^{-1}$) & (deg) & (deg)\\
\hline
NGC~3486 & 11:00:23.82 & +28:58:31.2 & 678 & 266 & 36.8 \\
NGC~3675 & 11:26:08.54 & +43:35:11.1 & 763 & 176 & 67.8 \\
NGC~3726 & 11:33:20.78 & +47:01:47.8 & 864 & 195 & 55.5 \\
NGC~3941 & 11:52:55.36 & +36:59:11.0 & 934 & 191 & 45.4 \\
NGC~3953 & 11:53:48.99 & +52:19:34.6 & 1050 & 13.4 & 55.8 \\
NGC~3972 & 11:55:45.69 & +55:19:18.6 & 845 & 305 & 64.4 \\
NGC~3992 & 11:57:35.94 & +53:22:28.0 & 1049 & 248 & 55.3 \\
NGC~3998 & 11:57:56.16 & +55:27:08.3 & 1062 & 72.5 & 66.3 \\
NGC~4051 & 12:03:09.85 & +44:31:53.4 & 704 & 310 & 44.6 \\
NGC~4138 & 12:09:29.75 & +43:41:07.6 & 888 & 141 & 36.0 \\
NGC~4389 & 12:25:34.93 & +45:41:03.5 & 731 & 296 & 47.1 \\
UGC~07639 & 12:29:52.69 & +47:31:48.3 & 378 & 96.0 & 52.5 \\
NGC~5033 & 13:13:27.44 & +36:35:38.3 & 875 & 350 & 67.3 \\
NGC~5055 & 13:15:49.14 & +42:01:44.1 & 512 & 106 & 58.5 \\
UGC~08839 & 13:55:25.41 & +17:47:46.3 & 966 & 230 & 46.1 \\
NGC~5608 & 14:23:18.54 & +41:46:35.2 & 670 & 92.0 & 45.0 \\
\hline
\end{tabular}
\begin{tablenotes}
\item[$\dagger$] Average value reported when adopted value varies with radius.
\end{tablenotes}
\end{threeparttable}
\end{table*}

\subsection{Rotation Curve Decompositions}
\label{sec:rcdecomp}

We use the method of rotation curve decomposition to uncover the distribution
of baryonic and non-baryonic mass in galaxies (e.g. \citealt{dmthings}). Estimates of stellar and
gas distributions are used in conjunction with the rotation curve derived from the neutral and 
ionized gas kinematics to constrain the distribution of dark matter as a function of radius
(e.g. \citealt{rcdecomp}). 

Total gas mass and stellar light surface density profiles are used to calculate the contributions
from gas and stars to the overall rotation curve. The atomic mass surface density was measured from
the HI integrated intensity maps for every galaxy. The molecular mass surface density was measured
in the same way when integrated CO intensity maps were available or was interpolated from individual
pointings. A CO/H$_2$ conversion factor of 2$\times$10$^{20}$ cm$^{-2}$ (K km s$^{-1}$)
was assumed when calculating molecular gas mass. The atomic and molecular gas distributions were derived 
separately and then added together to create a total gas mass surface density distribution. 
The total gas mass was multiplied by a factor of 1.4 to account for primordial helium.

The {\sc gipsy} task {\sc rotmod} was used to calculate model rotation curves for both the gas and stars 
as described in \cite{paperI}. For the gas, an infinitely thin disc is assumed. Model rotation
curves for the stellar distributions were derived from the 3.6$\mu$m light surface density profiles.
In most cases, this light could be well described as a disc distribution only. The disc potential was
constructed assuming a vertical mass density distribution of ${\rm sech^2}(z/z_0)$, where
$z_{0} = 2h_{\rm R}/q$. The oblateness parameter, $q$, is calculated using the fiducial 
relation derived in \cite{DMII}:
\begin{equation}
\log(q) = 0.367\log(h_{\rm R}/{\rm kpc}) + 0.708.
\end{equation}
However, for NGC~3953, NGC~5033 and NGC~5055, 
the bulge component was a non-negligible contributor to the distribution of the stellar mass. 
In those cases, model rotation curves were derived separately for the bulge and disc, based on a 
decomposition of the 3.6$\mu$m image. A spherical distribution was assumed for the bulge 
model rotational velocities.

As briefly described in Section~\ref{sec:kinematics}, it can be challenging 
to trace the central gravitational potential in galaxies with other dynamical contributions 
from bars or active nuclei. This is an issue for the rotation curve decomposition when
the stellar disc/bulge is assumed to contribute maximally. In this assumption, the
total mass of the stellar content is set by scaling it such that the inner dynamics
can be fully described by the gravitational influence of the baryons, without the need 
for dark matter. This breaks the degeneracy between the unknown absolute contribution
from the stars and the contribution from dark matter to the total mass of the galaxy
by setting an upper limit on the baryon content. However, if the inner rotational
velocities are being affected by dynamics other than the gravitational potential, then
the stellar contribution is essentially being scaled arbitrarily to something that does not depend
on the mass. This problem is avoided when the stellar disc/bulge M/L is fixed in the 
decomposition analysis.

\begin{figure*}
\includegraphics[width=\textwidth,height=0.9\textheight]{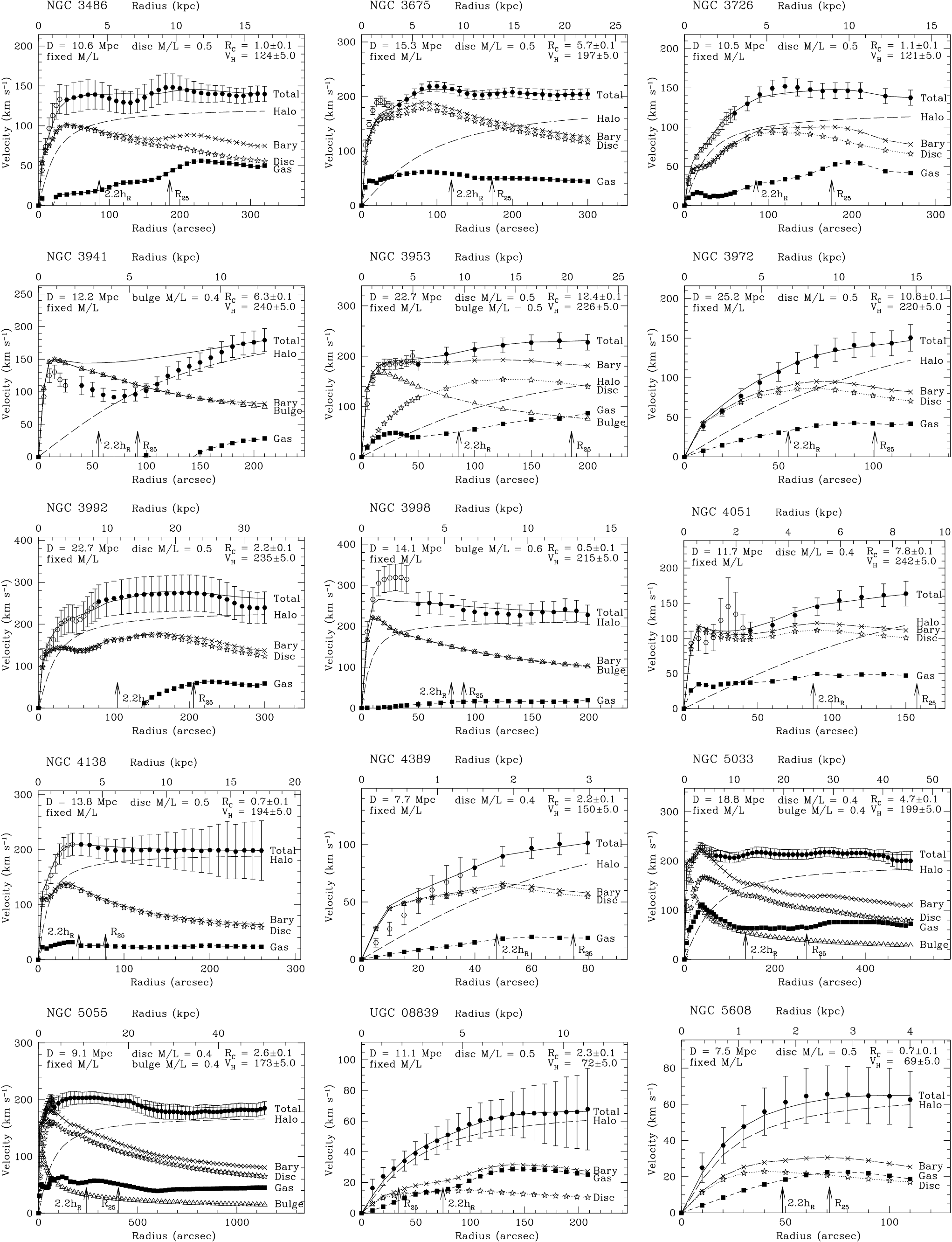}
\caption{Rotation curve decompositions assuming a fixed stellar disc/bulge
M/L = 0.5$\pm$0.1 for each galaxy.
The total gas ({\it filled squares}), stellar bulge ({\it open triangles}), stellar disc 
({\it open stars}), and dark matter halo model ({\it dashed line}) components are added 
in quadrature to achieve the best overall fit ({\it solid line}) to the observed rotation 
curve. Open circles represent circular rotation velocities derived from the ionized gas 
(SparsePak) observations and the filled circles are from the HI observations. 
Adding the baryonic components in quadrature without the dark matter results in the 
crosses.}
\label{fig:rcdecomp}
\end{figure*}

Rotation curve decomposition results for each galaxy are shown in Fig.~\ref{fig:rcdecomp}.
Circular rotation velocities for the total gas, stellar bulge and stellar disc were summed in 
quadrature and subtracted from the observed rotation curve. 
The stellar bulge and disc M/L ratios were assumed to be constant with radius 
and fixed at 0.5 with a $\pm$ 0.1 allowed variance to take into account distance 
uncertainties (See Section~\ref{sec:distances}). Both the bulge and disc were fixed to
the same M/L value.
A spherical pseudo-isothermal dark matter halo model was fit to the residuals 
\citep[e.g.][]{OstrikerDM, KentDM}. The halo parameters $R_{\rm C}$ and $V_{\rm H}$, 
where $R_{\rm C}$ is the halo core radius and $V_{\rm H}$ is the maximum velocity of the halo, 
were left as free parameters to be fit during the decomposition.

An interactive, iterative approach was used to determine a reasonable 
fit for each galaxy. The same intermediate radial range free from the influence of warps in
the outer rotation curves which was used to determine $V_{\rm flat}$ was used to 
determine the fit in the decomposition. Table~\ref{tab:dmprops} lists the dark matter halo
parameters fit in the decomposition. We do not try to model warps in the outer rotation
curves where it cannot be confirmed that the rotational velocities are tracing the
circular speed of the gravitational potential. The errors given for the fixed M/L derived 
$M_{\rm bary}$/$M_{\rm tot}$ at 2.2$h_{\rm R}$ and $R_{\rm trans}$ values are 
the computed average range of possible values assuming a $\pm$0.1 difference in M/L.

For comparison, a decomposition was additionally performed for each galaxy assuming 
the stellar disc and bulge contribute maximally to the inner observed rotation curve
on top of the contribution from the gas. 
A separate maximal M/L value was fit for the disc and bulge when both components are present.
The maximum disc/bulge M/L values and dark matter halo parameters for these fits 
are also listed in Table~\ref{tab:dmprops}. Based on the evidence
for the stellar M/L at 3.6$\mu$m $\sim$0.5 \citep[e.g.][]{LMCML,btfrml,ml5btfr}, the fixed
M/L decompositions appear to offer more astrophysically plausible results. It is important
to note, however, that without additional constraints on the true baryonic contribution,
it is not possible to distinguish either the maximal disc/bulge or fixed M/L decomposition
as the correct result. The magnitude of the differences between these two decompositions
is explored in Section~\ref{sec:discussion}.

\begin{landscape}
\begin{table}
\centering
\caption{Rotation Curve Decomposition Results}
\label{tab:dmprops}
\begin{threeparttable}
\begin{tabular}{lrrrrrrrccrrrr}
\hline
Galaxy & Dynamical\tnote{a} & \multicolumn{2}{c}{M/L$_{\rm disc}$, M/L$_{\rm bulge}$} & 
\multicolumn{2}{c}{$R_{\rm C}$} & \multicolumn{2}{c}{$V_{\rm H}$} & 
\multicolumn{2}{c}{$M_{\rm bary}$/$M_{\rm tot}$\tnote{b}} & 
\multicolumn{4}{c}{$R_{\rm trans}$\tnote{c}} \\
  & Mass & & & \multicolumn{2}{c}{(kpc)} & \multicolumn{2}{c}{(km s$^{-1}$)} & 
\multicolumn{2}{c}{(at 2.2$h_{\rm R}$)} & \multicolumn{2}{c}{(kpc)} & 
\multicolumn{2}{c}{($h_{\rm R}$)} \\
  & (10$^{10}$ M$_{\sun}$) & fixed & max & fixed & max & fixed & max & 
fixed & max & fixed & max & fixed & max\\
\hline
NGC~3486 & 7.48$\pm$0.04 & 0.5, -- & 1.0, -- & 1.0$\pm$0.1 & 7.9$\pm$0.1 & 124$\pm$5 & 158$\pm$5 &
0.46$\pm$0.08 & 0.82 & 3.4$\pm$2.9 & 13.7 & 1.7$\pm$1.4 & 6.8 \\
NGC~3675 & 21.6$\pm$0.1 & 0.5, -- & 0.7, -- & 5.7$\pm$0.1 & 16.8$\pm$0.1 & 197$\pm$5 & 268$\pm$5 &
0.69$\pm$0.12 & 0.88 & 15.1$\pm$3.7 & 21.1 & 3.8$\pm$0.9 & 5.3 \\
NGC~3726 & 6.05$\pm$0.03 & 0.5, -- & 1.1, -- & 1.1$\pm$0.1 & 12.7$\pm$0.1 & 121$\pm$5 & 181$\pm$5 &
0.46$\pm$0.09 & 0.95 & 1.1$\pm$0.4 & $>$13.7 & 0.5$\pm$2 & $>$7.0 \\
NGC~3941 & 9.28$\pm$0.09 & --, 0.4 & --, 0.3 & 6.3$\pm$0.1 & 7.5$\pm$0.1 & 240$\pm$5 & 266$\pm$5 &
1.7$\pm$0.4 & 1.3 & 7.8$\pm$0.9 & 7.0 & 5.2$\pm$0.6 & 4.7 \\
NGC~3953 & 26.5$\pm$0.1 & 0.5, 0.5 & 0.7, 0.5 & 12.4$\pm$0.1 & 20.0$\pm$0.1 & 226$\pm$5 & 231$\pm$5 &
0.82$\pm$0.15 & 0.94 & $>$22.0 & $>$22.0 & $>$5.1 & $>$5.1 \\
NGC~3972 & 7.71$\pm$0.10 & 0.5, -- & 0.6, -- & 10.8$\pm$0.1 & 9.8$\pm$0.1 & 220$\pm$5 & 199$\pm$5 &
0.63$\pm$0.11 & 0.73 & 9.6$\pm$1.7 & 10.7 & 3.1$\pm$0.6 & 3.5 \\
NGC~3992 & 44.0$\pm$1.1 & 0.5, -- & 0.8, -- & 2.2$\pm$0.1 & 3.0$\pm$0.1 & 235$\pm$5 & 198$\pm$5 & 
0.39$\pm$0.08 & 0.62 & 3.5$\pm$1.8 & 31.9 & 0.67$\pm$0.34 & 6.1 \\
NGC~3998 & 16.4$\pm$0.2 & --, 0.6 & --, 1.4 & 0.5$\pm$0.1 & 13.3$\pm$0.1 & 215$\pm$5 & 386$\pm$5 &
0.36$\pm$0.06 & 0.84 & 1.1$\pm$0.2 & 11.7 & 0.44$\pm$0.09 & 4.8 \\
NGC~4051 & 5.30$\pm$0.06 & 0.4, -- & 0.5, -- & 7.8$\pm$0.1 & 6.7$\pm$0.1 & 242$\pm$5 & 205$\pm$5 & 
0.72$\pm$0.15 & 0.84 & $\geq$7.8 & $>$8.5 & $\geq$3.5 & $>$3.8 \\
NGC~4138 & 15.9$\pm$1.2 & 0.5, -- & 1.1, -- & 0.7$\pm$0.1 & 3.8$\pm$0.1 & 194$\pm$5 & 213$\pm$5 &
0.39$\pm$0.07 & 0.85 & 1.7$\pm$0.9 & 7.1 & 1.2$\pm$0.6 & 4.9 \\
NGC~4389 & 0.71$\pm$0.01 & 0.4, -- & 0.1, -- & 1.8$\pm$0.1 & 1.3$\pm$0.1 & 133$\pm$5 & 135$\pm$5 &
0.55$\pm$0.13 & 0.23 & 2.0$\pm$0.4 & 0.7 & 2.5$\pm$0.4 & 0.8 \\
NGC~5033 & 42.6$\pm$0.4 & 0.4, 0.4 & 0.4, 0.4 & 4.7$\pm$0.1 & 3.7$\pm$0.1 & 199$\pm$5 & 194$\pm$5 &
0.53$\pm$0.11 & 0.50 & 13.6$\pm$5.6 & 12.3 & 2.4$\pm$1.0 & 2.2 \\
NGC~5055 & 40.1$\pm$0.2 & 0.4, 0.4 & 0.5, 0.4 & 2.6$\pm$0.1 & 5.1$\pm$0.1 & 173$\pm$5 & 177$\pm$5 &
0.51$\pm$0.11 & 0.64 & 11.2$\pm$4.9 & 16.8 & 2.3$\pm$1.0 & 3.5 \\
UGC~08839 & 1.20$\pm$0.18 & 0.5, -- & 3.0, -- & 2.3$\pm$0.1 & 5.6$\pm$0.1 & 72$\pm$5 & 85$\pm$5 &
0.18$\pm$0.02 & 0.59 & $<$0.5 & 6.0 & $<$0.3 & 3.3 \\
NGC~5608 & 0.36$\pm$0.02 & 0.5, -- & 1.8, -- & 0.7$\pm$0.1 & 1.5$\pm$0.1 & 69$\pm$5 & 71$\pm$5 &
0.23$\pm$0.03 & 0.59 & $<$0.4 & 2.3 & $<$0.5 & 2.8 \\
\hline
\end{tabular}
\begin{tablenotes}
\item[a] Calculated using the last measured rotation curve point.
\item[b] Relative baryonic contribution to the total observed gravitational potential measured 
at 2.2 times the average disc radial scale length, $h_{\rm R}$.
\item[c] The radius at which the gravitational potential transitions
from baryon-dominated to dark matter-dominated (where $M_{\rm bary}$/$M_{\rm tot}$ 
= 0.5) expressed in kpc and in terms of disc radial scale lengths ($h_{\rm R}$) measured
at 3.6$\mu$m.
\end{tablenotes}
\end{threeparttable}
\end{table}
\end{landscape}

\section{Discussion}
\label{sec:discussion}

\begin{figure}
\includegraphics[width=\columnwidth]{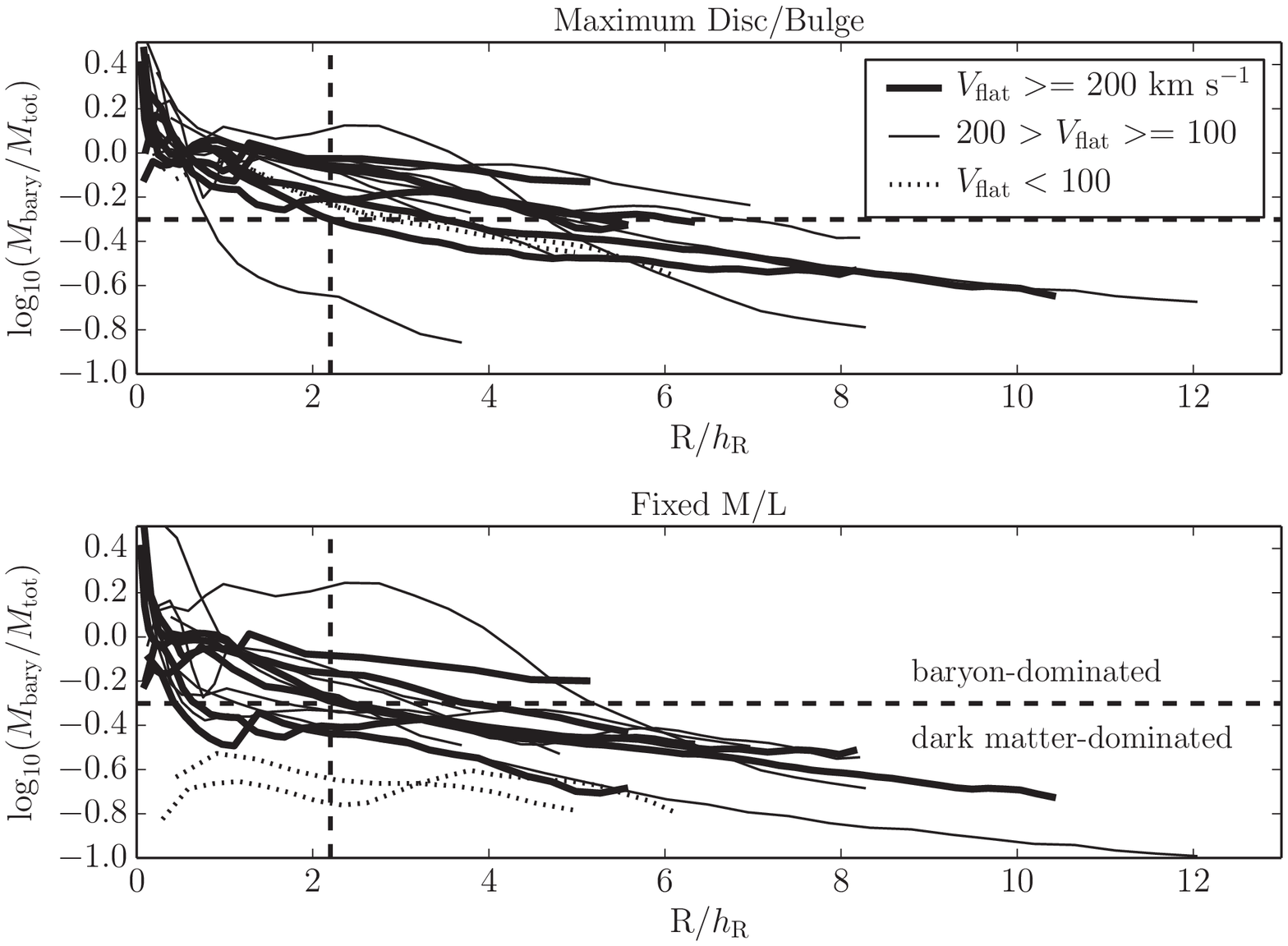}
\caption{Radial distribution of the mass in baryonic matter to total mass in each galaxy. 
The radius of each galaxy has been scaled by its average disc scale length measured at 3.6$\mu$m.
The most massive galaxies, as indicated by large $V_{\rm flat}$ through the BTFR (McGaugh et al. 2000), 
are plotted with a thick solid line, intermediate mass with a thin solid line, and the lowest mass galaxies
are plotted with a dotted line. {\it Top:} Baryon mass fraction distribution from the results
of the maximum stellar disc/bulge decompositions. {\it Bottom:} Baryon mass fraction
distribution from the results of the fixed stellar M/L = 0.5$\pm$0.1 decompositions.
The horizontal dashed line in each plot denotes where the mass fraction transitions from 
being baryon-dominated to dark matter-dominated (where $M_{\rm bary}$/$M_{\rm tot}$ = 0.5).
The vertical dashed line marks the location of 2.2$h_{\rm R}$.}
\label{fig:massfrac}
\end{figure}

With rotation curve decompositions in hand, we examine how strongly the inferred distribution
of mass is altered between fixed M/L and maximal disc/bulge assumptions.
Fig.~\ref{fig:massfrac} shows the radial distribution of $V_{\rm bary}^2$/$V_{\rm tot}^2$,
as determined from the model gas and star rotation curves and total observed rotation curve. 
The true quantity being traced is the baryonic contribution to the total potential of the galaxy, 
which is equivalent to the ratio of baryonic mass to total mass in the limit of spherical distributions,
so it is denoted as $M_{\rm bary}$/$M_{\rm tot}$. The radial extent of each galaxy is scaled by
its average disc scale length measured at 3.6$\mu$m.
The galaxies are separated by $V_{\rm flat}$ to distinguish the more massive spiral galaxies 
from the less massive dwarf galaxies. The top panel in Fig.~\ref{fig:massfrac} shows the 
baryon mass fraction distribution derived from the maximum stellar disc/bulge decompositions.
The bottom panel displays the baryon mass fraction distribution results from the fixed
M/L = 0.5$\pm$0.1 decompositions shown in Fig.~\ref{fig:rcdecomp}.
Note that $\log_{10}$($M_{\rm bary}$/$M_{\rm tot}$) ratios greater than zero occur when the 
estimated total baryonic contribution (the rotation curve labelled 'Bary' in Fig.~\ref{fig:rcdecomp}) 
over-estimates the observed rotational velocities in the centre. There is no noticeable
difference in the distribution of baryonic to total mass for galaxies of different $V_{\rm flat}$,
or baryonic mass, except for the lowest mass galaxies when a fixed M/L is used. A similar
result was found for the THINGS sample (\citealt{dmthings}).

\begin{figure}
\includegraphics[width=\columnwidth]{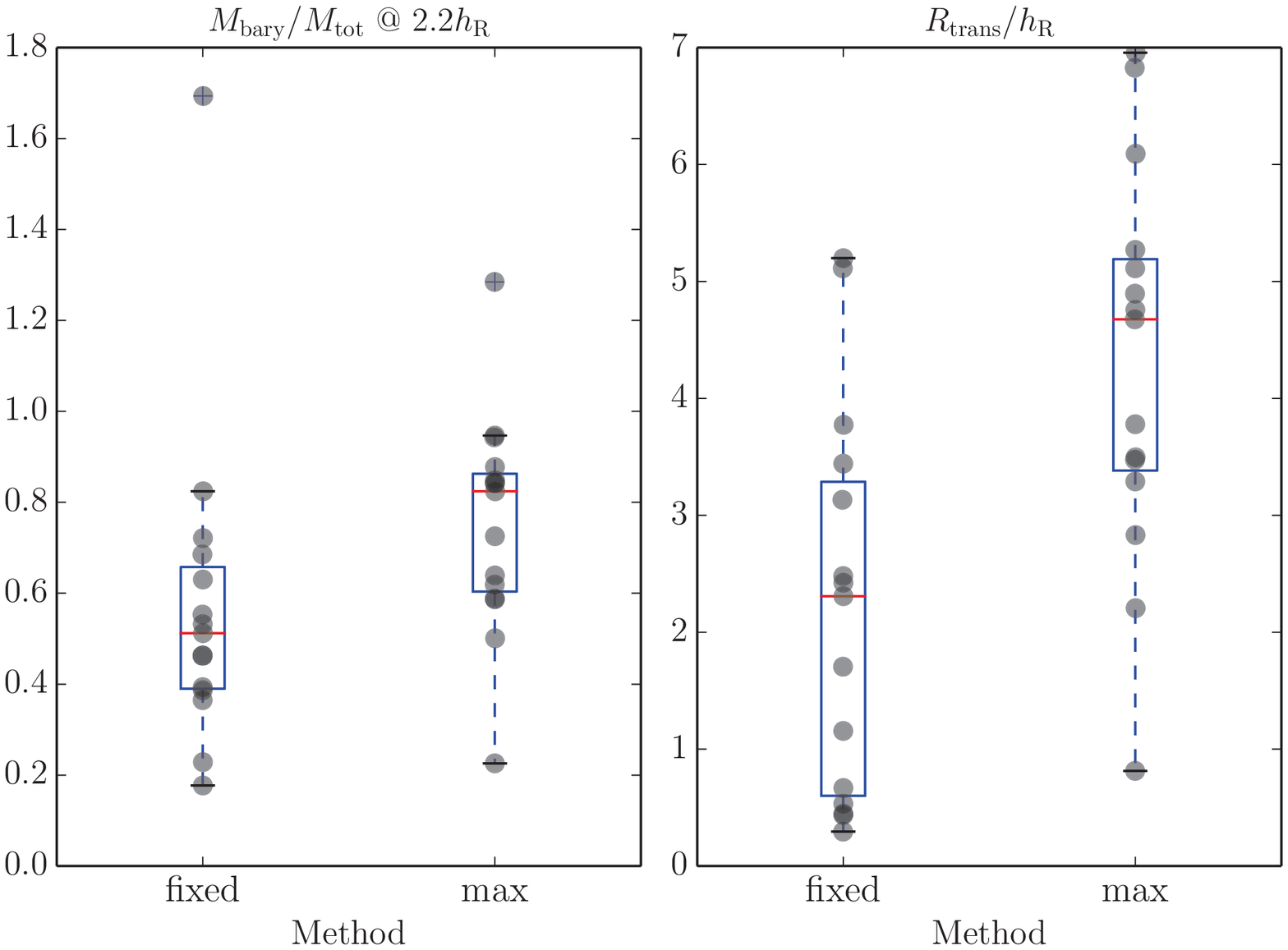}
\caption{{\it Left:} Boxplot comparison of $M_{\rm bary}$/$M_{\rm tot}$ 
at 2.2$h_{\rm R}$ from the fixed M/L and maximum disc/bulge decompositions. The 
median of the results assuming a fixed M/L = 0.5$\pm$0.1 is 0.51, while the median
for the maximum disc/bulge results is 0.82.
The large outlier with a non-physical baryon mass fraction $>$ 1 is NGC~3941 
whose up-bending rotation curve is likey not tracing the gravitational potential of the galaxy.
{\it Right:} Boxplot comparison of $R_{\rm trans}$/$h_{\rm R}$ from the fixed M/L and
maximum bulge/disc decompositions. Upper and lower limits on $R_{\rm trans}$ are included to illustrate the
range of the data. The median $R_{\rm trans}$/$h_{\rm R}$ values for the fixed M/L and
maximum disc/bulge results are 2.3 and 4.7, respectively.}
\label{fig:MLcomp}
\end{figure}

Dashed lines demarcating $M_{\rm bary}$/$M_{\rm tot}$ = 0.5  and 2.2$h_{\rm R}$
have been added to each plot of Fig.~\ref{fig:massfrac} to exemplify the differences between the 
two versions of the mass decomposition. By design, the baryon mass fraction at 2.2$h_{\rm R}$ 
is higher in the maximum disc/bulge decompositions with a mean and standard deviation of
0.75$\pm$0.24, and with most galaxies having baryon mass fractions greater than 0.5.
This can also be seen in the left panel of Fig.~\ref{fig:MLcomp} where all
but one galaxy lies at or above $M_{\rm bary}$/$M_{\rm tot}$ = 0.5 for the
maximum disc/bulge results. All but two galaxies have baryon mass fractions at 2.2$h_{\rm R}$
in their fixed M/L decompositions which are the same or less than the maximum disc/bulge results.
The two exceptions are NGC~3941 and NGC~4389, both of which are likely cases where the rotation curves
are not tracing the gravitational potential of the galaxy (see Sections~\ref{sec:n3941}
and~\ref{sec:n4389}). The baryon mass fractions at 2.2$h_{\rm R}$ for the fixed M/L 
decompositions have a mean and standard deviation of 0.57$\pm$0.35, or 0.49$\pm$0.19 if 
NGC~3941 and NGC~4389 are excluded. On average, the fixed M/L baryon mass fraction results are lower 
by 0.3$\pm$0.1 $M_{\rm bary}$/$M_{\rm tot}$ at 2.2$h_{\rm R}$ than the maximum disc/bulge results. 
The ratio between the maximum disc/bulge and fixed M/L baryon mass fraction results
scale directly with the ratio of M/L, assuming stellar mass is the dominant contributor 
to the total baryonic mass component at 2.2$h_{\rm R}$.

Another consequence of maximizing the baryonic contribution to the observed
rotation curve in the maximum disc/bulge scenario is that it pushes the baryon to dark
matter-dominated transition to a larger radius. The right panel of of Fig.~\ref{fig:MLcomp}
compares the transition radius, $R_{\rm trans}$, normalized by disc scale length for each
galaxy from the fixed M/L and maximum disc/bulge decompositions. Although the standard
deviations of the normalized $R_{\rm trans}$ values are similar in each case, the median 
value of the maximum disc/bulge results is double the median value from the fixed M/L results.
Understanding how $R_{\rm trans}$ scales with M/L is not as simple as for the baryon
mass fraction. The shape of the rotation curves can affect the exact location of $R_{\rm trans}$
causing some galaxies to have lower $R_{\rm trans}$ in the fixed M/L results than 
would be expected from the difference in the M/L assumptions.

Figs.~\ref{fig:massfrac} and~\ref{fig:MLcomp} illustrate the magnitude
of the effects the assumed stellar M/L has on the distribution of mass derived
from rotation curve decomposition. Even with this relatively small subset of galaxies
from the complete EDGES kinematic sample, it is apparent that comparisons
between the distribution of baryonic and non-baryonic matter derived in this 
manner will need to address the dependence on assumed M/L. In the future, we will
apply the methods used in the present study to the complete EDGES kinematic sample
to expore correlations between the distribution of dark matter and sturctural properties
of the stellar disc. This will include an investigation of how the results change
using an astrophysically-motivated fixed M/L and an extreme upper limit from the
maximum disc/bulge M/L. Comparison to the distributions of the various mass components
from cosmological hydrodynamical simulations \citep[e.g.][]{eaglesim} will
provide further insight into how baryons form and evolve inside dark matter haloes.

\section{Conclusion}
\label{sec:conclusion}

We have presented new and archival HI synthesis observations from the VLA and WSRT and new 
IFS observations from SparsePak on the WIYN 3.5 m telescope to constrain the neutral and ionized gas 
kinematics in sixteen galaxies. We additionally include {\it Spitzer} 3.6$\mu$m images to trace the 
extended stellar populations, as well as optical broadband $B$ and $R$ and narrowband H$\alpha$ from the 
WIYN 0.9 m telescope to measure properties of the dominant stellar populations. 
This multiwavelength
dataset was used to carry out rotation curve decomposition analysis and to illustrate how the assumed
stellar M/L effects the inferred distribution of mass. The main results are summarized below.

\begin{enumerate}
\item New HI radio synthesis observations were obtained with the VLA in C configuration for eight
galaxies, seven of which are presented in this study, to derive extended rotation curves from the neutral
gas kinematics. Parameters of the resulting data cubes are provided in Table~\ref{tab:cubeprops}.
In addition, new SparsePak IFS observations were acquired for tweleve galaxies to use
the ionized gas kinematics to constrain the rotational velocities in the central regions. Ionized
gas velocity fields and combined HI and ionized gas rotation curves are presented in Fig.~\ref{fig:spakvels}
and Fig.~\ref{fig:rc}, respectively. Kinematic properties are presented in Table~\ref{tab:kinprops}.

\item The ionized and neutral gas rotation curves were decomposed using model rotation curves to estimate
the stellar and gas contributions to observed dynamics. The contributions from the neutral and molecular
gas were determined from mass surface density measurements from HI synthesis and archival CO data. The
stellar bulge and disc contributions were derived from surface brightness measurements using the
{\it Spitzer} 3.6$\mu$m images. The distribution of dark matter was estimated by 
fitting a spherical pseudo-isothermal halo model. Rotation curve decomposition results using fixed
stellar disc and bulge M/L = 0.5$\pm$0.1 are presented in Fig.~\ref{fig:rcdecomp} and summarized in
Table~\ref{tab:dmprops}.

\item Rotation curve decompositions assuming maximum contribution from the stellar
disc and bulge were carried out for comparison. Results from these decompositions are presented
in Table~\ref{tab:dmprops}. We used the rotation curve decompositions for each galaxy 
to investigate how the baryon mass fraction is distributed as a function of radius 
(Fig.~\ref{fig:massfrac}) under the fixed M/L and maximum disc/bulge assumptions. 
The magnitude of the differences between the two methods can readily be seen in Fig.~\ref{fig:MLcomp}. 
By definition, the maximum disc/bulge method results in higher baryon mass fractions measured at
2.2 disc scale lengths and larger transition radii. The degree to which these values change is
directly related to how large the difference is in asumed M/L.
\end{enumerate}

\section*{Acknowledgements}

The authors acknowledge the referee for many helpful suggestions on
improving this manuscript. The authors thank J.M. van der Hulst and Martha P. Haynes 
for providing HI data cubes and moment maps of archival data.
The authors acknowledge observational and technical support from the National 
Radio Astronomy Observatory (NRAO). We acknowledge use of the WIYN 0.9 m telescope 
operated by WIYN Inc. on behalf of a Consortium of ten partner Universities 
and Organizations, and the WIYN 3.5 m telescope. The WIYN Observatory is a joint 
facility of the University of Wisconsin-Madison, Indiana University, the National 
Optical Astronomy Observatory and the University of Missouri. This work is based on observations made 
with the {\it Spitzer Space Telescope}, which is operated by the Jet Propulsion Laboratory, 
California Institute of Technology under a contract with NASA. EER 
acknowledges support from the Provost's Travel Award for Women in Science, a 
professional development fund supported through the Office of the Provost at 
Indiana University Bloomington, and the Indiana Space Grant Consortium. 
LvZ recognizes support from the Helena Kluyver Female Visitor programme
at ASTRON during which some of this analysis was completed.
This research has made use of the NASA/IPAC Extragalactic Database (NED) which is 
operated by the Jet Propulsion Laboratory, California Institute of Technology, 
under contract with the National Aeronautics and Space Administration. 




\bibliographystyle{mnras}
\bibliography{baryonsDMHalosI_rev2Clean}



\appendix

\section{Notes on Individual Galaxies}
\label{sec:notes}
In this section, we introduce each galaxy in the study and provide a brief description
of notable attributes. Comparison to other observations and analyses are discussed when available.
Figs.~\ref{fig:n3486sum}-\ref{fig:n5608sum} present a summary of the observational data for
each galaxy.

\subsection{NGC~3486}
\label{sec:n3486}

\begin{figure*}
\includegraphics[height=0.9\textheight]{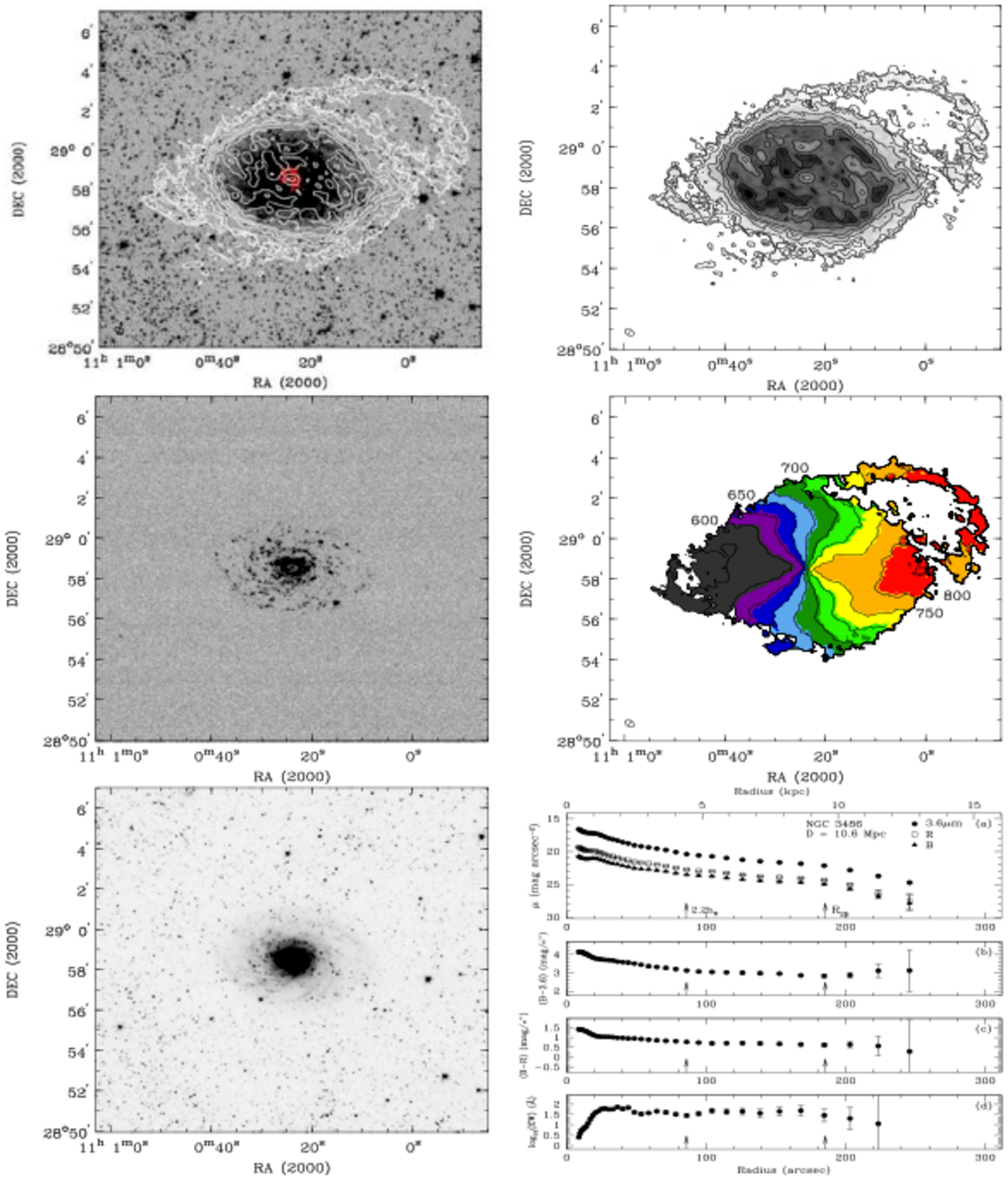}
\caption{{\bf NGC~3486:} {\it (top left)} Low spatial resolution HI integrated intensity contours 
from new VLA data {\it (white)} and CO intensity contours {\it (red)} from CARMA STING 
(\citealt{carmasting}) overlaid on the {\it Spitzer} 3.6$\mu$m image. 
{\it (top right)} Low spatial resolution HI intensity contours overlaid on the low 
spatial resolution HI intensity image. The first low resolution HI contour represents a 
column density of 4$\times$10$^{19}$ atoms cm$^{-2}$.
{\it (middle left)} Narrowband H$\alpha$ image from the WIYN 0.9 m telescope. 
{\it (middle right)} HI velocity field with isovelocity contours derived from the low spatial 
resolution HI data. The isovelocity contours are spaced every 25 km s$^{-1}$.
{\it (bottom left)} {\it Spitzer} 3.6$\mu$m image with a high surface brightness stretch.
{\it (bottom right)} Ellipse photometry results (see Sections~\ref{sec:3.6data} and~\ref{sec:optdata})
showing radial profiles of surface brightness, $B-3.6$ colour, $B-R$ colour, and equivalent width (EW).}
\label{fig:n3486sum}
\end{figure*}

New VLA HI observations of the SABc galaxy NGC~3486 presented in Fig.~\ref{fig:n3486sum} show stream-like features 
extending from the northwest and southeast sides of the HI disc. 
These features are low column-density, but appear to follow the rotation curve of the disc. 
There is no obvious nearby companion or signature of a recent merger. 
In the central region, the HI mass surface density shows a depression in the central 40 arcsec, then
remains constant before decreasing exponentially at $\sim$180 arcsec. The
molecular gas distribution is confined to the central 40 arcsec, but is only dominant within 6 arcsec,
as inferred from a CO map from the CARMA STING survey.

The neutral gas velocity field of NGC~3486 displays outer S-shaped contours 
characteristic of kinematic warps caused by a change in position and/or
inclination angle (e.g. M83; \citealt{velfields}). 
In the case of NGC~3486, kinematic models
with a changing position and inclination angle best match this outer warp. 
The ionized gas velocity field reveals solid-body rotation in the inner
30 arcsec (Fig.~\ref{fig:spakvels}). The outer HI rotation curve between
100-200 arcsec displays a wiggle likely due to the warp noted in the velocity field
(Fig.~\ref{fig:rc}). \cite{BroeilsPV} observed NGC~3486 with the WSRT and noted similar
structure in its position-velocity diagram.

The stellar light in NGC~3486 is mostly disc-dominated, with a slight
contribution from both a bulge and bar component. There is some evidence
of a small downbending in the NIR surface brightness profile at the $R_{25}$,
which also corresponds to a drop in $\log_{10}$(EW). The radial trend in $\log_{10}$(EW) is
inversely correlated with the $B-R$ colour, quickly rising from small EW in the centre
20 arcsec to a larger constant EW value across the disc. Although NGC~3486 has a relatively
small total SFR of just 0.50$\pm$0.36 M$_{\sun}$ yr$^{-1}$, it has a large EW of 
38.2$\pm$3.3~\AA, indicating that the current SFR is greater than the past average SFR. 

The rotation curve decomposition analysis was done using estimates of
the baryonic components from the stellar disc surface density and
a combination of the HI and CO for the total gas mass surface density (Fig.~\ref{fig:rcdecomp}).
The baryonic contribution to the total observed rotation curve is
stellar disc dominated at inner radii, while the gas contribution grows reaching a
peak near 200 arcsec. The fixed M/L = 0.5 decomposition requires a dark matter 
halo model rotation curve which is the dominant contributor to the observed
rotation curve at most radii. The total best-fitting rotation curve does
not follow the wiggles in the observed rotation curve, which is expected
assuming the structure is influenced by forces other than the gravitational potential.
The maximum disc assumption for this galaxy implies a much greater 
contribution from the stellar disc with a M/L of 1.0. This results in a baryon mass fraction
at 2.2$h_{\rm R}$ which differs by a factor of two between the fixed M/L and maximum
disc decompositions methods.

\subsection{NGC~3675}
\label{sec:n3675}

\begin{figure*}
\includegraphics[height=0.9\textheight]{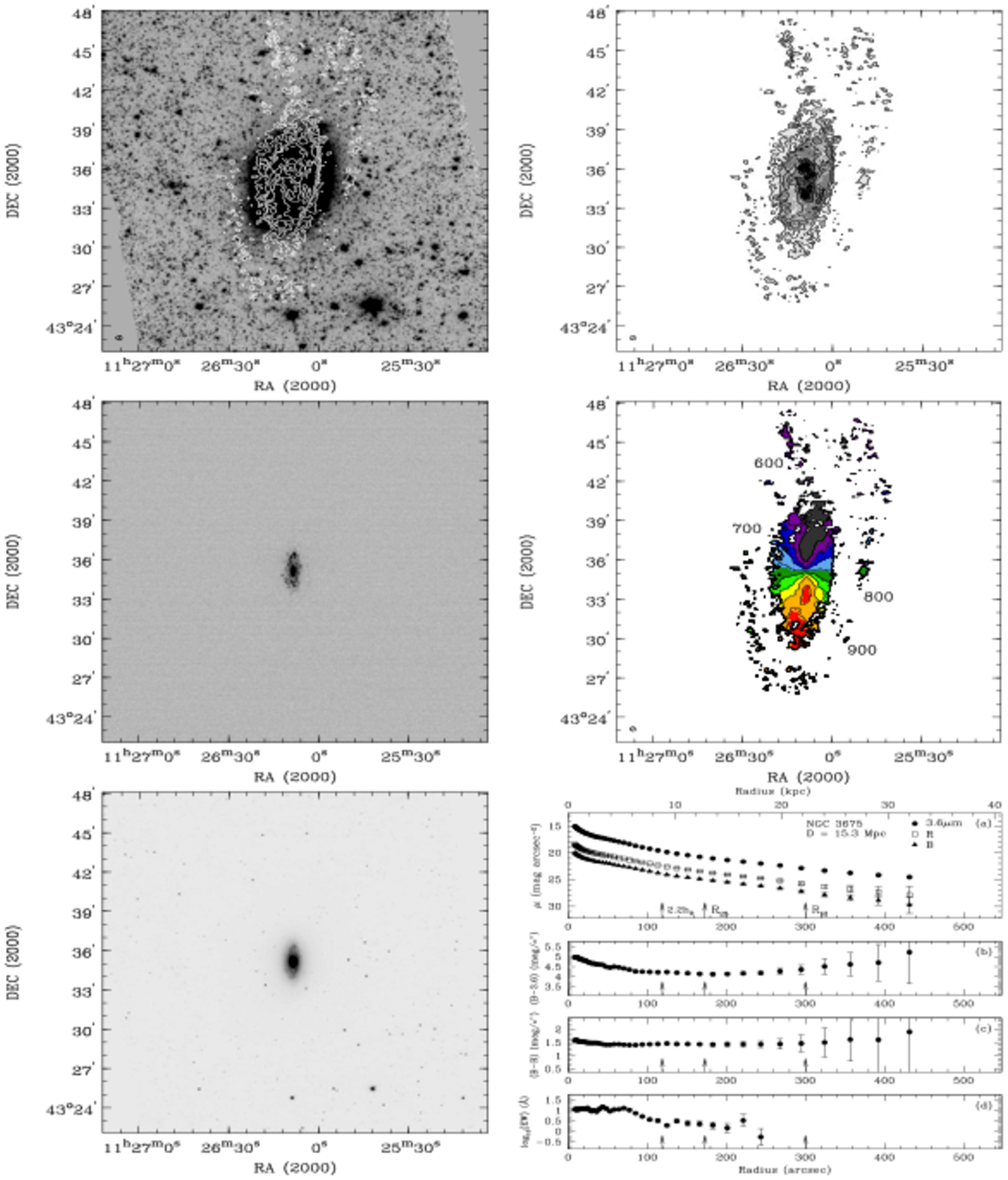}
\caption{{\bf NGC~3675:} {\it (top left)} Medium spatial resolution HI integrated intensity contours 
from archival VLA data overlaid on the {\it Spitzer} 3.6$\mu$m image. 
{\it (top right)} Medium spatial resolution HI intensity contours overlaid on the medium
spatial resolution HI intensity image. The first medium resolution HI contour represents a 
column density of 6$\times$10$^{19}$ atoms cm$^{-2}$. 
{\it (middle left)} Narrowband H$\alpha$ image from the WIYN 0.9 m telescope.
{\it (middle right)} HI velocity field with isovelocity contours derived from the medium spatial 
resolution HI data. The isovelocity contours are spaced every 50 km s$^{-1}$.
{\it (bottom left)} {\it Spitzer} 3.6$\mu$m image with a high surface brightness stretch.
{\it (bottom right)} Ellipse photometry results (see Sections~\ref{sec:3.6data} and~\ref{sec:optdata})
showing radial profiles of surface brightness, $B-3.6$ colour, $B-R$ colour, and equivalent width (EW).}
\label{fig:n3675sum}
\end{figure*}

Archival VLA observations of this SAb galaxy shown in Fig.~\ref{fig:n3675sum} 
reveal an HI disc 10 arcmin in diameter measured
at 10$^{20}$ atoms cm$^{-2}$ with patchy diffuse emission that extends out to a diameter
of 21.5 arcmin. There appears to be a gap between the main HI disc and 
this outer diffuse neutral gas ring. There is no obvious extended diffuse stellar light
at 3.6$\mu$m to match the extended component of the HI disc.
The main HI disc features two dense clumps of HI emission 
directly north and south of the galaxy's nucleus. NGC~3675 was observed by \cite{BroeilsPV}
with the WSRT, but the observations were low signal-to-noise and did not detect the diffuse HI ring.  
The HI mass surface density remains roughly constant in the inner 80 arcsec before slowly decreasing. 
The molecular gas distribution dominates over the HI gas within radii less than 100 arcsec, based
on CO emission detected in four pointings out to a radius of 135 arcsec in the FCRAO survey.

The neutral gas kinematics in Fig.~\ref{fig:rc} show a flat rotation curve with a 
slight warp in the outer disc at a radius 
of $\sim$120 arcsec that can be modeled as a change in position angle. Although the diffuse outer ring 
appears to follow the rotation curve of the galaxy, it was not included in the mass decomposition analysis.
The ionized gas rotation curve rises more steeply than the HI derived rotation curve in the centre
20 arcsec, an effect that can be attributed to beam smearing in the lower resolution HI data. The
ionized gas rotation curve peaks at a velocity of 193 km s$^{-1}$ at a radius of 25 arcsec before
turning over and decreasing out to 40 arcsec, where it joins the HI rotation curve.

The distribution of stellar light in NGC~3675 is primarily disc-like with a small pseudo-bulge
component with a radius of 8.5 arcsec (\citealt{stellarkin}). There is no strong colour gradient;
the $B-R$ colour remains fairly constant at about 1.44, which is normal for an intermediate
type spiral galaxy (\citealt{galbr}). NGC~3675 has a large roughly constant EW value within 80 arcsec which
declines before leveling out again around 120 arcsec. The small star formation rate is consistent
with the relatively little amount of H$\alpha$ emission visible in the narrowband image.

The stellar contribution to the total observed rotation curve was modeled using a disc only
stellar light distribution with a M/L = 0.5 as shown in Fig.~\ref{fig:rcdecomp}. The model rotation curve 
for this stellar disc rises steeply in the inner 20 arcsec like the ionized gas rotation curve, 
but then the slope appears to flatten out slightly before reaching a peak velocity at 80 arcsec. 
This unusual feature could be attributed to dust extinction near the centre within 
NGC~3675 (\citealt{n3675dust}). The shape of the observed ionized and neutral gas rotation curve
mimics the shape of the model stellar disc rotation curve, and is well-described by
the maximum disc model with M/L = 0.7.

\subsection{NGC~3726}
\label{sec:n3726}

\begin{figure*}
\includegraphics[height=0.9\textheight]{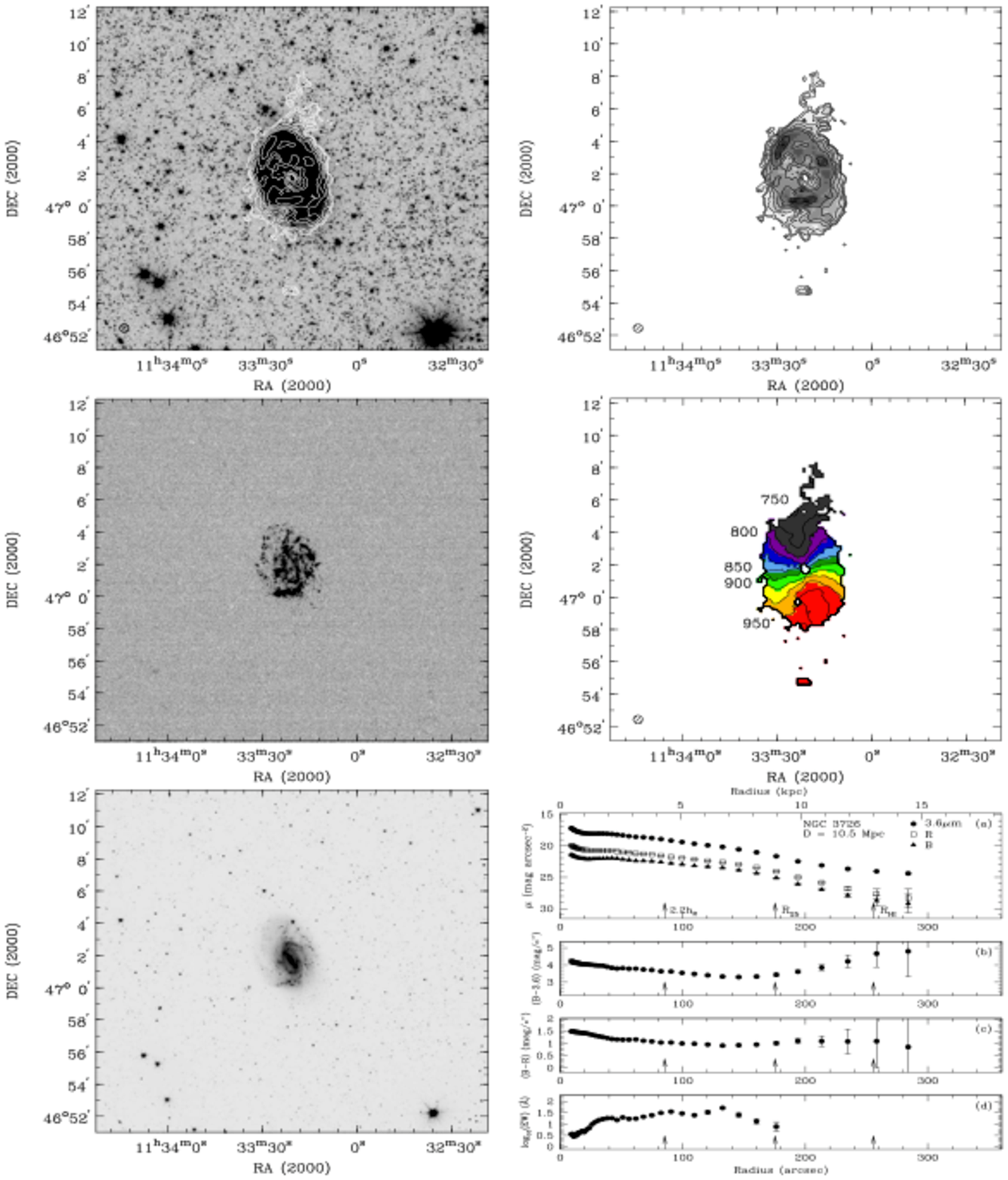}
\caption{{\bf NGC~3726:} {\it (top left)} 30\arcsec~spatial resolution HI integrated intensity contours 
from the WSRT overlaid on the {\it Spitzer} 3.6$\mu$m image. 
{\it (top right)} 30\arcsec~spatial resolution HI intensity contours overlaid on the same
spatial resolution HI intensity image. The first HI contour represents a 
column density of 6$\times$10$^{19}$ atoms cm$^{-2}$. 
{\it (middle left)} Narrowband H$\alpha$ image from the WIYN 0.9 m telescope.
{\it (middle right)} HI velocity field with isovelocity contours derived from the 30\arcsec~spatial 
resolution HI data. The isovelocity contours are spaced every 25 km s$^{-1}$. 
{\it (bottom left)} {\it Spitzer} 3.6$\mu$m image with a high surface brightness stretch.
{\it (bottom right)} Ellipse photometry results (see Sections~\ref{sec:3.6data} and~\ref{sec:optdata})
showing radial profiles of surface brightness, $B-3.6$ colour, $B-R$ colour, and equivalent width (EW).}
\label{fig:n3726sum}
\end{figure*}

NGC~3726 is an SABc galaxy in the Ursa Major cluster. HI synthesis data from the WHISP survey
displayed in Fig.~\ref{fig:n3726sum} shows that the neutral gas is distributed in a ring-like
structure with decreasing column density towards the centre. There are hints of extended
diffuse gas to the north (approaching side) and south (receding side) along the major axis.
The HI velocity field and isovelocity contours hint that the outer disc may be warped.
This warp manifests itself as a dip in the rotation curve around 180 arcsec, which doesn't seem
to be well-modeled by the baryonic contributions (Fig.~\ref{fig:rcdecomp}). This can also
be seen in the decompositions in \cite{verheijenphd} using WSRT observations.
The kinematic signature of a bar can be observed in the S-shaped contours in the ionized
gas velocity field from SparsePak (Fig.~\ref{fig:spakvels}). The ionized gas rotation 
curve and full rotation curve decomposition agrees well with that presented in \cite{ghaspdm}
from Fabry-Perot H$\alpha$ observations.

The 3.6$\mu$m surface brightness profile of NGC~3726 shows that the bulge component is relatively
small and does not contribute greatly to the total light distribution. The shape of the surface
brightness profile at radii within $\sim$50 arcsec displays a dip in brightness, reflecting the 
presence of a ring-like structure in the stellar component (see the bottom left panel in 
Fig.~\ref{fig:n3726sum}). Beyond the ring, the disc light decreases exponentially with a corresponding
flattening of the $B-R$ colour and $\log$(EW) profiles.

The observed ionized and neutral gas rotation curve was decomposed using a total gas contribution
which includes CO observations for the molecular gas and a stellar disc M/L = 0.5 using the 3.6$\mu$m
stellar disc distribution. In this case, CO observations from the FCRAO survey were used in place
of the BIMA SONG map to better recover spatially extended flux.
The maximum disc decomposition results in a disc M/L = 1.1 and a stellar disc 
model rotation curve which accounts for 95 per cent of the observed potential at 
2.2 disc scale lengths versus 46 per cent for the fixed M/L = 0.5 model. This is one of
the largest discrepancies between the fixed M/L and maximum disc models for this sample.

\subsection{NGC~3941}
\label{sec:n3941}

\begin{figure*}
\includegraphics[height=0.9\textheight]{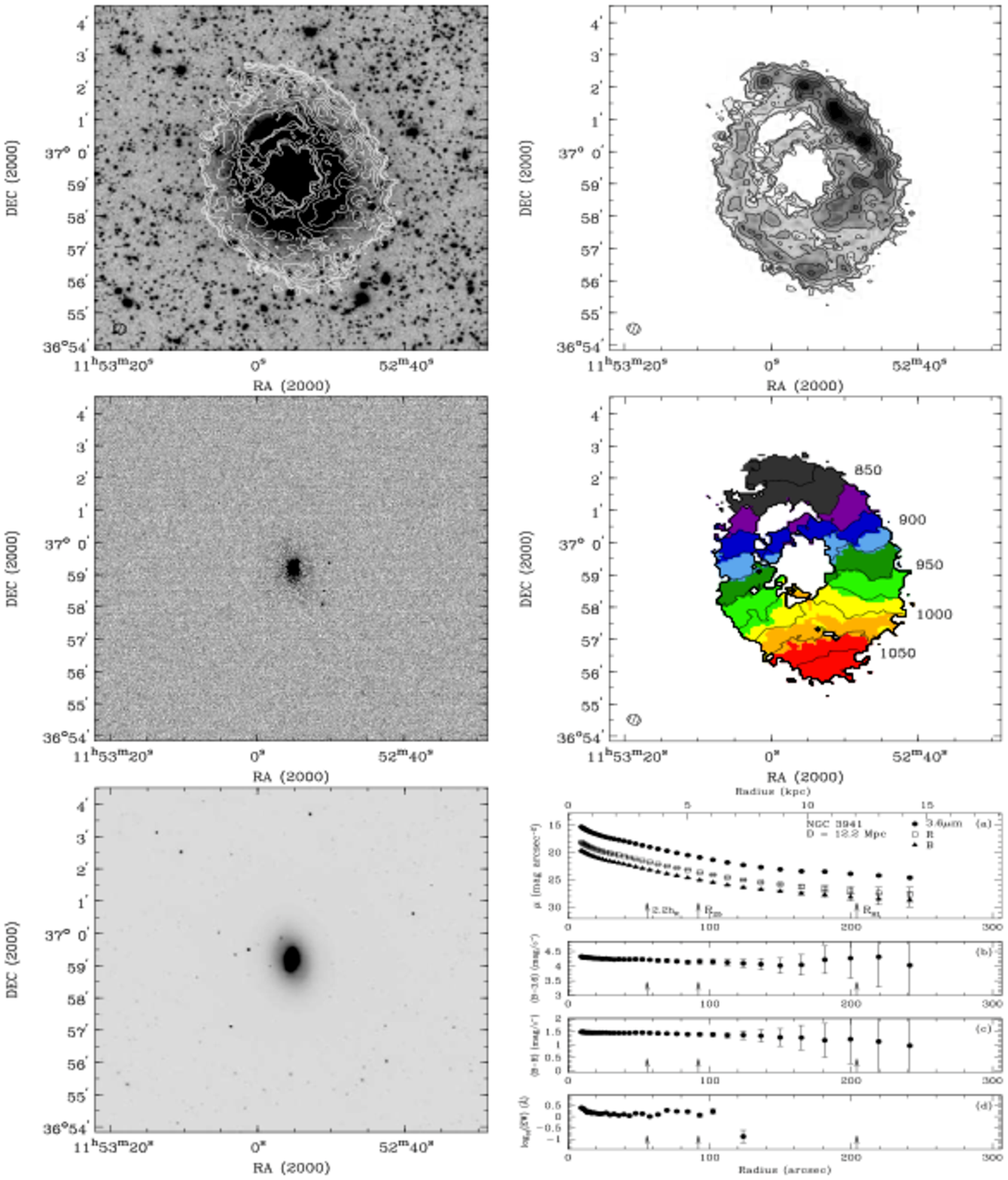}
\caption{{\bf NGC~3941:} {\it (top left)} Low spatial resolution HI integrated intensity contours 
from new VLA data overlaid on the {\it Spitzer} 3.6$\mu$m image. 
{\it (top right)} Low spatial resolution HI intensity contours overlaid on the low 
spatial resolution HI intensity image. The first low resolution HI contour represents a 
column density of 2$\times$10$^{19}$ atoms cm$^{-2}$. 
{\it (middle left)} Narrowband H$\alpha$ image from the WIYN 0.9 m telescope. 
{\it (middle right)} HI velocity field with isovelocity contours derived from the low spatial 
resolution HI data. The isovelocity contours are spaced every 25 km s$^{-1}$. 
{\it (bottom left)} {\it Spitzer} 3.6$\mu$m image with a high surface brightness stretch.
{\it (bottom right)} Ellipse photometry results (see Sections~\ref{sec:3.6data} and~\ref{sec:optdata})
showing radial profiles of surface brightness, $B-3.6$ colour, $B-R$ colour, and equivalent width (EW).}
\label{fig:n3941sum}
\end{figure*}

NGC~3941 is a barred S0 galaxy with a known counterrotating gas disc relative to the stellar 
rotation (\citealt{S0kin}). The low velocity dispersion of this gas disc as measured by
\cite{S0kin} implies it has settled into an equilibrium distribution despite having a likely 
origin due to a merger event or accretion of gas with oppositely directed angular momentum. 
Indeed, the neutral gas appears to have settled in a disc, although it is not a
cohesive structure (Fig.~\ref{fig:n3941sum}). NGC~3941 is known to have a complex 
inner structure involving an inner disc with spiral arms and bar with an oval distortion 
(\citealt{S0phot}). There is no HI detected in the central 40 arcsec, which is a common 
feature of barred galaxies in this sample. The ionized gas emission detected
with SparsePak only fills the inner 25 arcsec. NGC~3941 was also observed with the WSRT as
part of the ATLAS$^{\rm 3D}$ survey. \cite{atlas3dhi} found a similar distribution of 
neutral gas and velocity field.

The average rotation curve derived from the ionized gas velocity field agrees with that derived
by \cite{S0kin}, reaching a peak velocity near 15 arcsec before falling back down to lower
velocities. The neutral gas rotation curve picks up where the ionized gas left off and bends
upwards out to the last measured point at 210 arcsec. 

The stellar light in NGC~3941 is best modeled as a bulge-like distribution with a slight break
at a radius of 132 arcsec where the high surface brightness component ends. The colour and EW
radial trends are remarkably flat, with only a slight increase in EW in the central 
$\sim$20 arcsec where there is the most ionized gas emission.

Given the unusual shape of its rotation curve, it is possible that the 
neutral gas in NGC~3941 is not tracing the gravitational potential. An attempt was made
to decompose the rotation curve into separate mass components, anyways. Fig.~\ref{fig:rcdecomp}
shows the decomposition assuming a bulge distribution for the stellar component with the
bulge M/L fixed to 0.4. NGC~3941 is one of two cases in the present sample where the
fixed M/L model over-estimates the observed rotation curve.
The maximum M/L that the stellar bulge can have without over-shooting the peak ionized
gas velocity is 0.3. In both cases, the model total rotation curve provides a poor fit
to the observed rotation curve, particularly between 25-150 arcsec, causing the baryon
mass fraction at 2.2$h_{\rm R}$ to be $>$1. A distance of 12.2 Mpc estimated from
surface brightness fluctuations (\citealt{sbfdist}) is adopted for NGC~3941. It is
difficult to compare this distance to the one estimated from the LTFR due to the
lack of a suitable radial range to measure $V_{\rm flat}$.

\subsection{NGC~3953}
\label{sec:n3953}

\begin{figure*}
\includegraphics[height=0.9\textheight]{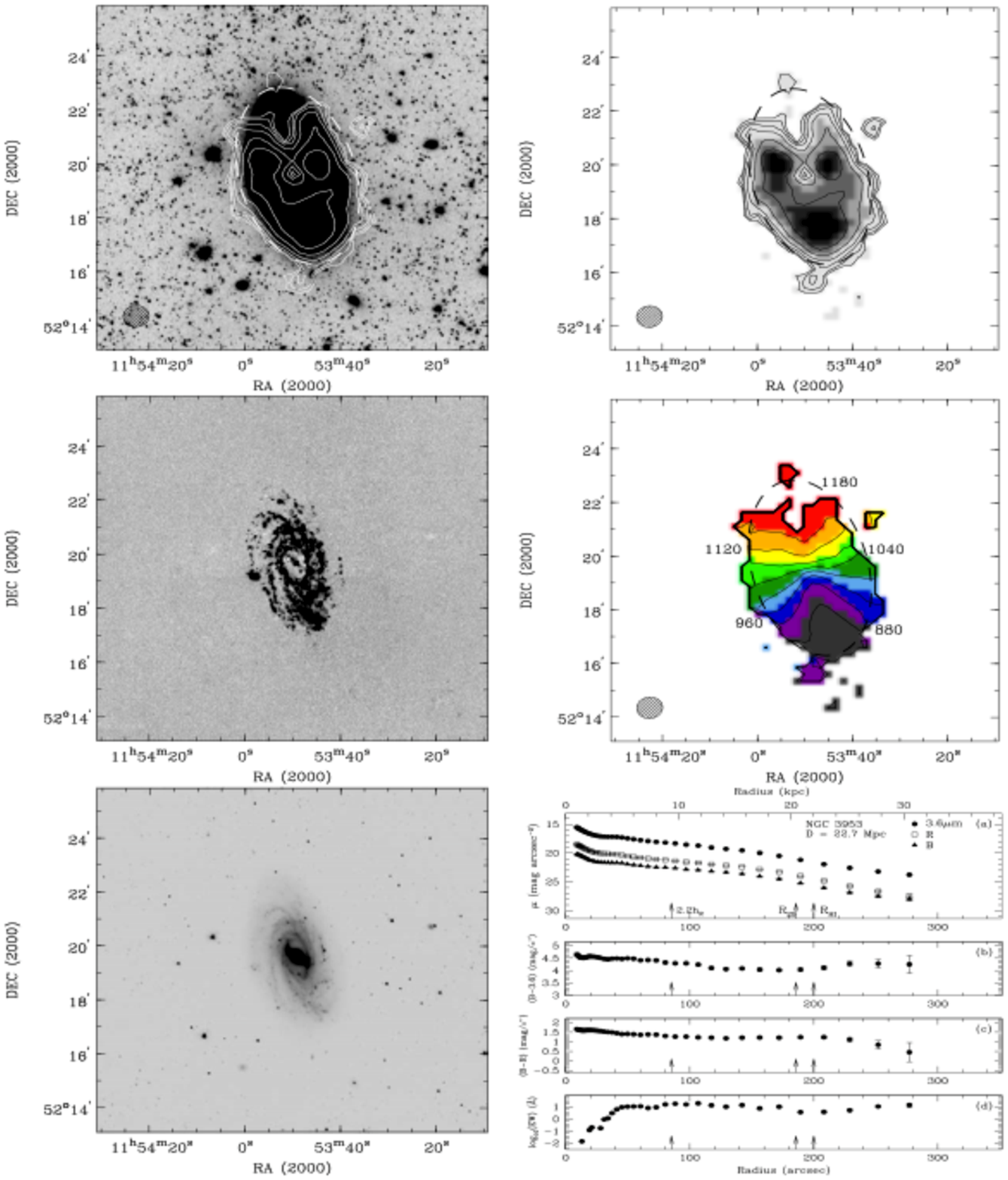}
\caption{{\bf NGC~3953:} {\it (top left)} 30\arcsec~spatial resolution HI integrated intensity contours 
from the WSRT and CO intensity contours {\it (red)} from BIMA SONG (\citealt{bimasongII}) overlaid on the 
{\it Spitzer} 3.6$\mu$m image. 
{\it (top right)} 30\arcsec~spatial resolution HI intensity contours overlaid on the same
spatial resolution HI intensity image. The first HI contour represents a 
column density of 2$\times$10$^{20}$ atoms cm$^{-2}$. 
{\it (middle left)} Narrowband H$\alpha$ image from the WIYN 0.9 m telescope.
{\it (middle right)} HI velocity field with isovelocity contours derived from the 30\arcsec~spatial 
resolution HI data. The isovelocity contours are spaced every 50 km s$^{-1}$.
{\it (bottom left)} {\it Spitzer} 3.6$\mu$m image with a high surface brightness stretch.
{\it (bottom right)} Ellipse photometry results (see Sections~\ref{sec:3.6data} and~\ref{sec:optdata})
showing radial profiles of surface brightness, $B-3.6$ colour, $B-R$ colour, and equivalent width (EW).}
\label{fig:n3953sum}
\end{figure*}

NGC~3953 is an SBbc galaxy in the Ursa Major cluster. Although, one of the most massive galaxies
in the Ursa Major cluster and in the present sample, it is relatively deficient of HI.
The neutral gas does not extend much beyond the stellar disc, and the surface density decreases
towards the centre where there is a prominent bar visible in the H$\alpha$ image in 
Fig.~\ref{fig:n3953sum}. The archival VLA observations did not have
a bandpass wide enough to include the full velocity range of NGC~3953, as is 
evident in the HI figures. The receding side of the galaxy is missing about 
77 km s$^{-1}$, or 15 channels at a velocity resolution of 5.2 km s$^{-1}$. This
corresponds to approximately 35 per cent of the velocity coverage of the receding side,
assuming W$_{20}$ = 441.9 km s$^{-1}$ (\citealt{ursamajhi}). Despite this, the archival
VLA data have significantly higher S/N than the WHISP data in which NGC~3953 was detected
beyond WSRT's primary beam half-power width. Previous study of atomic gas in
NGC~3953 by \cite{ursamajhi} shows that both the distribution and kinematics of the HI
are symmetric. Therefore, we use the rotation curve and gas distribution derived from
the approaching side only as indicative of the global rotation curve and HI mass surface
density distribution. Values for the integrated HI flux density, W$_{20}$, diameter
of the HI disc, dynamical centre and systemic velocity are all adopted from \cite{ursamajhi}.
A dashed ellipse indicating the full diameter of the HI disc is overlaid on the 
HI figures in Fig.~\ref{fig:n3953sum} for reference.

Fig.~\ref{fig:spakfibers} shows that the ionized gas emission observed 
in the SparsePak field of view for NGC~3953 is largely from a bar-like feature. 
The signature S-shaped isovelocity contours from the bar dynamics can be observed
in the ionized gas velocity field in Fig.~\ref{fig:spakvels} and in the H$\alpha$ velocity 
field of \cite{bhabar}. The rotation curve derived from the ionized gas shows
a steep inner rise followed by a plateau at 180 km s$^{-1}$ (Fig.~\ref{fig:rc}). The
rotation curve derived from the approaching side of the HI velocity field agrees well
with \cite{ursamajhi}.

The shape of the ionized gas rotation curve is somewhat reproduced in the model baryonic
rotation curve components (Fig.~\ref{fig:rcdecomp}). The steep rise 
and plateau can be described by a combination of a broad bulge component and slowly 
rising disc model rotation curve both with fixed M/L = 0.5. An equally good fit 
to the observed rotation curve can be achieved by maximizing the disc contribution
with a M/L = 0.7. The bulge contribution appears to be maximal at M/L = 0.5.
The model rotation curve for the gas includes a contribution from molecular gas
estimated from FCRAO observations.
The mass at 2.2 disc scale lengths is  baryon-dominated in both the fixed M/L 
and maximum disc/bulge decomposition models.

\subsection{NGC~3972}
\label{sec:n3972}

\begin{figure*}
\includegraphics[height=0.9\textheight]{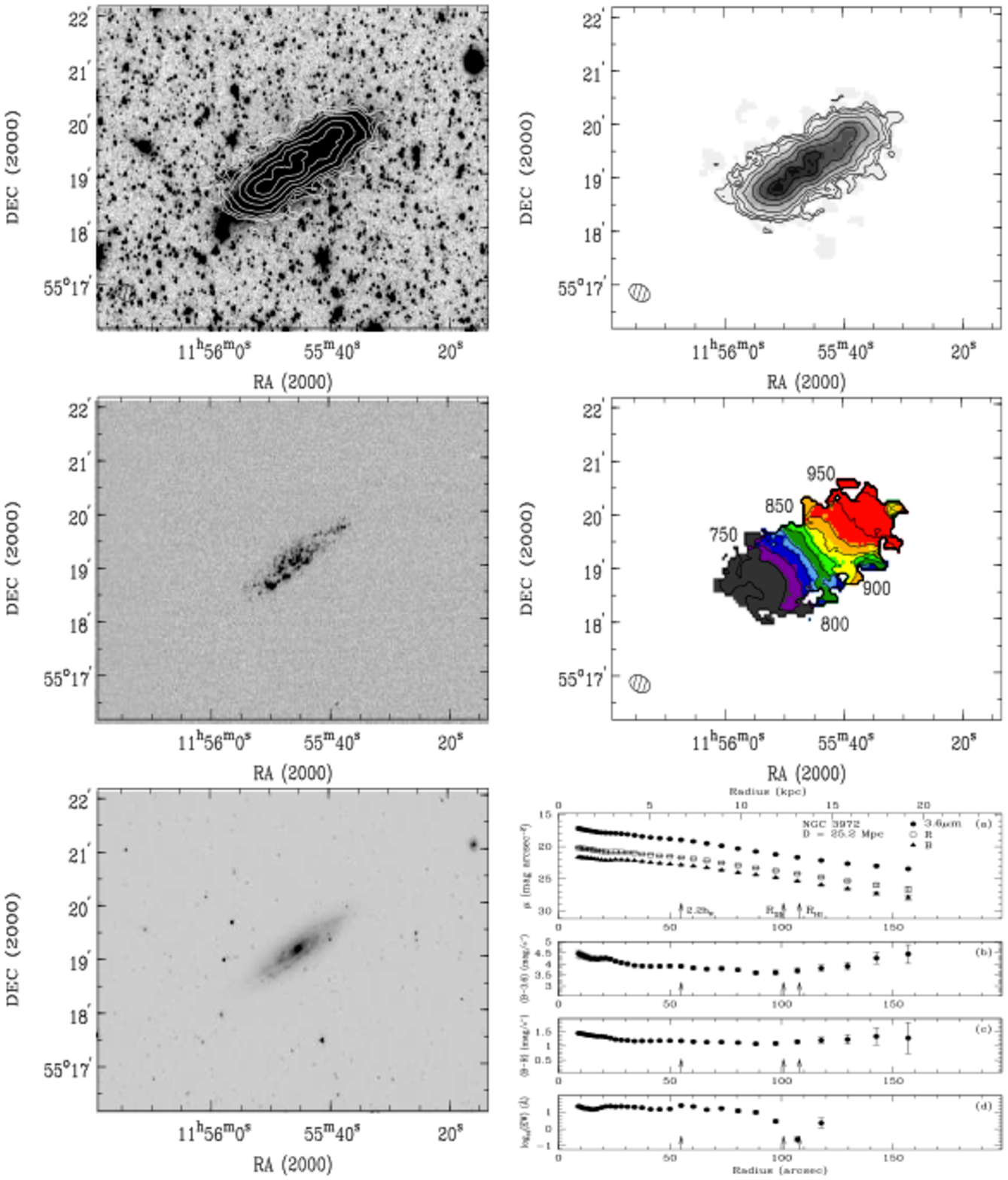}
\caption{{\bf NGC~3972:} {\it (top left)} Low spatial resolution HI integrated intensity contours 
from new VLA data overlaid on the {\it Spitzer} 3.6$\mu$m image. 
{\it (top right)} Low spatial resolution HI intensity contours overlaid on the same
spatial resolution HI intensity image. The first HI contour represents a 
column density of 2$\times$10$^{19}$ atoms cm$^{-2}$. 
{\it (middle left)} Narrowband H$\alpha$ image from the WIYN 0.9 m telescope.
{\it (middle right)} HI velocity field with isovelocity contours derived from the low spatial 
resolution HI data. The isovelocity contours are spaced every 25 km s$^{-1}$.
{\it (bottom left)} {\it Spitzer} 3.6$\mu$m image with a high surface brightness stretch.
{\it (bottom right)} Ellipse photometry results (see Sections~\ref{sec:3.6data} and~\ref{sec:optdata})
showing radial profiles of surface brightness, $B-3.6$ colour, $B-R$ colour, and equivalent width (EW).}
\label{fig:n3972sum}
\end{figure*}

NGC~3972 is an SAbc galaxy detected in the same field of view as NGC~3998 in new VLA HI observations.
Although NGC~3972 lies beyond the primary beam half-power radius, the recovered integrated flux is
comparable to that in \cite{ursamajhi}. The HI morphology matches the stellar morphology and appears
as a highly inclined disc (Fig.~\ref{fig:n3972sum}). The approaching (east) side has a higher column
density of neutral gas as well as a higher density of star formation as seen in the narrowband
H$\alpha$ image. The straight isovelocity contours in the HI velocity field indicate solid body
rotation, and the derived rotation curve in Fig.~\ref{fig:rc} shows as much.

Fig.~\ref{fig:rcdecomp} shows the rotation curve decomposition. The model baryonic rotational velocities
turn over and begin to decrease at a radius of about 80 arcsec, whereas the observed rotation curve
is still rising at the last measured point. The stellar disc is maximal at a
M/L of 0.6 to avoid over-estimating the inner rotation velocities. It is possible that the
slope of the inner rotation curve appears shallow due to the effect of beam smearing. In
this case, the stellar disc could be maximized further. The estimated LTFR distance of
22.2 Mpc is in close agreement with the adopted SNIa distance of 25.2 Mpc (\citealt{n3972distref}).

\subsection{NGC~3992}
\label{sec:n3992}

\begin{figure*}
\includegraphics[height=0.9\textheight]{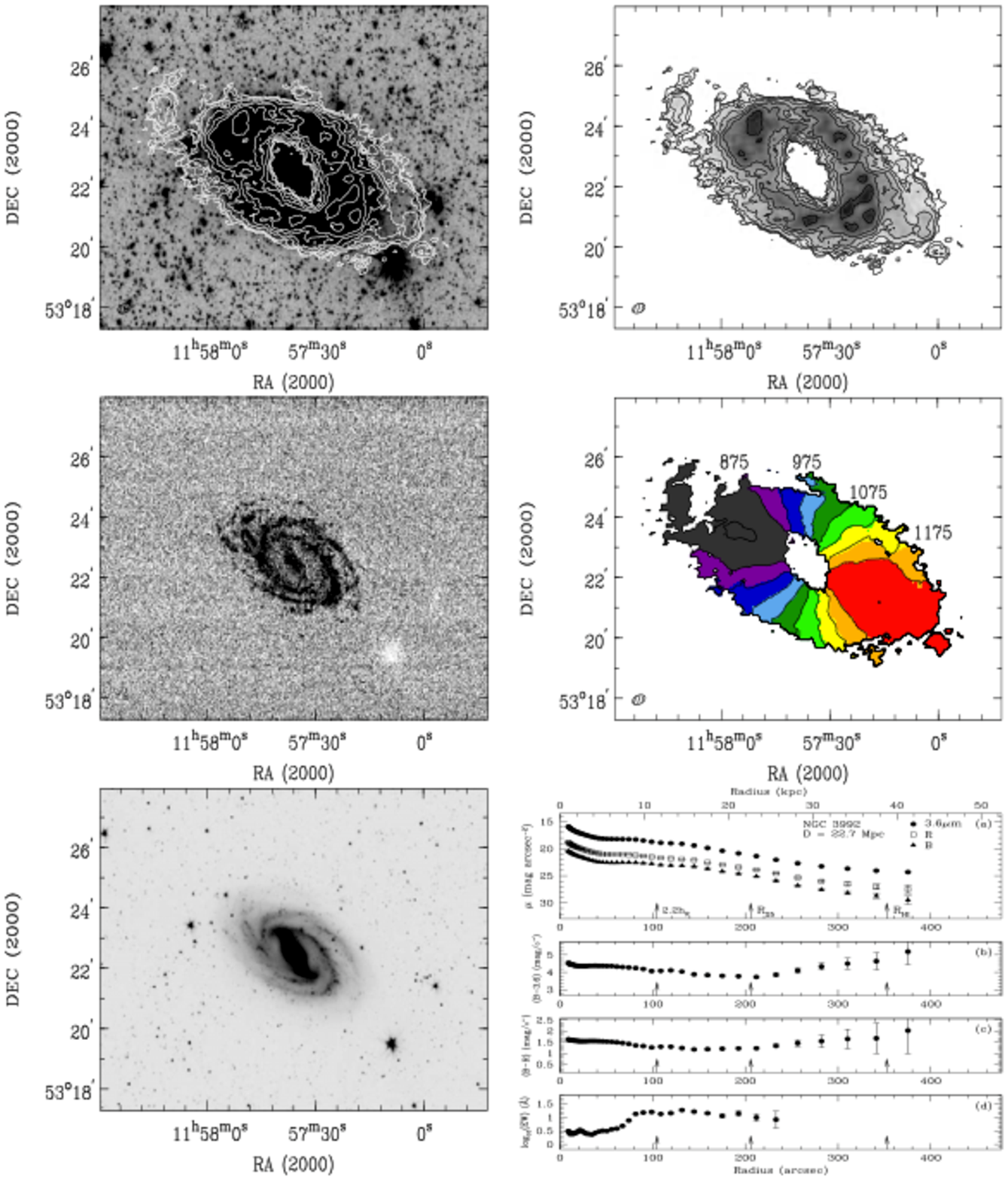}
\caption{{\bf NGC~3992:} {\it (top left)} Low spatial resolution HI integrated intensity contours 
from archival VLA data overlaid on the {\it Spitzer} 3.6$\mu$m image. 
{\it (top right)} Low spatial resolution HI intensity contours overlaid on the low 
spatial resolution HI intensity image. The first low resolution HI contour represents a 
column density of 4$\times$10$^{19}$ atoms cm$^{-2}$. 
{\it (middle left)} Narrowband H$\alpha$ image from the WIYN 0.9 m telescope.
{\it (middle right)} HI velocity field with isovelocity contours derived from the low spatial 
resolution HI data. The isovelocity contours are spaced every 50 km s$^{-1}$.  
{\it (bottom left)} {\it Spitzer} 3.6$\mu$m image with a high surface brightness stretch.
{\it (bottom right)} Ellipse photometry results (see Sections~\ref{sec:3.6data} and~\ref{sec:optdata})
showing radial profiles of surface brightness, $B-3.6$ colour, $B-R$ colour, and equivalent width (EW).}
\label{fig:n3992sum}
\end{figure*}

NGC~3992 is a strongly barred Sbc galaxy with prominent spiral features visible
in the stellar disc. The archival VLA HI observations reveal a complete absence of detectable neutral
gas within the central 80 arcsec -- about the length of the bar (Fig.~\ref{fig:n3992sum}). 
Even the narrowband H$\alpha$ image shows little ionized gas emission around the bar. 
Likewise, \cite{bimasongII} find a remarkable lack of molecular gas in the central region.
Beyond the bar, the HI disc has approximately the same radial extent as the high surface 
brightness stellar disc observed at 3.6$\mu$m. Three dwarf galaxies, 
UGC~06969, UGC~06940, and UGC~06923, are also detected in the HI data. Properties of 
these galaxies and other known companions to NGC~3992 are presented in Table~\ref{tab:n3992comps}.

The outer kinematics of the neutral gas disc in NGC~3992 are regular with only a slight 
warp at the outermost radii. The resultant HI rotation curve is relatively flat, but appears 
to decrease at a radii greater than 220 arcsec. This same decrease was observed by
\cite{n3992vlahi}, although they trace the HI distribution out to larger radii than
the present data. The inner kinematics as measured by the ionized gas
emission are affected by the strong bar (Fig.~\ref{fig:spakvels}). The rotational 
velocities derived from the ionized gas shown in Fig.~\ref{fig:rc} increase from the centre up to the
neutral gas rotational velocities, but display a kink between 
$\sim$40--60 arcsec which can most likely be attributed to the bar. A similar
structure can be seen in the H$\alpha$ position velocity diagram of \cite{bhabar}.

The stellar distribution of NGC~3992 is disc dominated, but with contributions from
bulge and bar components. Decomposition of the 3.6$\mu$m image with {\sc DiskFit} 
(\citealt{diskfit}) revealed that the bulge contributes less than 5 per cent to the total 
luminosity. At larger radii, the 3.6$\mu$m surface brightness profile shows a change in slope around 
$\sim$250 arcsec where the high surface brightness disc component ends. The colour
gradient appears flat across radii within $R_{25}$, but then slowly increases out
to larger radii. The EW distribution, on the other hand, shows a sharp increase beyond
the bar at the start of the star forming disc near 80 arcsec. 

The rotation curve decomposition analysis was carried out using neutral gas
and stellar disc components for the baryonic contributions to the overall
observed rotation curve and is presented in Fig.~\ref{fig:rcdecomp}. The structure in the ionized gas
rotational velocities is reproduced in the stellar disc contribution, likely due
the influence of the bar. The lack of detectable atomic or molecular gas in the centre
results in negative rotational velocities within 140 arcsec for the gas component.
We adopt the Type Ia supernova distance of 22.7 Mpc (\citealt{n3992distref}) 
for this galaxy. The LTFR distance is considerably larger at 36.4 Mpc using
a value for $V_{\rm flat}$ estimated from an intermediate radial range which avoids
the downbending portion of the outer warped rotation curve. The dark matter halo 
model rises rapidly in the fixed M/L = 0.5 decomposition resulting in a small
transition radius and, therefore, mass distribution which is dark matter-dominated
at nearly all measured radii. The decomposition presented here agrees well with 
the best-fitting stellar disc only decomposition presented in \cite{n3992vlahi}.
The maximum disc decomposition with M/L = 0.8 provides an equally
reasonable fit to the observed rotation curve. The mass distribution derived
from the maximum disc model is quite different and does not become
dark matter-dominated until over 6 disc scale lengths.

\begin{landscape}
\begin{table}
\tiny
\centering
\caption{Properties of Known Companions to NGC~3992}
\label{tab:n3992comps}
\begin{threeparttable}
\begin{tabular}{lcccccrcccccc}
\hline
Galaxy & RA, Dec & Systemic\tnote{a} & Angular & Physical\tnote{b} & Size\tnote{c} & Position & 
$m_{B}$ & $m_{R}$ & $m_{\rm 3.6}$ & ($B-R$)$_{\rm 0}$ & $M_{B}$\tnote{b} & HI \\
  & (J2000) & Velocity & Separation & Separation & (a$\times$b) & Angle &  &  &  &  &  & Flux \\
  &  & (km s$^{-1}$) & (arcmin) & (kpc) & (arcsec$\times$arcsec) & (deg) &  &  &  &  &  & (Jy km s$^{-1}$) \\
\hline
UGC~06940 & 11:57:48.0,+53:14:03 & 1113(1) & 8.58 & 56.7 & 70.0$\times$35.6 & -39.6 &
16.27$\pm$0.04 & 15.59$\pm$0.04 & 13.62$\pm$0.13 & 0.64$\pm$0.05 & -15.61$\pm$0.04 & 2.5 \\
SDSS J115834.34+532043.6 & 11:58:34.4,+53:20:44 & 1150(2) & 8.88 & 58.6 & 101.2$\times$62.2 & 76.3 &
17.27$\pm$0.11 & 16.05$\pm$0.11 & 13.97$\pm$0.15 & 1.17$\pm$0.16 & -14.62$\pm$0.11 & - \\
UGC~06969 & 11:58:46.8,+53:25:31 & 1115(1) & 11.10 & 73.3 & 124.6$\times$67.2 & -34.1 &
15.07$\pm$0.03 & 14.23$\pm$0.03 & 12.13$\pm$0.06 & 0.80$\pm$0.05 & -16.82$\pm$0.03 & 6.2 \\
UGC~06923 & 11:56:49.7,+53:09:37 & 1069(1) & 14.61 & 96.5 & 163.2$\times$71.1 & -13.2 &
13.90$\pm$0.02 & 12.96$\pm$0.02 & 10.78$\pm$0.04 & 0.90$\pm$0.03 & -17.99$\pm$0.02 & 11.4 \\
\hline
\end{tabular}
\begin{tablenotes}
\item {\it Note.} -- The apparent magnitudes are measured values and are not corrected for 
extinction. $B-R$ and $M_{B}$ are extinction corrected assuming $A_{B}$=0.106 and 
$A_{R}$=0.063 (\citealt{galextinct}). The extinction correction for the NIR is 
assumed to be negligible.
\item[a] Sources: (1) this work; (2) Sloan Digital Sky Survey Data Release 2.
\item[b] Assuming a distance of 22.7 Mpc.
\item[c] Sizes indicate the diameters in arcseconds of the apertures used to 
measure the reported magnitudes.
\end{tablenotes}
\end{threeparttable}
\end{table}
\end{landscape}

\subsection{NGC~3998}
\label{sec:n3998}

\begin{figure*}
\includegraphics[height=0.9\textheight]{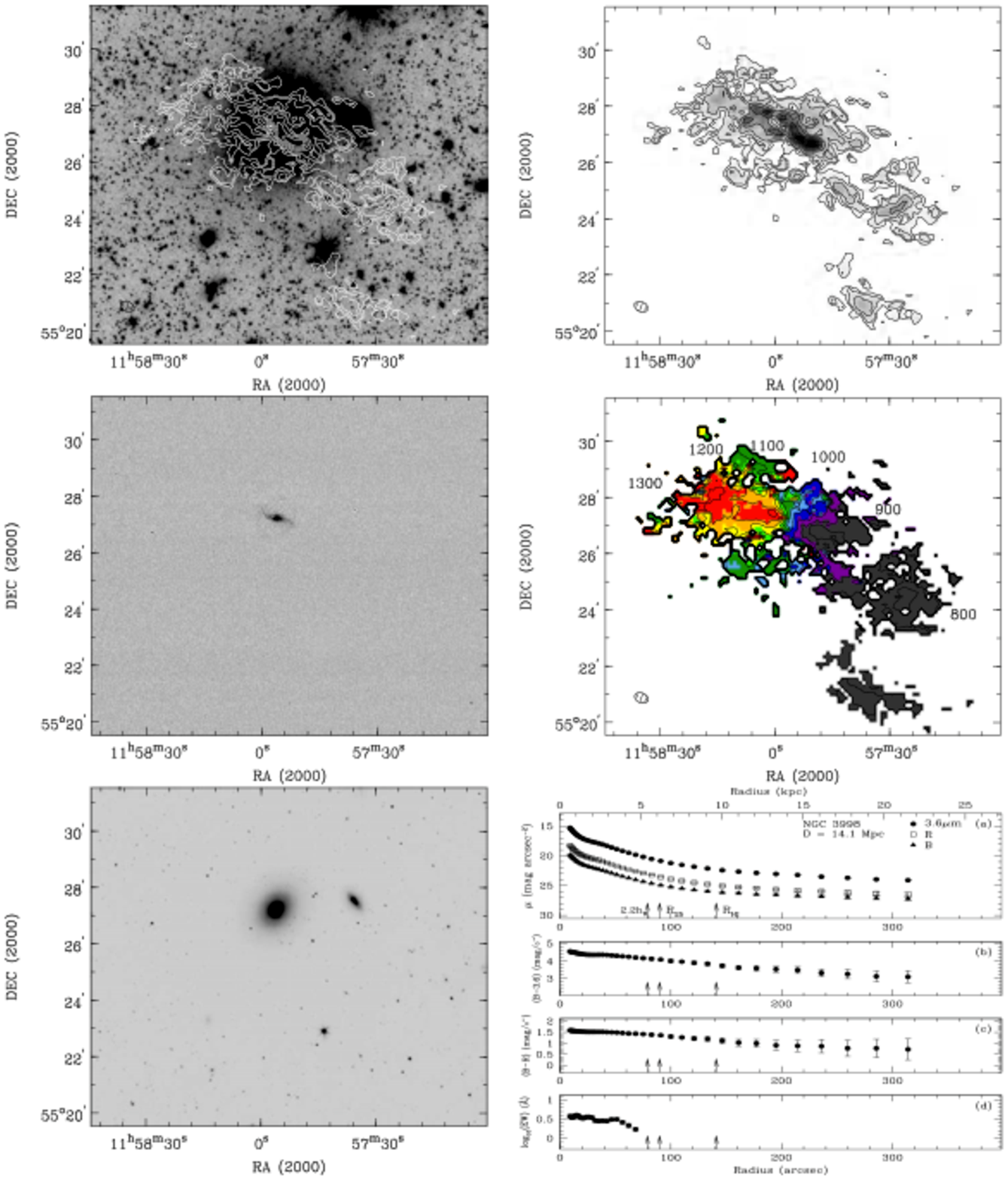}
\caption{{\bf NGC~3998:} {\it (top left)} Spectral binned low spatial resolution HI integrated 
intensity contours from new VLA data overlaid on the {\it Spitzer} 3.6$\mu$m image. 
{\it (top right)} Spectral binned low spatial resolution HI intensity contours overlaid on the 
spectral binned low spatial resolution HI intensity image. The first low resolution HI contour represents a 
column density of 2$\times$10$^{19}$ atoms cm$^{-2}$. 
{\it (middle left)} Narrowband H$\alpha$ image from the WIYN 0.9 m telescope. 
{\it (middle right)} HI velocity field with isovelocity contours derived from the spectral binned low spatial 
resolution HI data. The isovelocity contours are spaced every 50 km s$^{-1}$.  
{\it (bottom left)} {\it Spitzer} 3.6$\mu$m image with a high surface brightness stretch.
{\it (bottom right)} Ellipse photometry results (see Sections~\ref{sec:3.6data} and~\ref{sec:optdata})
showing radial profiles of surface brightness, $B-3.6$ colour, $B-R$ colour, and equivalent width (EW).}
\label{fig:n3998sum}
\end{figure*}

NGC~3998 is a non-barred S0 and well-studied low-ionization nuclear emission-line region (LINER)
galaxy in a relatively populated group.
In addition to NGC~3998, four other galaxies are detected in the new HI observations from the VLA 
including NGC~3972, NGC~3982, UGC~06919, and UGC~06988. Table~\ref{tab:n3998comps} presents properties of
the known companions to NGC~3998. The neutral gas in NGC~3998 is diffuse and not 
well organized. Fig.~\ref{fig:n3998sum} shows that it lies in a roughly disc-like distribution with an 
apparent position angle that is perpendicular to the position angle of the stellar distribution. 
\cite{atlas3dhi} observed NGC~3998 with the WSRT as part of the ATLAS$^{\rm 3D}$ survey and noted 
the same lopsided and misaligned morphology of the HI. Their HI map displays a similar distribution 
of HI as we observed with the VLA, with two distinct HI emission components on the approaching side 
to the west and south of the primary HI disc. The ionized
gas velocity field in Fig.~\ref{fig:spakvels} indicates that the ionized gas shares the same orientation
as the HI. However, the ionized gas shows a steeper rise and greater peak rotational velocities 
than the neutral gas (see Fig.~\ref{fig:rc}). In addition, the ionized gas emission lines in the SparsePak
spectra display broad FWHM values and non-thermal emission line flux ratios. \cite{S0kin} found that
the velocity dispersion profiles of both the gas and stars in NGC~3998 rise sharply towards the
centre, reaching values near 320 km s$^{-1}$. 

Moving beyond the centre, we find a bulge-like stellar distribution in the 3.6$\mu$m surface
brightness profile. Therefore, a spherical distribution was adopted for the model stellar
rotation curve. This model stellar bulge rotation curve falls off too quickly to accurately
describe the observed ionized gas rotation curve in the decomposition analysis (Fig.~\ref{fig:rcdecomp}). 
The decomposition does, however, adequately fit the outer HI rotation
curve points. Model stellar rotation curves using a bulge and disc as well as disc only distributions
were tried, but were unable to fit both the inner and outer observed rotational velocities concurrently.
The stellar bulge rotation curve model requires a M/L = 1.4 in order to match the large rotational 
velocities of the ionized gas in the inner 40 arcsec at the adopted distance of 14.1 Mpc derived
from surface brightness fluctuations (\citealt{sbfdist}). The LTFR distance estimate is more than
double at 30.1 Mpc based on $V_{\rm flat}$ measured from the HI rotational velocities.
A decomposition using a fixed bulge M/L = 0.6 can adequately describe the HI rotation
curve, but not the ionized gas. It is possible that the ionized gas is not in gravitational
equilibrium and cannot be described by the model rotation curves.

\begin{landscape}
\begin{table}
\tiny
\centering
\caption{Properties of Known Companions to NGC~3998}
\label{tab:n3998comps}
\begin{threeparttable}
\begin{tabular}{lcccccrcccccc}
Galaxy & RA, Dec & Systemic\tnote{a} & Angular & Physical\tnote{b} & Size\tnote{c} & Position & 
$m_{B}$ & $m_{R}$ & $m_{\rm 3.6}$ & ($B-R$)$_{\rm 0}$ & $M_{B}$\tnote{b} & HI \\
  & (J2000) & Velocity & Separation & Separation & (a$\times$b) & Angle &  &  &  &  &  & Flux \\
  &  & (km s$^{-1}$) & (arcmin) & (kpc) & (arcsec$\times$arcsec) & (deg) &  &  &  &  &  & (Jy km s$^{-1}$) \\
\hline
NGC3990 & 11:57:35.4,+55:27:33 & 696(1) & 2.93 & 12.0 & 110.1$\times$92.2 & 28.9 &
13.57$\pm$0.02 & 12.10$\pm$0.02 & 9.39$\pm$0.02 & 1.45$\pm$0.03 & -17.24$\pm$0.02 & - \\
SDSS J115813.68+552316.5 & 11:58:13.6,+55:23:18 & 966(2) & 4.66 & 19.1 & 90.3$\times$58.7 & -48.5 &
16.61$\pm$0.06 & 15.41$\pm$0.06 & 13.03$\pm$0.10 & 1.18$\pm$0.09 & -14.19$\pm$0.06 & - \\
SDSS J115701.86+552511.1 & 11:57:01.6,+55:25:11 & 1215(2) & 7.96 & 32.6 & 87.2$\times$55.8 & -85.5 &
16.53$\pm$0.05 & 15.37$\pm$0.05 & 13.19$\pm$0.10 & 1.14$\pm$0.08 & -14.27$\pm$0.05 & - \\
SDSS J115703.08+553612.3 & 11:57:03.0,+55:35:14 & 763(2) & 10.97 & 45.0 & 62.4$\times$38.7 & -49.6 &
16.88$\pm$0.04 & 15.72$\pm$0.04 & 13.67$\pm$0.13 & 1.14$\pm$0.06 & -13.92$\pm$0.04 & - \\
SDSS J115849.17+551824.7 & 11:58:49.2,+55:18:25 & 939(2) & 11.59 & 47.5 & 93.7$\times$76.1 & -43.7 &
16.91$\pm$0.10 & 15.69$\pm$0.10 & 13.41$\pm$0.11 & 1.19$\pm$0.15 & -13.90$\pm$0.10 & - \\
UGC06919 & 11:56:37.6,+55:37:58 & 1283(2) & 15.49 & 63.5 & 137.4$\times$50.3 & 85.8 &
15.19$\pm$0.02 & 13.85$\pm$0.02 & -- & 1.32$\pm$0.04 & -15.61$\pm$0.02 & 0.4 \\
UGC~06988 & 11:59:51.7,+55:39:55 & 745(3) & 20.70 & 84.9 & -- & -- &
-- & -- & -- & -- & -- & 3.2 \\
NGC3982 & 11:56:28.1,+55:07:31 & 1102(3) & 23.35 & 95.8 & 175.3$\times$172.0 & 68.2 &
12.20$\pm$0.02 & 11.18$\pm$0.02 & -- & 1.01$\pm$0.03 & -18.60$\pm$0.02 & 30.9 \\
SDSS J115356.95+551017.3 & 11:53:57.0,+55:10:16 & 1249(2) & 38.01 & 155.9 & 38.2$\times$29.6 & 9.2 &
17.79$\pm$0.04 & 16.65$\pm$0.04 & -- & 1.11$\pm$0.06 & -13.02$\pm$0.04 & - \\
\hline
\end{tabular}
\begin{tablenotes}
\item{\it Note.} -- The apparent magnitudes are measured values and are not corrected for 
extinction. $B-R$ and $M_{B}$ are extinction corrected assuming $A_{B}$=0.059 and 
$A_{R}$=0.036 (\citealt{galextinct}). The extinction correction for the NIR is 
assumed to be negligible.
\item[a] Sources: (1) RC3 (\citealt{rc3}); (2) Sloan Digital Sky Survey Data Release 3;
(3) this work.
\item[b] Assuming a distance of 14.1 Mpc.
\item[c] Sizes indicate the diameters in arcseconds of the apertures used to 
measure the reported magnitudes.
\end{tablenotes}
\end{threeparttable}
\end{table}
\end{landscape}

\subsection{NGC~4051}
\label{sec:n4051}

\begin{figure*}
\includegraphics[height=0.9\textheight]{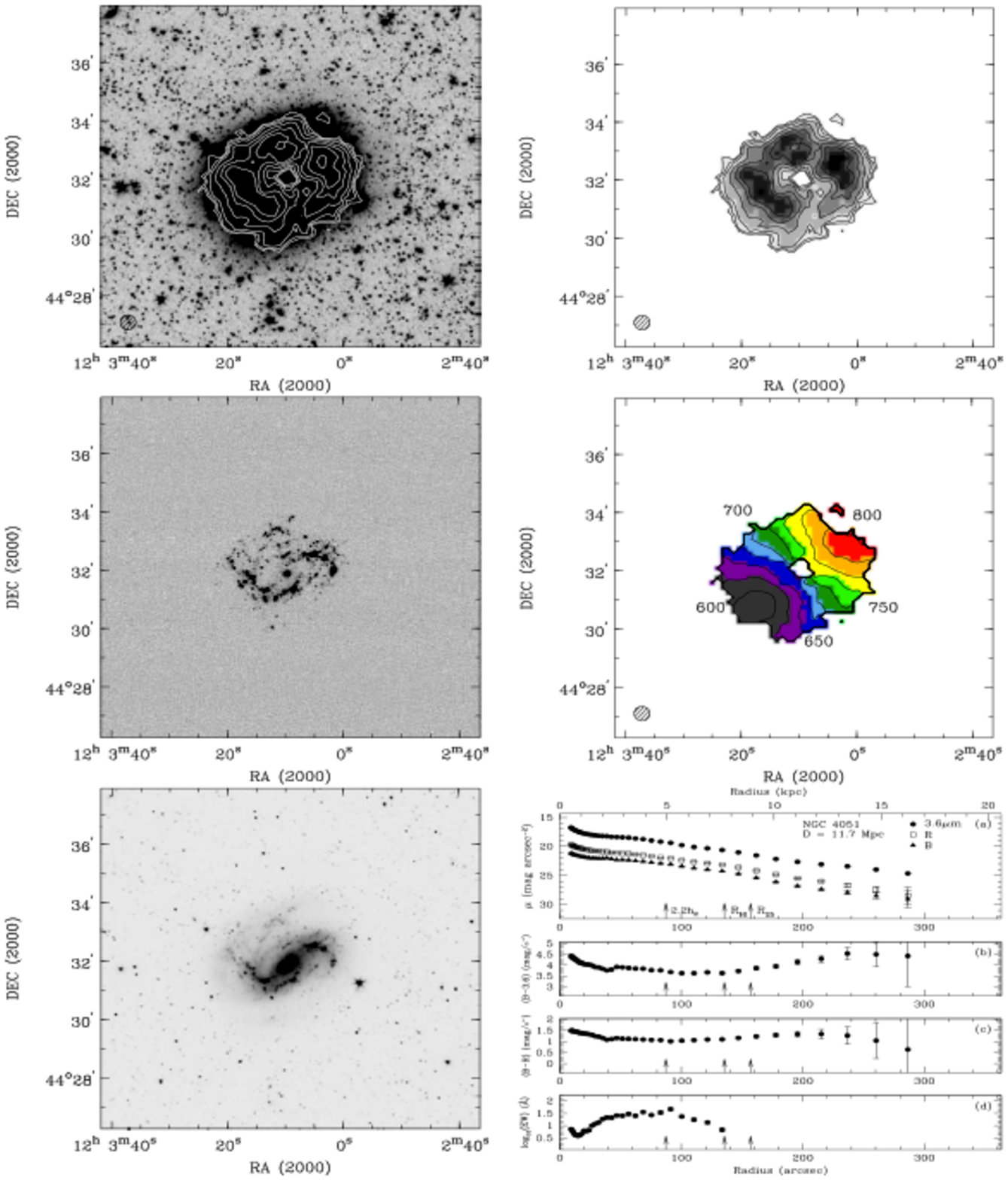}
\caption{{\bf NGC~4051:} {\it (top left)} 30\arcsec~spatial resolution HI integrated intensity contours 
from the WSRT overlaid on the {\it Spitzer} 3.6$\mu$m image. 
{\it (top right)} 30\arcsec~spatial resolution HI intensity contours overlaid on the same
spatial resolution HI intensity image. The first HI contour represents a 
column density of 6$\times$10$^{19}$ atoms cm$^{-2}$. 
{\it (middle left)} Narrowband H$\alpha$ image from the WIYN 0.9 m telescope. 
{\it (middle right)} HI velocity field with isovelocity contours derived from the 30\arcsec~spatial 
resolution HI data. The isovelocity contours are spaced every 25 km s$^{-1}$.  
{\it (bottom left)} {\it Spitzer} 3.6$\mu$m image with a high surface brightness stretch.
{\it (bottom right)} Ellipse photometry results (see Sections~\ref{sec:3.6data} and~\ref{sec:optdata})
showing radial profiles of surface brightness, $B-3.6$ colour, $B-R$ colour, and equivalent width (EW).}
\label{fig:n4051sum}
\end{figure*}

NGC~4051 is an SABbc galaxy and well-studied AGN. HI synthesis observations from WHISP
show gas distributed in a ring around a central hole where the active nucleus lies
(Fig.~\ref{fig:n4051sum}). The neutral gas morphology is similar to the ionized gas seen
in the narrowband H$\alpha$ image which displays a bright central nucleus and two
star-forming spiral arms wrapped around the centre. The radial
distribution of $B-R$ colour reveals the central concentration to be redder than the
star-forming disc. The EW gradient matches this interpretation with small values
in the centre and higher in the disc. The ionized gas velocity field
in Fig.~\ref{fig:spakvels} shows twisted isovelocity contours indicating that it is likely
being influenced by the nuclear activity or possible presence of a bar.
As a result, the ionized gas rotation curve is uncertain and displays
structure that cannot be described by the model rotation curves in the decomposition.
The HI rotation curve rises slowly and flattens out around 120 arcsec.

The rotation curve decomposition using a disc only stellar distribution is shown in Fig.~\ref{fig:rcdecomp}.
The total gas contribution to the observed rotation curve includes CO observations for the molecular 
gas from the FCRAO survey, which were used in place of the BIMA SONG map to better recover spatially 
extended flux. The stellar disc model rotation curve has an unusual shape
with a sharp peak at 10 arcsec followed by a downwards dip before rising and then turning
over again. The sharp peak is likely due to the bright central nucleus. The combination
of the complex distribution of stellar light and ionized gas rotation curve makes it
difficult to constrain the stellar M/L in the maximum disc scenario. However, it appears
that a stellar disc M/L greater than 0.5 will over-estimate the first few HI rotation
curve points beyond the errors. The mass distribution in NGC~4051 is baryon-dominated 
until at least the last measured point.

\subsection{NGC~4138}
\label{sec:n4138}

\begin{figure*}
\includegraphics[height=0.9\textheight]{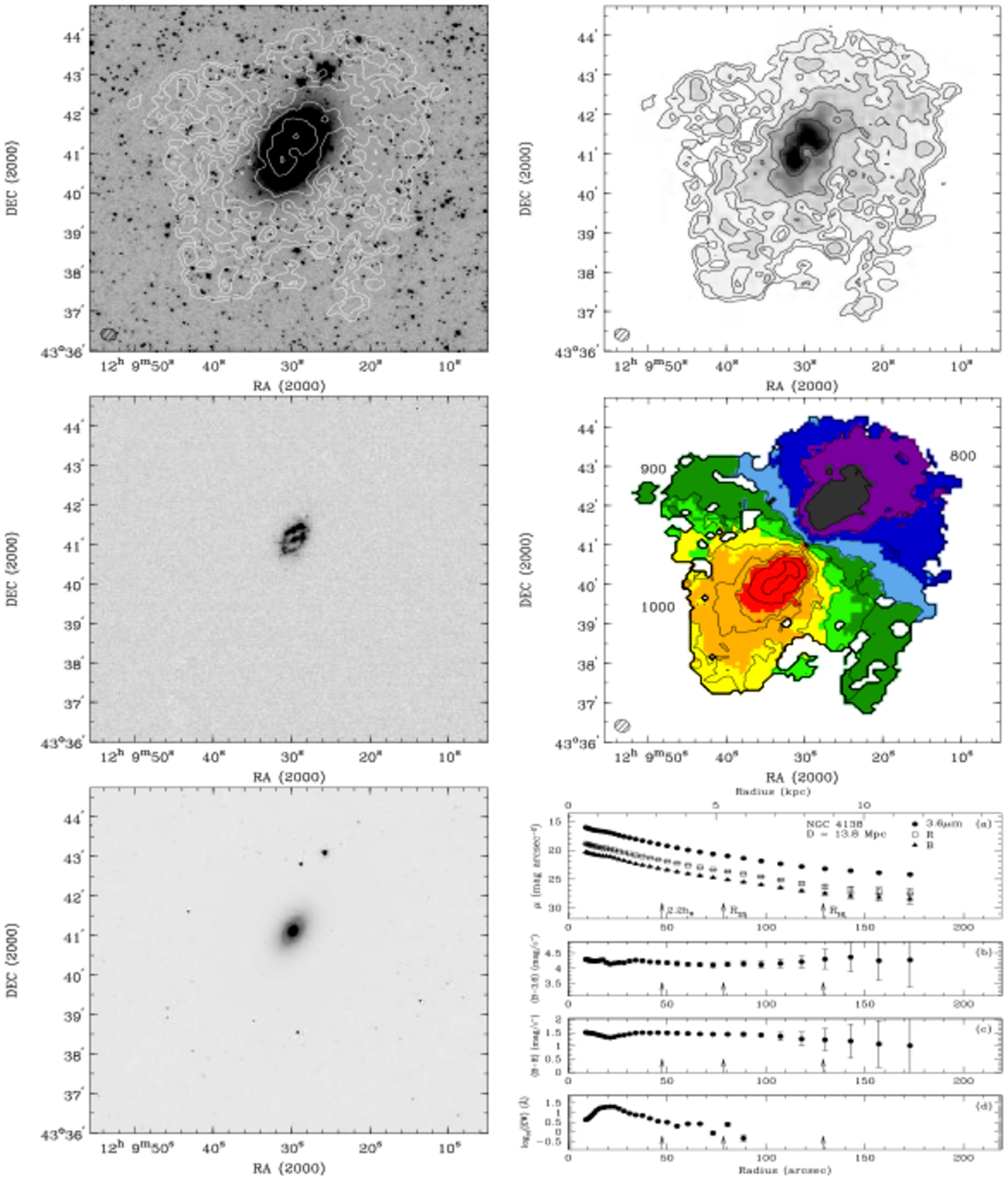}
\caption{{\bf NGC~4138:} {\it (top left)} Low spatial resolution HI integrated intensity contours 
from archival VLA data overlaid on the {\it Spitzer} 3.6$\mu$m image. 
{\it (top right)} Low spatial resolution HI intensity contours overlaid on the low 
spatial resolution HI intensity image. The first low resolution HI contour represents a 
column density of 3.4$\times$10$^{19}$ atoms cm$^{-2}$. 
{\it (middle left)} Narrowband H$\alpha$ image from the WIYN 0.9 m telescope. 
{\it (middle right)} HI velocity field with isovelocity contours derived from the low spatial 
resolution HI data. The isovelocity contours are spaced every 30 km s$^{-1}$.
{\it (bottom left)} {\it Spitzer} 3.6$\mu$m image with a high surface brightness stretch.
{\it (bottom right)} Ellipse photometry results (see Sections~\ref{sec:3.6data} and~\ref{sec:optdata})
showing radial profiles of surface brightness, $B-3.6$ colour, $B-R$ colour, and equivalent width (EW).}
\label{fig:n4138sum}
\end{figure*}

NGC~4138 is an SA0 galaxy in the Ursa Major cluster. The kinematics and distribution
of mass in NGC~4138 have been studied extensively using HI synthesis observations from both
the WSRT (\citealt{ursamajhi}) and VLA (\citealt{n4138jbh}). The HI observations presented
in Fig.~\ref{fig:n4138sum} are taken from \cite{n4138jbh}. In \cite{n4138jbh}, they
derive two rotation curves: one which rapidly declines (more so than is typically observed) 
and one which they fix to be flat, which then implies a strong outer warp with a decreasing inclination
angle. It's not possible to confirm which model is correct based on the velocity field, but the 
diffuse minor axis gas appears as if it has face-on orbits, which supports the possibility
of a smaller inclination angle at large radii. In the present analysis, we did not fix the 
rotation curve to be flat, but we used the results of \cite{n4138jbh} to force the 
inclination angle to decrease (see Fig.~\ref{fig:rc}). 

\cite{n4138jbh} additionally acquired optical spectroscopic observations of NGC~4138 and 
discovered a counterrotating stellar disc which appears to coincide with the ring of
H$\alpha$ emission visible in Fig.~\ref{fig:n4138sum}. Ionized gas kinematics were additionally 
derived from the spectroscopic observations and used to constrain rotational velocities in the 
central 40 arcsec. The ionized gas velocity field (Fig.~\ref{fig:spakvels}) and rotation
curve presented here agree well with the results of \cite{n4138jbh} and appear to match the 
neutral gas rotational velocities (Fig.~\ref{fig:rc}). 

The flat rotation curve can be described easily by the baryonic and dark matter halo contributions
as shown in Fig.~\ref{fig:rcdecomp}. The gas model rotation curve includes an estimate of the 
molecular gas from a single pointing CO observation in the FCRAO survey. The stellar distribution 
measured at 3.6$\mu$m is well-described by a disc only component. Assuming
a fixed stellar disc M/L = 0.5 results in a distribution of mass that is dark matter-dominated
beyond about one disc scale length. The flat rotation curve can be fit just as easily
by maximizing the stellar disc contribution with M/L = 1.1. This pushes the transition
radius out to almost 5 disc scale lengths. NGC~4138 is another case where the LTFR distance
of 28.4 Mpc is significantly larger than the adopted surface brightness fluctuation distance
of 13.8 (\citealt{sbfdist}). The value of $V_{\rm flat}$ is uncertain in this case since
it was essentially forced to the adopted value, so it is not necessarily surprising if the 
exact value is incorrect. The relative differences between the fixed M/L and maximum disc 
decomposition results are still valid.

\subsection{NGC~4389}
\label{sec:n4389}

\begin{figure*}
\includegraphics[height=0.9\textheight]{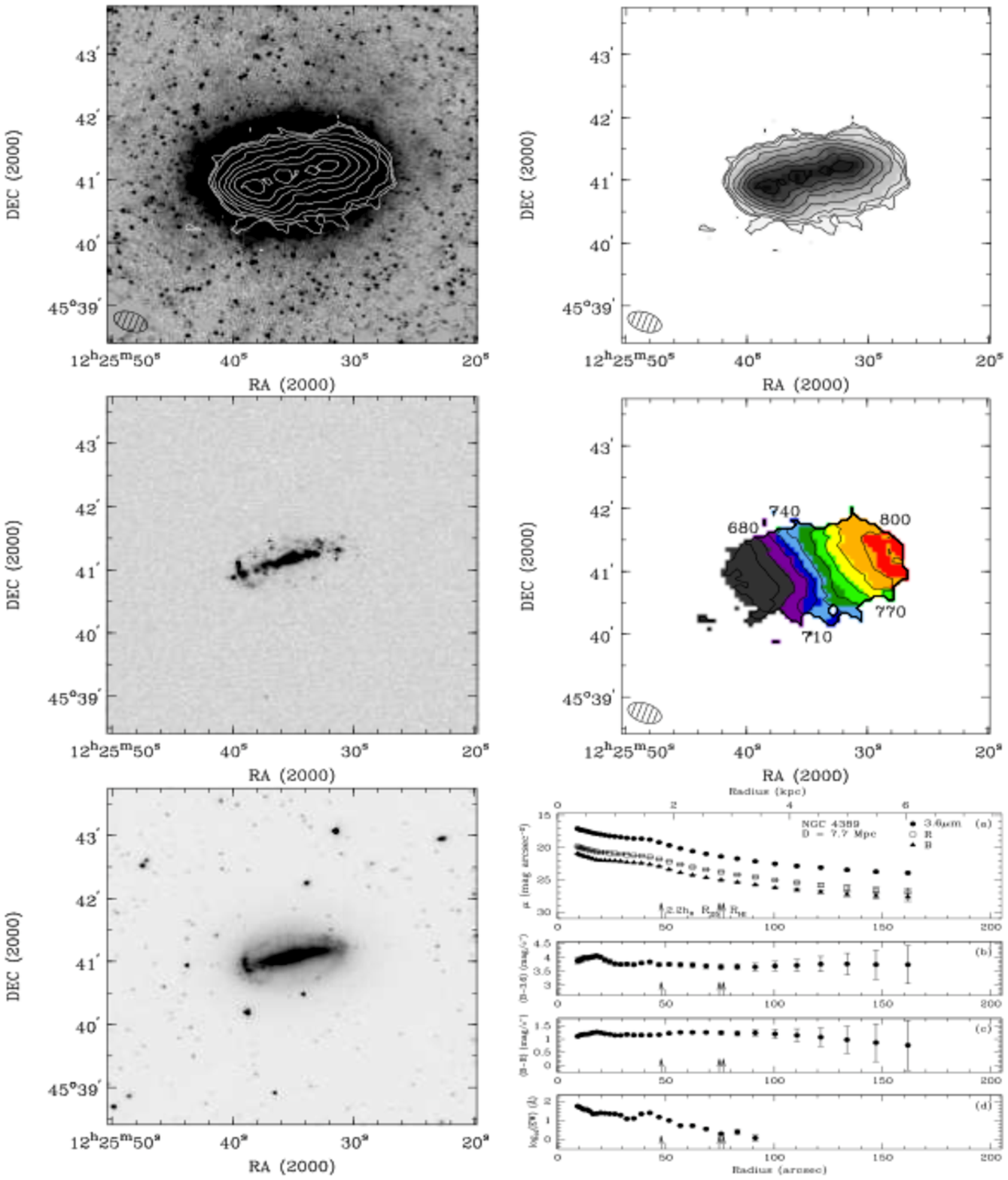}
\caption{{\bf NGC~4389:} {\it (top left)} Low spatial resolution HI integrated intensity contours 
from new VLA data overlaid on the {\it Spitzer} 3.6$\mu$m image. 
{\it (top right)} Low spatial resolution HI intensity contours overlaid on the low 
spatial resolution HI intensity image. The first low resolution HI contour represents a 
column density of 4$\times$10$^{19}$ atoms cm$^{-2}$. 
{\it (middle left)} Narrowband H$\alpha$ image from the WIYN 0.9 m telescope.
{\it (middle right)} HI velocity field with isovelocity contours derived from the low spatial 
resolution HI data. The isovelocity contours are spaced every 15 km s$^{-1}$.  
{\it (bottom left)} {\it Spitzer} 3.6$\mu$m image with a high surface brightness stretch.
{\it (bottom right)} Ellipse photometry results (see Sections~\ref{sec:3.6data} and~\ref{sec:optdata})
showing radial profiles of surface brightness, $B-3.6$ colour, $B-R$ colour, and equivalent width (EW).}
\label{fig:n4389sum}
\end{figure*}

NGC~4389 is an SBbc galaxy in the Ursa Major cluster. The distribution of the H$\alpha$
emission and of the HI from new VLA observations shown in Fig.~\ref{fig:n4389sum} suggests
that the morphology is dominated by a bar. The rotation curves derived from the ionized
and neutral gas suggest that the kinematics are also dominated by this bar (Fig.~\ref{fig:rc}).
\cite{ursamajhi} came to the same conclusion using WSRT observations of the HI. The
observations presented here trace the neutral gas out to a larger radius than \cite{ursamajhi}, 
but show a similar slowly rising rotation curve which only reaches maximum rotational velocities 
near 100 km s$^{-1}$. 

A rotation curve for NGC~4389 has also been derived by \cite{n4389hiirc} using
the distribution of H II regions. They find a slowly rising rotation curve wich has a peak
velocity around 90 km s$^{-1}$ at the last measured point near 60 arcsec. This is in excellent
agreement with both the ionized and neutral gas rotation curves derived in the present study.

The stellar distribution in NGC~4389 displays a change in slope around 45 arcsec where the
stellar light begins to drop off more quickly. This corresponds to the radius of the ring-like
emission which emanates from the ends of the bar, as can be seen in the H$\alpha$ and
high surface brightness 3.6$\mu$m images.
This sharp transition is reflected in the stellar contribution to the observed rotation curve
in Fig.~\ref{fig:rcdecomp}. The rotation curve derived from the distribution of the HI gas
displays the same feature, indicating that the morphology of both the gas and stars is
indeed dominated by the bar.

The structure in the distribution of the gas and stars caused by the prominent bar means
the mass distribution in NGC~4389 would need to be dominated by dark matter at almost all
radii in order to produce a smooth, slowly rising rotation curve. 
A M/L larger than 0.1 introduces too much structure into the total rotation
curve to describe the observed one, as can be observed in the fixed M/L = 0.4 decomposition
shown in Fig.~\ref{fig:rcdecomp}. It is likely that the gas dynamics are being influenced by
the presence of the bar. It is not straightforward to interpret the results of the rotation curve
decomposition in the context of a bar potential.

\subsection{UGC~07639}
\label{sec:u7639}

\begin{figure*}
\includegraphics[height=0.9\textheight]{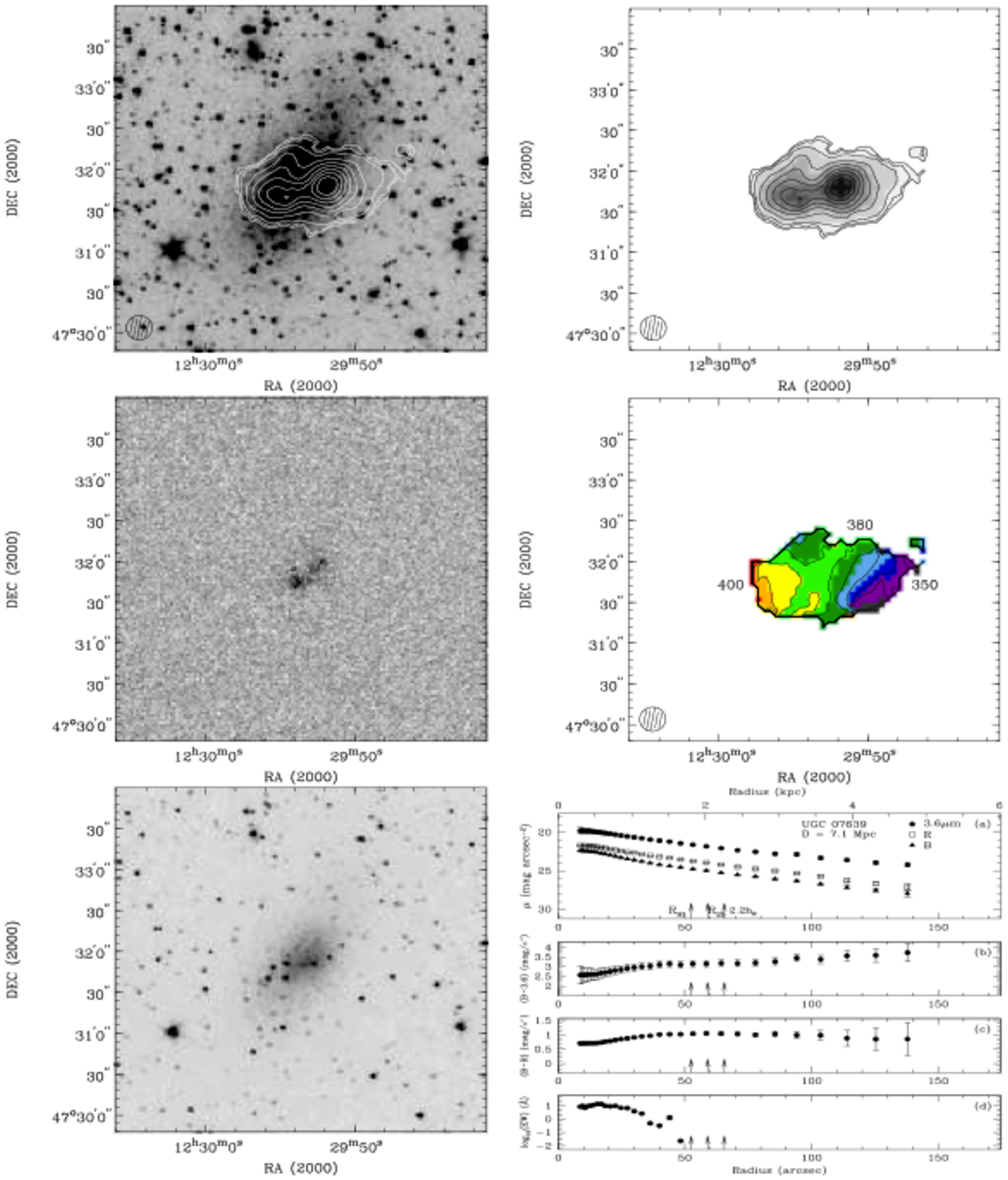}
\caption{{\bf UGC~07639:} {\it (top left)} Low spatial resolution HI integrated intensity contours 
from new VLA data overlaid on the {\it Spitzer} 3.6$\mu$m image. 
{\it (top right)} Low spatial resolution HI intensity contours overlaid on the low 
spatial resolution HI intensity image. The first low resolution HI contour represents a 
column density of 4$\times$10$^{19}$ atoms cm$^{-2}$. 
{\it (middle left)} Narrowband H$\alpha$ image from the WIYN 0.9 m telescope.
{\it (middle right)} HI velocity field with isovelocity contours derived from the low spatial 
resolution HI data. The isovelocity contours are spaced every 5 km s$^{-1}$.  
{\it (bottom left)} {\it Spitzer} 3.6$\mu$m image with a high surface brightness stretch.
{\it (bottom right)} Ellipse photometry results (see Sections~\ref{sec:3.6data} and~\ref{sec:optdata})
showing radial profiles of surface brightness, $B-3.6$ colour, $B-R$ colour, and equivalent width (EW).}
\label{fig:u07639sum}
\end{figure*}

UGC~07639 is an irregular dwarf galaxy with very little HI detected in new VLA observations
(Fig.~\ref{fig:u07639sum}). The HI is actually distributed off-centre and nearly perpendicular
to the stellar distribution in the 3.6$\mu$m image. Furthermore, the HI velocity field appears
disorganized with only a hint of circular rotation. UGC~07639 was not targeted for SparsePak
ionized gas observations due to the lack of visible H$\alpha$ emission in the narrowband image.
The stellar distribution at 3.6$\mu$m is purely exponential, which is typical for dwarf galaxies.
However, the integrated ($B-R$)$_{\rm 0}$ colour of 0.86 is somewhat redder than for typical Im type
galaxies (\citealt{galbr}), which is consistent with the lack of recent star formation activity.

An attempt was made to extract a rotation curve from the velocity field and
to decompose it into separate mass components. The rotation curve was found to be linearly 
rising out to 50 arcsec and only reaching a velocity of about 20 km s$^{-1}$. Such small
rotational velcoities could be described entirely by the total baryonic contribution alone
without the need for dark matter. The addition of a dark matter halo model causes the
total modeled rotation curve to over estimate the observed velocities at all inner radii. 
This results in an unexpected (and likely unreasonable) scenario where a dwarf galaxy 
is baryon-dominated at all measured radii.

To make matters worse, an unrealistically small stellar M/L must be assumed to prevent the total
baryonic contribution from drastically over estimating the observed rotation curve, even without
the presence of dark matter. We have adopted the distance of 7.1 Mpc estimated from surface brightness
fluctuations (\citealt{u07639distref}). Other literature distance estimates to UGC~07639 only
vary by $\la$2 Mpc \citep[e.g.][]{u07639brightstard,nedhlg}, so it is not likely that the
unpyhiscal stellar M/L is due to incorrect normalization of the baryonic contributions caused by
inaccurate distance estimates. Given the morphology of the gas and somewhat chaotic 
appearance of the velocity field, it is likely the gas is dynamically unstable and is not 
tracing the gravitational potential of the galaxy. For these reasons, we do not include 
UGC~07639 in the discussion of the results.

\subsection{NGC~5033}
\label{sec:n5033}

\begin{figure*}
\includegraphics[height=0.9\textheight]{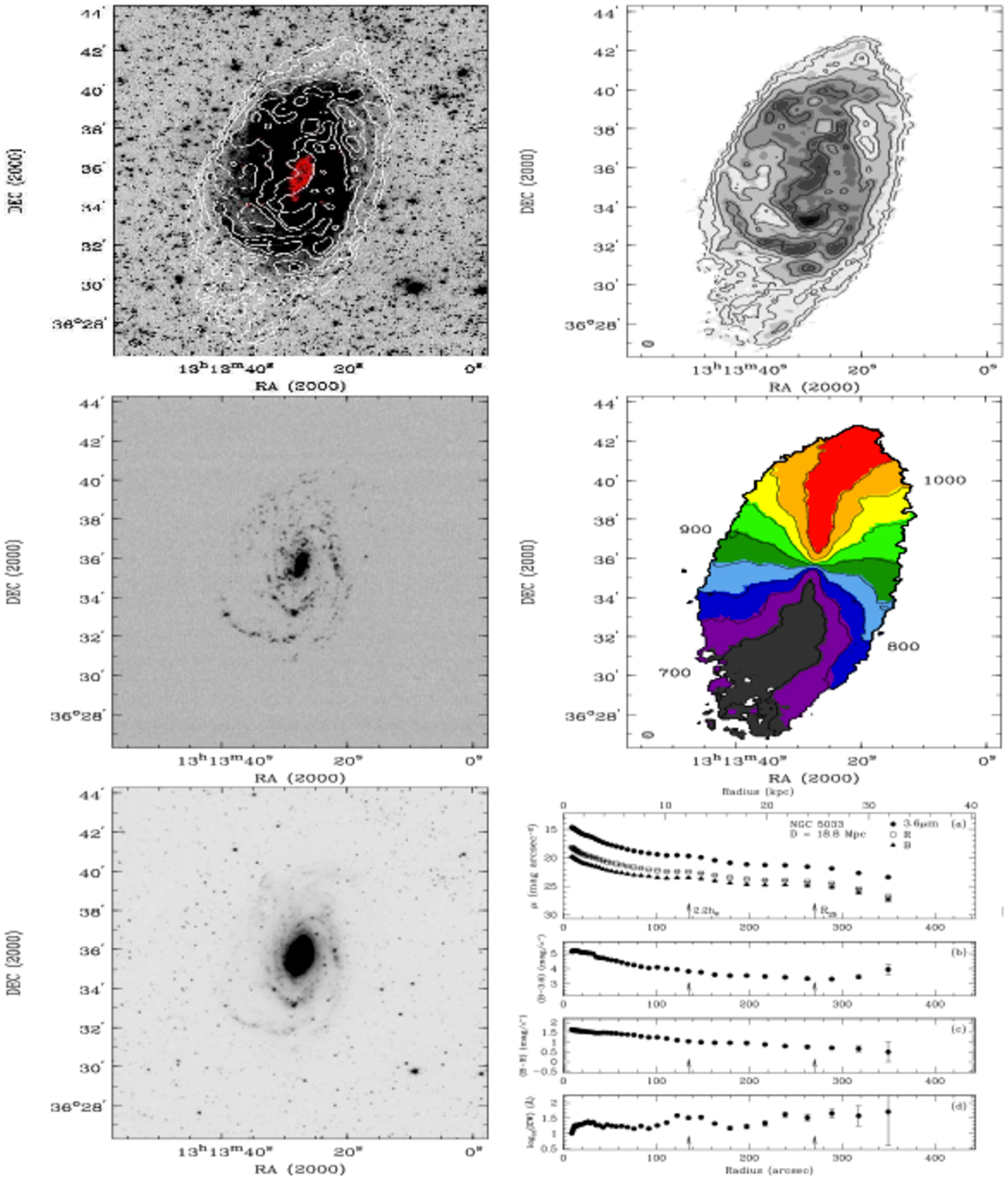}
\caption{{\bf NGC~5033:} {\it (top left)} Medium spatial resolution HI integrated intensity contours 
from archival VLA data {\it (white)} and CO intensity contours {\it (red)} from BIMA SONG 
(\citealt{bimasongII}) overlaid on the {\it Spitzer} 3.6$\mu$m image. 
{\it (top right)} Medium spatial resolution HI intensity contours overlaid on the medium 
spatial resolution HI intensity image. The first low resolution HI contour represents a 
column density of 4$\times$10$^{19}$ atoms cm$^{-2}$. 
{\it (middle left)} Narrowband H$\alpha$ image from the WIYN 0.9 m telescope.
{\it (middle right)} HI velocity field with isovelocity contours derived from the medium spatial 
resolution HI data. The isovelocity contours are spaced every 50 km s$^{-1}$.  
{\it (bottom left)} {\it Spitzer} 3.6$\mu$m image with a high surface brightness stretch.
{\it (bottom right)} Ellipse photometry results (see Sections~\ref{sec:3.6data} and~\ref{sec:optdata})
showing radial profiles of surface brightness, $B-3.6$ colour, $B-R$ colour, and equivalent width (EW).}
\label{fig:n5033sum}
\end{figure*}

NGC~5033 is a massive SAc galaxy whose kinematics have been well-studied in the literature.
The HI data presented in Fig.~\ref{fig:n5033sum} were processed from archival VLA observations
which were originally published in \cite{n5033hipub}. Two dwarf companion galaxies, UGC~08314
and UGC~08303, are also detected in these observations (see Table~\ref{tab:n5033comps}). 
The distribution of HI in NGC~5033
displays a lot of structure. In general, regions of greater HI surface density correspond to
regions with strong H$\alpha$ emission, which is mostly in the centre and along the spiral
arms. The HI velocity field of NGC~5033 is well-ordered with no signatures of barred or non-circular
orbits in the centre. The outer isovelocity contours along the major axis deviate away from
the line of nodes forming an S-shape which is the canonical sign of an outer disc warp. The
smaller wiggles in the disc are likely due to streaming motions along the spiral arms.

\cite{n5033n5055hakin} and \cite{n5033n5055hikin} were the first to perform an in-depth
study of the emission line and HI kinematics in NGC~5033. \cite{n5033n5055hikin} was able
to model the warped outer disc using a changing position angle. The results in \cite{n5033hipub}
and shown in Fig.~\ref{fig:rc} agree well with \cite{n5033n5055hikin}. The ionized gas velocity
in Fig.~\ref{fig:spakvels} confirm the well-behaved central gas kinematics seen in the HI.
The rotation curve derived from the ionized gas velocity field rises more steeply than the HI
rotation curve and peaks near 230 km s$^{-1}$ at 35 arcsec before coming back down. 
The shape of the ionized gas rotational velocities derived here are in 
agreement with those presented in \cite{n5033singshakin} within the central 60 arcsec.

The shape of the ionized and neutral gas rotation curves are well-described by the distributions
of the baryonic components as demonstrated in Fig.~\ref{fig:rcdecomp}. CO observations from
BIMA SONG (\citealt{bimasongII}) were included in the total gas estimate to account for the considerable
amount of molecular gas in NGC~5033. The stellar distribution was decomposed into bulge and disc
components using {\sc DiskFit} (\citealt{diskfit}). The contributions from both
the stellar bulge and disc are maximal at M/L = 0.4. The mass distribution transitions from
baryon-dominated to dark matter-dominated at 2.2$h_{\rm R}$.

\begin{landscape}
\begin{table}
\tiny
\centering
\caption{Properties of Known Companions to NGC~5033}
\label{tab:n5033comps}
\begin{threeparttable}
\begin{tabular}{lcccccrcccccc}
Galaxy & RA, Dec & Systemic\tnote{a} & Angular & Physical\tnote{b} & Size\tnote{c} & Position & 
$m_{B}$ & $m_{R}$ & $m_{\rm 3.6}$ & ($B-R$)$_{\rm 0}$ & $M_{B}$\tnote{b} & HI \\
  & (J2000) & Velocity & Separation & Separation & (a$\times$b) & Angle &  &  &  &  &  & Flux \\
  &  & (km s$^{-1}$) & (arcmin) & (kpc) & (arcsec$\times$arcsec) & (deg) &  &  &  &  &  & (Jy km s$^{-1}$) \\
\hline
SDSS J131420.58+363407.9 & 13:14:20.8,+36:34:13 & 1737(1)\tnote{d} & 10.77 & 58.9 & 66.9$\times$48.4 & 22.7 &
17.08$\pm$0.07 & 16.40$\pm$0.07 & 14.32$\pm$0.17 & 0.67$\pm$0.10 & -14.29$\pm$0.07 & 0.6 \\
UGC08314 & 13:14:00.6,+36:19:01 & 935(2) & 17.83 & 97.5 & 106.0$\times$82.0 & 52.1 &
15.93$\pm$0.06 & 15.12$\pm$0.06 & 13.13$\pm$0.10 & 0.79$\pm$0.08 & -15.44$\pm$0.06 & 2.9 \\
SDSS J131152.03+362858.2 & 13:11:52.0,+36:28:58 & -- & 20.31 & 111 & -- & -- &
-- & -- & -- & -- & -- & 2.5 \\
UGC08303 & 13:13:18.3,+36:12:35 & 945(3) & 22.67 & 124 & 183.4$\times$155.0 & -102.5 &
13.73$\pm$0.03 & 12.96$\pm$0.03 & 10.87$\pm$0.04 & 0.76$\pm$0.05 & -17.64$\pm$0.03 & 14.0 \\ 
\hline
\end{tabular}
\begin{tablenotes}
\item {\it Note.} -- The apparent magnitudes are measured values and are not corrected for 
extinction. $B-R$ and $M_{B}$ are extinction corrected assuming $A_{B}$=0.042 and 
$A_{R}$=0.025 (\citealt{galextinct}). The extinction correction for the NIR is 
assumed to be negligible.
\item[a] Sources: (1) Sloan Digital Sky Survey Data Release 6; 
(2) \cite{arecibohi}; (3) this work.
\item[b] Assuming a distance of 18.8 Mpc.
\item[c] Sizes indicate the diameters in arcseconds of the apertures used to 
measure the reported magnitudes.
\item[d] SDSS DR12 provides an updated measurement of z=0.00405 (v=1215 km s$^{-1}$), 
which is consistent with the velocity range in which this galaxy is detected in the HI data.
\end{tablenotes}
\end{threeparttable}
\end{table}
\end{landscape}

\subsection{NGC~5055}
\label{sec:n5055}

\begin{figure*}
\includegraphics[height=0.9\textheight]{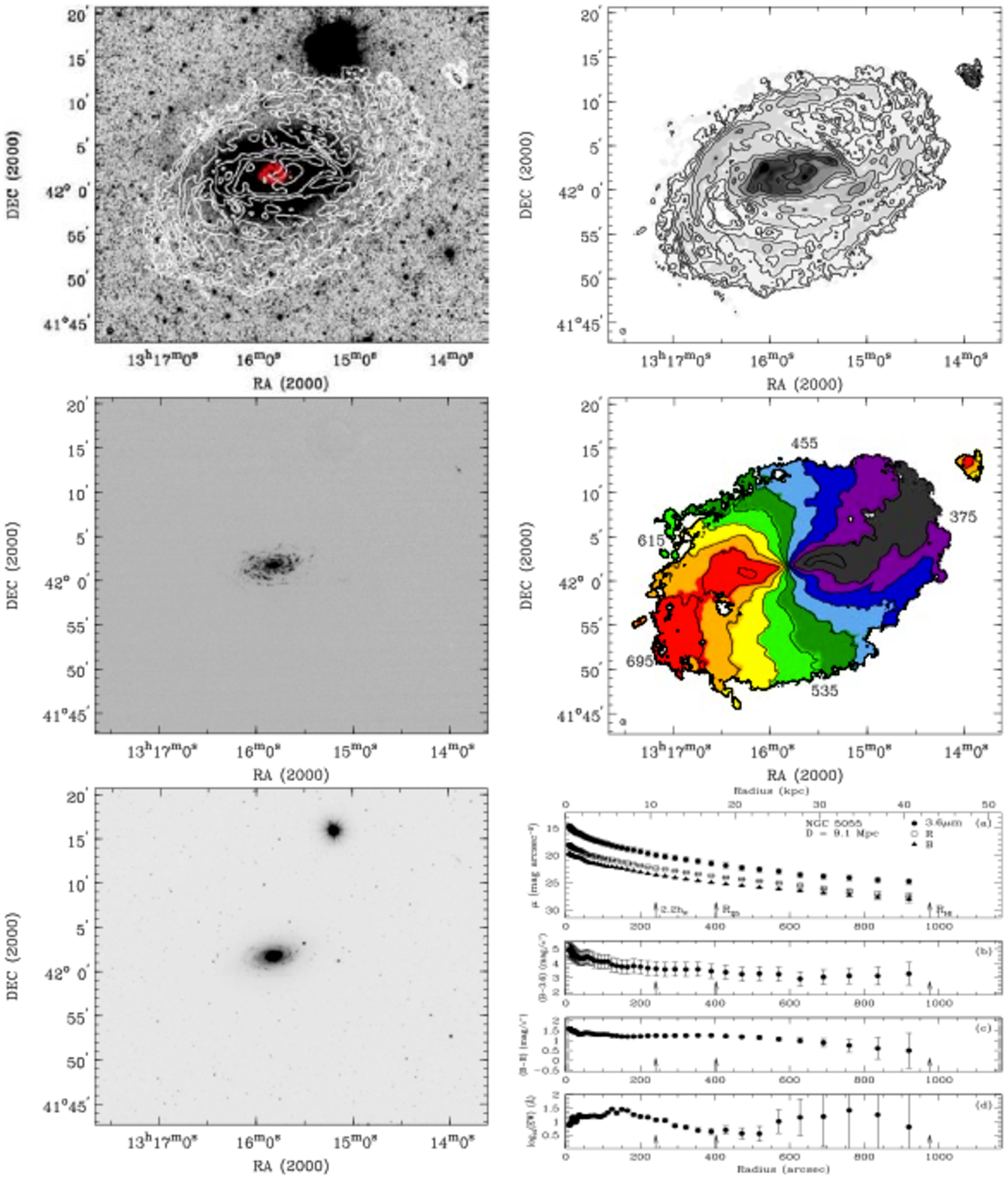}
\caption{{\bf NGC~5055:} {\it (top left)} HI integrated intensity contours from the tapered image
from archival VLA data {\it (white)} and CO intensity contours {\it (red)} from BIMA SONG 
(\citealt{bimasongII}) overlaid on the {\it Spitzer} 3.6$\mu$m image. 
{\it (top right)} HI intensity contours from the tapered image overlaid on the medium 
spatial resolution HI intensity image. The first low resolution HI contour represents a 
column density of 4$\times$10$^{19}$ atoms cm$^{-2}$. 
{\it (middle left)} Narrowband H$\alpha$ image from the WIYN 0.9 m telescope.
{\it (middle right)} HI velocity field with isovelocity contours derived from the tapered spatial 
resolution HI data. The isovelocity contours are spaced every 40 km s$^{-1}$.
{\it (bottom left)} {\it Spitzer} 3.6$\mu$m image with a high surface brightness stretch.
{\it (bottom right)} Ellipse photometry results (see Sections~\ref{sec:3.6data} and~\ref{sec:optdata})
showing radial profiles of surface brightness, $B-3.6$ colour, $B-R$ colour, and equivalent width (EW).}
\label{fig:n5055sum}
\end{figure*}

NGC~5055 (M63) is a well-studied, massive SAbc galaxy. Archival VLA observations originally
published in \cite{n5055hipub} are presented in Fig.~\ref{fig:n5055sum}. The distribution of
neutral gas follows the distribution of the stellar light, but appears to change position
angle slightly beyond the stellar disc. This is more apparent in the HI velocity field where
the isovelocity contours become distorted in the outer disc, especially along the major axis.
Deviations from circular motion within the stellar disc are likely due to streaming motions
along the spiral arms (\citealt{n5055hipub}). The dwarf companion galaxy UGC~08313 to the 
northwest of NGC~5055 is detected in the 3.6$\mu$m  image and the HI observations (see 
Table~\ref{tab:n5055comps}). CO observations from BIMA SONG (\citealt{bimasongII}) display a 
strong central concentration 
surrounded by more diffuse emission which closely follows the distribution of H$\alpha$ emission. 

The central ionized gas kinematics in the SparsePak velocity field (Fig.~\ref{fig:spakvels}) do not 
show any clear deviations from circular motion. The derived ionized gas rotation curve peaks near
200 km s$^{-1}$ at 60 arcsec, then shows about a 20 km s$^{-1}$ drop at 70 arcsec before
the HI rotation curve comes in and rises back near 200 km s$^{-1}$. The average H$\alpha$ rotation 
curve from Fabry-Perot observations in \cite{n5055hakin} shows a similar feature, and agrees well 
with our results. However, \cite{n5055hakin} detected a separate velocity component in the inner 
8 arcsec, which could be associated with a bipolar outflow or almost counter-rotating disc.

Fig.~\ref{fig:rc} shows the rotation curve derived from the HI velocity field. The inclination and 
position angles for the HI vary with radius. The warped outer HI disc can be modeled using an 
increasing position angle. The resulting HI rotation curve shows a drop in rotational velocities 
around 500 arcsec, in agreement with the results of \cite{n5033n5055hikin} and \cite{n5055hipub}.

The observed rotation curve in NGC~5055 is decomposed into stellar bulge and disc and total
gas disc components for baryonic mass (Fig.~\ref{fig:rcdecomp}). The stellar
disc and bulge are near maximal at a fixed M/L = 0.4. The stellar disc M/L can go up to 0.5
without over-shooting the observed inner rotation curve. \cite{dmthings} also came up with
M/L = 0.5 at 3.6$\mu$m for the outer stellar disc in NGC~5055. Like NGC~5033, the mass distribution
in NGC~5055 transitions from baryon-dominated to dark matter-dominated near 2.2$h_{\rm R}$.

\begin{landscape}
\begin{table}
\scriptsize
\centering
\caption{Properties of Known Companions to NGC~5055}
\label{tab:n5055comps}
\begin{threeparttable}
\begin{tabular}{lcccccrcccccc}
Galaxy & RA, Dec & Systemic\tnote{a} & Angular & Physical\tnote{b} & Size\tnote{c} & Position & 
$m_{B}$ & $m_{R}$ & $m_{\rm 3.6}$ & ($B-R$)$_{\rm 0}$ & $M_{B}$\tnote{b} & HI \\
  & (J2000) & Velocity & Separation & Separation & (a$\times$b) & Angle &  &  &  &  &  & Flux \\
  &  & (km s$^{-1}$) & (arcmin) & (kpc) & (arcsec$\times$arcsec) & (deg) &  &  &  &  &  & (Jy km s$^{-1}$) \\
\hline
UGC~8313 & 13:13:53.9,+42:12:34 & 632(1) & 23.96 & 63.4 & 180.0$\times$83.5 & 31.9 &
14.45$\pm$0.03 & 13.49$\pm$0.03 & 11.32$\pm$0.04 & 0.93$\pm$0.04 & -15.35$\pm$0.03 & 5.5 \\ 
LEDA 166159 & 13:13:40.0,+42:02:38 & 371(2) & 24.08 & 63.7 & 60.7$\times$55.8 & -25.6 &
18.51$\pm$0.14 & 17.22$\pm$0.14 & 15.47$\pm$0.29 & 1.27$\pm$0.20 & -11.29$\pm$0.14 & -- \\ 
\hline
\end{tabular}
\begin{tablenotes}
\item {\it Note.} -- The apparent magnitudes are measured values and are not corrected for 
extinction. $B-R$ and $M_{B}$ are extinction corrected assuming $A_{B}$=0.064 and 
$A_{R}$=0.038 (\citealt{galextinct}). The extinction correction for the NIR is 
assumed to be negligible.
\item[a] Sources: (1) this work; (2) \cite{hidist}.
\item[b] Assuming a distance of 9.1 Mpc.
\item[c] Sizes indicate the diameters in arcseconds of the apertures used to 
measure the reported magnitudes.
\end{tablenotes}
\end{threeparttable}
\end{table}
\end{landscape}

\subsection{UGC~08839}
\label{sec:u8839}

\begin{figure*}
\includegraphics[height=0.9\textheight]{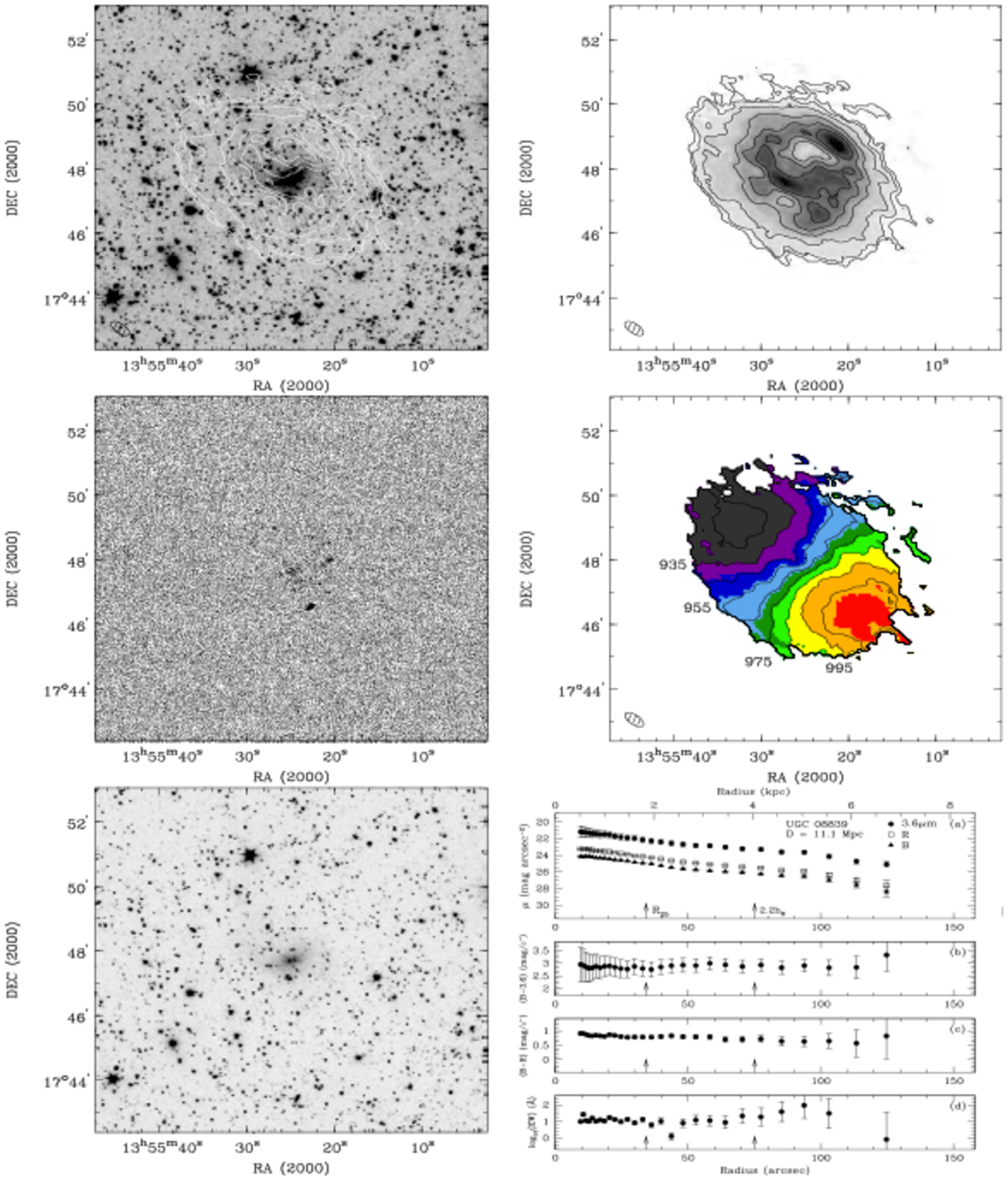}
\caption{{\bf UGC~08839:} {\it (top left)} Low spatial resolution HI integrated intensity contours 
from new VLA data overlaid on the {\it Spitzer} 3.6$\mu$m image. 
{\it (top right)} Low spatial resolution HI intensity contours overlaid on the low 
spatial resolution HI intensity image. The first low resolution HI contour represents a 
column density of 4$\times$10$^{19}$ atoms cm$^{-2}$. 
{\it (middle left)} Narrowband H$\alpha$ image from the WIYN 0.9 m telescope. 
{\it (middle right)} HI velocity field with isovelocity contours derived from the low spatial 
resolution HI data. The isovelocity contours are spaced every 10 km s$^{-1}$. 
{\it (bottom left)} {\it Spitzer} 3.6$\mu$m image with a high surface brightness stretch.
{\it (bottom right)} Ellipse photometry results (see Sections~\ref{sec:3.6data} and~\ref{sec:optdata})
showing radial profiles of surface brightness, $B-3.6$ colour, $B-R$ colour, and equivalent width (EW).}
\label{fig:u08839sum}
\end{figure*}

UGC~08839 is a dwarf irregular galaxy with an extended HI disc 
($D_{\rm HI}$/$D_{\rm 25}$ = 5). The HI integrated intensity map in Fig.~\ref{fig:u08839sum} 
from new VLA observations shows a slight central depression surrounded
by two peaks in intensity. Very little H$\alpha$ emission was detected in the narrowband
image, so UGC~08839 was not targeted for SparsePak observations. The velocity field of 
UGC~08839 shows solid body rotation in with a slight warp at the outer edges where the isovelocity
contours turn over. This warp is modeled as a decrese in inclination angle beyond a radius of 
130 arcsec. The resulting rotation curve rises slowly in the central $\sim$150 arcsec after
which it flattens out to a radius of 210 arcsec (Fig.~\ref{fig:rc}).

The rotation curve decomposition analysis was done using estimates of the
baryonic components from the stellar disc surface density and the atomic HI
gas surface density distributions. The baryonic contributions to the total
observed rotation curve shown in Fig.~\ref{fig:rcdecomp} assuming a fixed
M/L = 0.5 are less than the dark matter halo model contribution at all measured radii.
The contribution from gas overtakes the stellar disc contribution at about 70 arcsec
in the fixed M/L model. This is not unexpected given the visible dominance of the
HI over the stars in terms of radial extent. However, the stellar disc contribution 
can be maximized with a M/L up to 3.0 to describe the observed rotation at inner radii. 
In this maximum disc model, the stellar contribution is greater than or equal to the gas 
contribution at all measured radii. The distribution of mass in this context seems
much less astrophysically plausible, but it provides an upper limit on the total
baryon content in UGC~08839. The maximum baryon mass fraction at 2.2$h_{\rm R}$
derived from the maximum disc decomposition is 0.59, compared to the fixed
M/L decomposition which is only 0.18.

\subsection{NGC~5608}
\label{sec:n5608}

\begin{figure*}
\includegraphics[height=0.9\textheight]{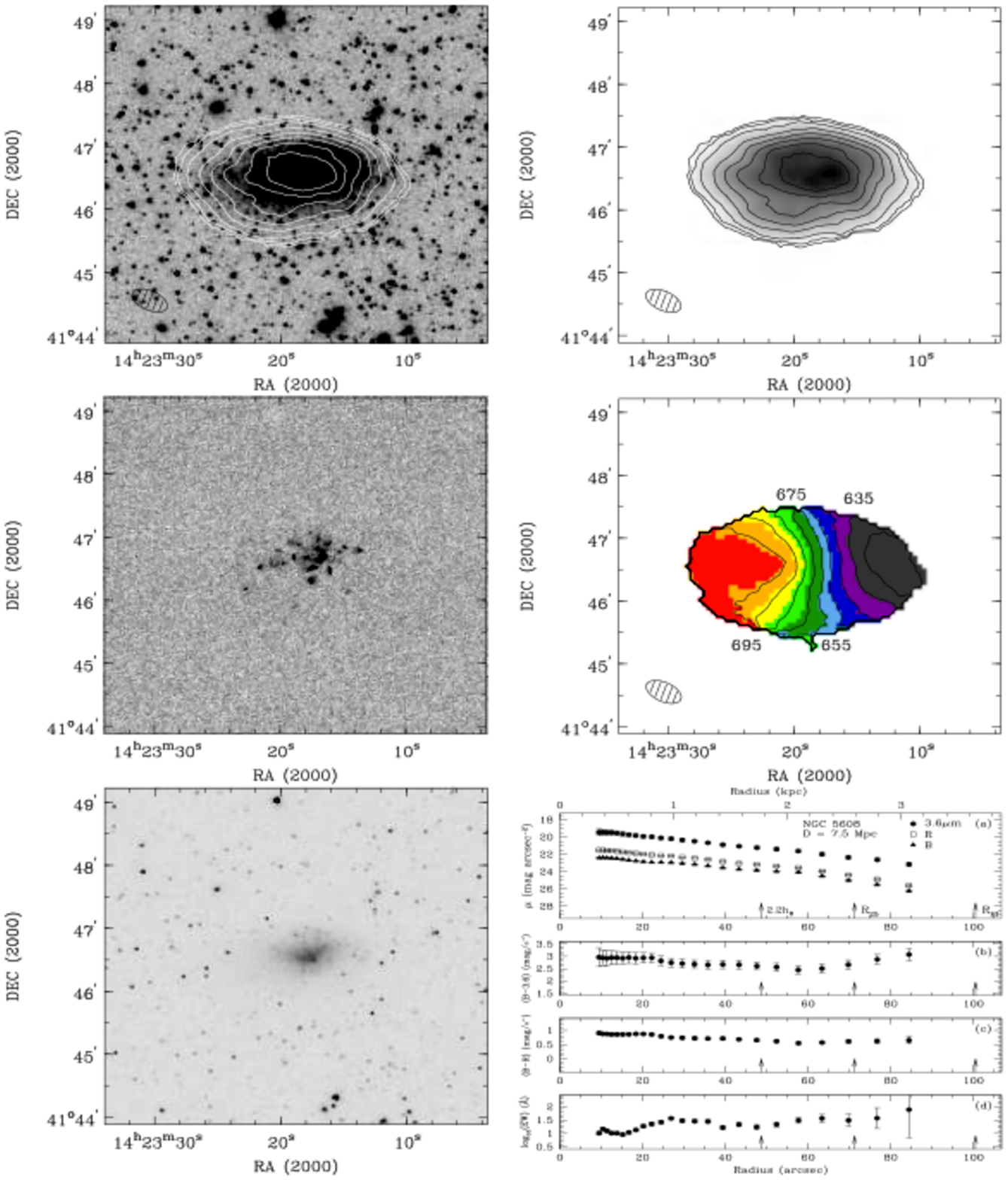}
\caption{{\bf NGC~5608:} {\it (top left)} Low spatial resolution HI integrated intensity contours 
from new VLA data overlaid on the {\it Spitzer} 3.6$\mu$m image. 
{\it (top right)} Low spatial resolution HI intensity contours overlaid on the low 
spatial resolution HI intensity image. The first low resolution HI contour represents a 
column density of 4$\times$10$^{19}$ atoms cm$^{-2}$. 
{\it (middle left)} Narrowband H$\alpha$ image from the WIYN 0.9 m telescope.
{\it (middle right)} HI velocity field with isovelocity contours derived from the low spatial 
resolution HI data. The isovelocity contours are spaced every 10 km s$^{-1}$. 
{\it (bottom left)} {\it Spitzer} 3.6$\mu$m image with a high surface brightness stretch.
{\it (bottom right)} Ellipse photometry results (see Sections~\ref{sec:3.6data} and~\ref{sec:optdata})
showing radial profiles of surface brightness, $B-3.6$ colour, $B-R$ colour, and equivalent width (EW).}
\label{fig:n5608sum}
\end{figure*}

NGC~5608 is a dwarf irregular galaxy for which we acquired new VLA HI observations. The
new data reveal an HI disc with a smooth, featureless morphology similar to the stellar
disc at 3.6$\mu$m (Fig.~\ref{fig:n5608sum}). The highest density of HI gas additionally
appears to coincide with location of the strongest H$\alpha$ observed in the narrowband
image. NGC~5608 was not targeted for ionized gas kinematics from SparsePak observations 
due to the limited spatial extent of the H$\alpha$ emission. The distribution of HI
is slightly asymmetric, which can best be seen in the velocity field where the receding
(east) side of the galaxy appears to extend farther than the approaching (west) side.

Similar to UGC~08839, Fig.~\ref{fig:rc} demonstrates that the rotation curve in NGC~5608
is solid body in the inner parts and then turns over and remains flat beyond 60 arcsec.
The decomposition of the observed HI rotation curve was performed assuming disc
potentials for both the gas and star rotation curve models. Like
UGC~08839, the decomposition for NGC~5608 using fixed M/L = 0.5 results in a total
mass distribution that is dark matter-dominated at all radii and has a significant
contribution from gas at large radii. The maximum stellar disc contribution requires
a M/L = 1.8 and gives a baryon mass fraction at 2.2$h_{\rm R}$ of 0.59, which is the
same as was found for UGC~08839.

\bsp	
\label{lastpage}
\end{document}